\documentclass[11pt,a4paper]{report}

\usepackage{fancyhdr}
\usepackage[hmargin=3cm,vmargin=3.7cm]{geometry}
\usepackage[dvips]{graphicx}
\usepackage{ifthen}
\usepackage{enumerate}
\usepackage{amsmath}
\usepackage{amssymb}
\usepackage[mathscr]{eucal}
\usepackage[errorshow]{tracefnt}
\usepackage{xypic}
\usepackage{amsmath}
\usepackage{amssymb}
\usepackage{array}
\usepackage{amsthm}
\usepackage{eucal}
\usepackage{epsf}
\usepackage{epsfig}
\usepackage[apple mac]{inputenc}

\usepackage{psfrag}
\usepackage{subfigure}
\usepackage{bbm}
\usepackage[dvipsnames,table]{pstricks}
\usepackage[table,dvipsnames]{xcolor}

\RequirePackage{hyperref}

\numberwithin{equation}{section}

\renewcommand{\thesection}{\arabic{section}}

\theoremstyle{definition}
\newtheorem{defi}{Definition}[section]
\newtheorem{ex}{Example}[section]
\newtheorem{theor}{Theorem}[section]
\theoremstyle{plain}
\newtheorem{exc}{Exercise}[section]

\newcommand{\eq}[1]{\begin{equation}#1\end{equation}}
\newcommand{\spl}[1]{\begin{split}#1\end{split}}
\newcommand{\al}[1]{\begin{align}#1\end{align}}
\newcommand{\subeq}[1]{\begin{subequations}#1\end{subequations}}

\def\d{\text{d}}

\newcommand{\slashchar}[1]{\underline{#1}}

\def\Re           {{\rm Re\hskip0.1em}}
\def\Im           {{\rm Im\hskip0.1em}}

\newcommand{\E}{\text{\tiny E}}
\newcommand\bbone{\ensuremath{\mathbbm{1}}}
\newcommand{\ul}{\underline}
\newcommand{\tr}{\text{tr}\,}

\def\calf         {{\cal F}}
\def\caln         {{\cal N}}
\def\cali         {{\cal I}}

\begin{document}

\fancypagestyle{beginsection}{%
\fancyhead{}
}

\thispagestyle{empty}
\begin{flushright}
KUL-TF-10/04
\end{flushright}

\begin{center}

\vspace*{2cm}

\noindent
{\LARGE\textsf{\textbf{Lectures on Generalized Complex Geometry for Physicists}}}
\vskip 2truecm

\begin{center}
{\large \textsf{\textbf{Paul Koerber}}} \\
\vskip 1truecm
        {\it   {Instituut voor Theoretische Fysica, K.U.\ Leuven\\
                Celestijnenlaan 200D, 3001 Leuven, Belgium}\\
        [3mm]e-mail:} {\tt koerber at itf.fys.kuleuven.be} \\
\end{center}
\vskip .5 cm

\small{Based on lectures given by the author
at the Center for Quantum Space-Time at Sogang University in Seoul in August 2007,
and at the Fourth International Modave Summer School on Mathematical Physics, held in Modave, Belgium, in
September 2008.}

\vskip 1 cm
\centerline{\sffamily\bfseries Abstract}
\end{center}

\noindent
In these lectures we review Generalized Complex Geometry and discuss two main applications
to string theory: the description of supersymmetric flux compactifications and the supersymmetric
embedding of D-branes. We start by reviewing $G$-structures, and in particular SU(3)-structure and
its torsion classes, before extending to Generalized Complex Geometry. We then discuss the supersymmetry
conditions of type II supergravity in terms of differential conditions on pure spinors, and
finally introduce generalized calibrations to describe D-branes. As examples we discuss in some detail
AdS$_4$ compactifications, which play a role as the geometric duals in the AdS$_4$/CFT$_3$-correspondence.

\clearpage

\tableofcontents
\clearpage

\pagestyle{fancy}
\fancyhead{}
\fancyhead[c]{\bf \rightmark}


\thispagestyle{beginsection}
\section{Introduction}

\subsection{Motivation}

\subsubsection*{Flux compactifications}

To connect string theory to real-world physics we must compactify it from ten dimensions (10D)
to four dimensions (4D). The low-energy theory of string theory is supergravity, which brings us
to the following simplification, namely to address the compactification at the level of supergravity.
One has then to take care that the compactifications that are constructed fall in a regime where the supergravity description
is valid. Later on this can be (partially) amended by considering various perturbative and non-perturbative corrections.

So we are looking for a solution of supergravity whose geometry
looks like a product of four uncompactified external dimensions and six compact internal dimensions.
In fact, for two reasons it makes sense to look for such a solution that is in addition {\em supersymmetric}.
The first is that the supersymmetry conditions provide a comparatively easy way to obtain solutions to
the full equations of motions. This is because the supersymmetry conditions are much easier than
the supergravity equations of motion themselves, while at the same time it can be shown that solutions to these supersymmetry conditions ---
if completed with the Bianchi identities for the form fields of supergravity --- automatically provide solutions
to the full equations of motion. The second reason is that for phenomenological reasons, like the hierarchy
problem, supersymmetry is expected to be broken at a much lower scale than the compactification scale. Of course,
after one has constructed a supersymmetric solution to the supergravity equations, the breaking of
supersymmetry remains an important problem, which is under intensive study and for which multiple scenario's are
proposed. This also includes trying to construct models with positive cosmological constant, which automatically
breaks supersymmetry.

Now, it turns out that putting the form fields of supergravity to zero the class of six-dimensional (6D) manifolds that satisfies the supersymmetry conditions
is exactly the class of Calabi-Yau manifolds (including trivial ones like tori). Therefore, much research went in the study
of compactifications on Calabi-Yau manifolds. It turns out that such compactifications have a lot of massless scalar fields
in the 4D effective theory, which roughly describe the shape and size of the
Calabi-Yau or the value of the form potentials integrated over cycles in the Calabi-Yau manifold.
These massless fields, called {\em moduli}, are unobserved in nature. A fruitful mechanism to render these fields very massive, so that
they cannot be excited anymore in the low-energy 4D theory, called
{\em lifting the moduli}, is to turn on expectation values of the various form field strengths of supergravity, that is
to introduce {\em fluxes}\footnote{For reviews on compactifications with fluxes see \cite{fluxrev1,fluxrev2,fluxrev3,fluxrev4,fluxrev5}.}. This
however also complicates the supersymmetry conditions. In a specific case, type IIB with D3/D7-branes and O3-planes,
the geometry is still a Calabi-Yau manifold, although only up to a rescaling of the metric (one calls this a {\em
conformal Calabi-Yau manifold}). See e.g.~\cite{gkp,kklt} for seminal work on this configuration. There are also related models in F-theory compactified
on an eight-dimensional Calabi-Yau manifold.
Because this is the setup that is closest to the fluxless case, it is the one that so far has been most intensively researched and best understood. In order to obtain an understanding of the whole
string theory landscape, however, it is important to study the general case. It turns out that generically
solutions with fluxes are far removed from the Calabi-Yau limit. This is where Generalized Complex Geometry, pioneered in \cite{hitchinGCY,gualtieri}, comes in.
It is a unification and generalization of the language of complex and symplectic geometry --- a Calabi-Yau manifold is both --- which seems to be natural to describe
supersymmetric compactifications of supergravity with fluxes \cite{granaN1}. It also naturally describes the supersymmetric embedding of D-branes
with world-volume gauge flux into these backgrounds. The goal of this review is to describe both of these applications, which will be the topic of
respectively chapter \ref{sugrasusycond} and chapter \ref{branesection}.

Apart from the supergravity approach, there has also been a lot of work applying Generalized Complex Geometry
to describe the world-sheet theory of a string in a curved target space with non-trivial NSNS three-form, see
e.g.~\cite{lindstromWS1,lindstromproof,lindstromgenkal,lindstromgenpot,tomasiellogenpot}, making the connection
with the bihermitian geometry of \cite{ghr}. For work on supersymmetric D-branes in this setup see \cite{alexDbranes1,alexDbranes2},
and for work on a topological world-sheet model with three-form flux \cite{kapustinlitopo,zucchinitopo,pestuntopo,hulltopo}.
Necessarily, the world-sheet approach, which is based on the RNS action, cannot take
into account RR-fields. In this review, we will not focus on this aspect, but instead refer to the review \cite{zabzinereview}.

\subsubsection*{AdS/CFT}

Generalized Complex Geometry also proves to be fruitful in constructing and studying AdS-solutions,
which can be considered as the geometric duals in the AdS/CFT-correspondence. The construction of AdS compactifications
with fluxes is easier than Minkowski or dS compactifications, because in this case the
no-go theorem of Maldacena-N\'u\~nez \cite{malnun} (see theorem \ref{nogomalnun}) does not apply. There is therefore no need
to introduce localized orientifold planes, which complicate the compactification. We will discuss examples in the context of the AdS$_4$/CFT$_3$
correspondence in section \ref{sec:AdS4SU3ex}. For work in the context of AdS$_5$/CFT$_4$ see~\cite{gengeomAdSCFT1,gengeomAdSCFT2,gauntlettAdSCFT}.

\subsubsection*{Structure of the review}

In the next subsection we briefly setup the stage by discussing the supersymmetry conditions in the fluxless case.
To discuss these same conditions in full generality in the presence of fluxes we will first have to delve more deeply
into mathematics and introduce the formalism of Generalized Complex Geometry. It will be useful
to start with reviewing the most important concepts of ordinary complex geometry, $G$-structures, torsion classes and Calabi-geometry
in chapter \ref{SU3}. Albeit we will introduce these concepts in perhaps a slightly unfamiliar way such that they can be readily generalized,
the reader familiar with them can quickly glance over chapter \ref{SU3}, and move to chapter \ref{gengeomfund}, where we
introduce the fundamentals of Generalized Complex Geometry. In the next two chapters we will discuss two main applications to string theory:
the study of the background supersymmetry conditions with fluxes in chapter~\ref{sugrasusycond},
and the D-brane supersymmetry conditions in chapter~\ref{branesection}.

Each chapter concludes with a number of exercises. It provides the readers with a chance to get their hands dirty and
also releases me from the burden of giving many proofs. They are classified from ``easy'' (just toying around with the equations)
to ``hard'' (can be more extensive and/or contain new conceptual insights). For the ``intermediate'' and ``hard'' exercises there
is usually a reference to the literature where the solution can be found.

\subsection{Invitation: supersymmetry conditions in the fluxless case}

As a warm-up we will study the conditions for unbroken supersymmetry in compactifications without fluxes and see how
the condition for a Calabi-Yau geometry comes about. This is also excellently reviewed in a lot more detail in \cite[Chapter 15]{GSWII}.

The bosonic sector of type II supergravity (see appendix \ref{sugra} for a brief review) contains, next to
the metric and the dilaton, a bunch of form fields, which come from both the NSNS- and the RR-sector of
string theory. Putting the vacuum expectation values of these fields --- the so-called {\em fluxes} --- to zero,
it turns out that a state with some unbroken supersymmetry solves the equations of motion. This is well-known
for theories with global supersymmetry, where the Hamiltonian can be written as a sum of squares and possibly
a topological term. The squares vanish precisely when there is unbroken supersymmetry so that the supersymmetric
configuration is a global minimum within its topological class (since the topological term does not vary within this
class). The statement is more subtle for local supersymmetry, but it still holds for supergravity in the fluxless case. For
the case with fluxes we will provide the exact statement in chapter \ref{sugrasusycond} (theorem \ref{theor:integr}).
The bottom line is that one can construct solutions to the full supergravity equations of motion by solving the relatively
simpler supersymmetry conditions.

Let us proceed with such fluxless compactifications to 4D Minkowski space.
This means we split the total 10D
space-time in a 4D uncompactified part with flat Minkowski metric and an internal part with --- to be determined --- curved metric.
Then we follow the strategy of constructing solutions by imposing unbroken supersymmetry. This means that
there must be some fraction of the supersymmetry generators for which the supersymmetry variations of all the fields vanish.

It turns out that the variations
of the bosonic fields always contain a fermionic field. Therefore, if we put the vacuum expectation values of all the fermionic
fields to zero in our background, the variations of the bosonic fields automatically vanish. In fact, we must put the
vacuum expectation values of all fermionic fields to zero, since they
would not be compatible with the compactification ansatz, which requires
4D Poincar\'e symmetry in the four uncompactified dimensions. To understand this let us consider how the structure
group reduces under the compactification ansatz. We will give a precise definition of the structure group in section \ref{langstruc},
but for now let us mention that the structure group is the group of transformations of the fields between different local patches of the manifold.
The important point is that in the presence of a (Minkowskian) metric the structure group reduces to Spin(9,1) (the universal cover
of SO(9,1)), and with our compactification ansatz further to Spin(3,1)$\times$Spin(6). Now, the spinor representation of Spin(9,1)
reduces into a sum of two products of a 4D and a 6D spinorial representation (we display this reduction explicitly in eq.~\eqref{adskilling}).
The important part is that the reduction does not contain an invariant under Spin(3,1), so that
a vacuum expectation value for a spinorial field would transform non-trivially
under Spin(3,1) and thus break the Poincar\'e invariance.

What remains to consider are the supersymmetry variations of the fermions, which are the two gravitino's and the two dilatino's.
These variations are displayed in \eqref{susyvar} and in the absence of fluxes become simply
\al{
\delta\psi^{1,2}_M & = \nabla_M \epsilon^{1,2} \,  , \\
\delta\lambda^{1,2} & = \slashchar{\partial}\Phi \, \epsilon_{1,2} \, .
}
The supersymmetry conditions require that there exist $\epsilon^{1,2}$ so that $\delta\psi^{1,2}_M=\delta\lambda^{1,2}=0$.
The number of such $\epsilon^{1,2}$ determines the number of supercharges and thus the amount of 4D supersymmetry of the
configuration. Looking at it more closely, these supersymmetry conditions imply conditions of topological and differential type.

The {\em topological condition} requires that there exist globally defined everywhere non-vanishing spinors in the first place.
This is not a problem for the flat uncompactified 4D part, but
becomes a non-trivial condition on the structure group of the 6D internal part. Now suppose that the
internal part of $\epsilon^{1}$ and $\epsilon^2$ is the same and call it $\eta$. Since $\eta$ is globally defined it must
be the same in two different local patches, and thus invariant under the transition functions making up
the structure group. Suppose $\eta$ is a Weyl spinor and for definiteness suppose it is of positive chirality, $\eta=\eta_+$.
Now, Spin(6) is isomorphic to SU(4), and the positive chirality spinor representation is the fundamental $\mathbf{4}$ of the latter.
If we choose a basis so that our invariant spinor takes the form,
\eq{
\eta_+= \left(\begin{array}{c} 0\\0\\0\\\eta_0\end{array}\right) \, ,
}
then we immediately read off that the transformations leaving $\eta$ invariant form the SU(3)
subgroup of SU(4) of the form
\eq{
\left( \begin{array}{cc} U & \mathbf{0}_{3\times 1} \\ \mathbf{0}_{1\times 3} & 1 \end{array} \right) \, , \qquad U \in \text{SU(3)} \, .
}
We conclude that the topological condition boils to down to the requirement that the structure group of the internal space
is reduced from Spin(6) to SU(3).

The {\em differential part} of the supersymmetry conditions follows from putting $\delta\psi^{1,2}_M=\delta\lambda^{1,2}=0$
for our globally defined spinors. The first equation implies that $\eta_+$ is covariantly constant so that the internal space not only has SU(3)-structure,
but in fact SU(3)-{\em holonomy}. Indeed, if we parallel transport $\eta_+$ around a loop in the internal space it
is constant. Now the holonomy group is the group of transformations that one gets after parallel transport around all loops.
It follows that this group must leave $\eta_+$ invariant and, following the same reasoning as above, thus reduces to SU(3). A manifold
with SU(3)-holonomy is called a {\em Calabi-Yau} manifold. We stress
that the holonomy group depends on the metric, since parallel transport and covariant derivatives depend on the metric, while
the structure group is independent of the metric. The second equation implies that the dilaton is constant.

As for the 4D part of the supersymmetry generator, the supersymmetry conditions only impose that it be constant,
so we have the freedom of choosing a constant 4D spinor for both $\epsilon^{1,2}$, which leads, after properly taking into account
the 10D Majorana condition, to eight real supercharges. This implies $\mathcal{N}=2$ supersymmetry in four dimensions. From \eqref{susyvar}
one can see that in the presence of RR-fluxes the supersymmetry conditions relate $\epsilon^{1}$ and $\epsilon^2$ so that the 4D
parts cannot be chosen independently anymore. Therefore, a compactification with RR-fluxes has only $\mathcal{N}=1$ supersymmetry in four dimensions.
There can be more supersymmetry, but this leads to stronger constraints.

We can characterize an SU(3)-structure in an alternative way. The existence of a globally defined nowhere-vanishing spinor $\eta_+$ allows for the construction of a globally defined real two-form and complex three-form as follows
\eq{
|\eta_+|^2 \omega_{ij} = i \eta_+^\dagger \gamma_{ij} \eta_+ \, , \qquad |\eta_+|^2 \Omega_{i_1i_2i_{3}} = \eta_-{}^\dagger \gamma_{i_1i_2i_{3}} \eta_+ \, ,
}
where $\eta_-$ is the complex conjugate and we normalized using $|\eta_+|^2=\eta_+^\dagger \eta_+$. These forms satisfy the compatibility and normalization condition
\eq{
\label{compat}
\omega \wedge \Omega =0 \, , \qquad \Omega \wedge \bar{\Omega} = \frac{8 i}{3!} \omega^3 = 8 i \, \text{vol}_6 \, .
}
Furthermore $\Omega$ is a {\em decomposable} form, which means that it can be written as a wedge product of three one-forms.
Conversely, a real two-form and a decomposable complex three-form satisfying \eqref{compat} define a metric,
and if this metric is positive-definite they also completely define an SU(3)-structure.

The differential requirement of SU(3)-holonomy is in this alternative picture simply
\eq{
\d \omega = \d \Omega = 0 \, .
}
$\omega$ and $\Omega$ are then called the K\"ahler form and the holomorphic $(3,0)$-form.

It turns out that if we want to embed supersymmetric D-branes (without world-volume gauge flux) in such a
Calabi-Yau background they must wrap complex or special Lagrangian submanifolds. These are both examples of {\em calibrated submanifolds}
with respect to respectively
\eq{
 \frac{1}{l!} \omega^l \, , \qquad \Re \left(e^{i \alpha} \Omega\right) \, ,
}
where $0\le l \le 3$, and $e^{i\alpha}$ is a constant phase.

The goal of these lectures is to extend this analysis to the case with NSNS- and RR-fluxes as well as world-volume
fluxes on the D-branes. If one wants to study the most general case (where there are different internal spinors $\eta^{(1,2)}$
in $\epsilon^{1,2}$) the appropriate framework is Generalized Complex Geometry. But also in the less general case it provides a unifying
framework, putting complex geometry (based on $\Omega$) and symplectic geometry (based on $\omega$) on the same footing.

\clearpage

\thispagestyle{beginsection}
\section{From $G$-structures to Calabi-Yau geometry}
\label{SU3}

The following two chapters will be devoted to introducing the mathematical
machinery, which we will apply for describing the supersymmetry conditions
of supergravity and D-branes in the presence of fluxes
in chapters \ref{sugrasusycond} and \ref{branesection}. The supersymmetry conditions of supergravity split into a topological
condition, the existence of a certain structure, and a differential condition, related to the integrability of this structure.

In this chapter we review $G$-structures. More in particular we will work our way up towards
describing a SU($d/2$)-structure, and eventually we will put $d=6$. Our angle is perhaps
a little non-standard, but will be geared towards the generalization to
Generalized Complex Geometry, which is the subject of the next chapter.

What makes Generalized Complex Geometry a little confusing is that there are
three pictures in which one can express roughly the same object. For instance,
a generalized complex structure that is part of a SU($d/2$)$\times$SU($d/2$)-structure
can alternatively be described as a polyform, and also as a bispinor. We will first explain these three points of view ---
structures, polyforms, and spinors --- for the simpler case of a SU($d/2$)-structure.

\subsection{The language of structures}
\label{langstruc}

To explain what structures are, we start with the definition of the structure group. But
before we come to that we need to introduce some concepts from the theory of fiber bundles.
We will be rather sketchy and refer for more details to e.g.~\cite{nashsen,nakahara} (physics-oriented)
or \cite{kobanomi1,joycebook} (mathematics-oriented).

Let us consider a compact manifold $M$, later on it will take the role of the internal manifold
of the supergravity compactification. For now, we keep the dimension $d$ of $M$ general. A bundle $E$
on $M$ with fiber $F$ is a manifold itself, which looks locally like a product of the base, which is $M$, and the fiber.
It comes equipped with a smooth projection $\pi$ to the base. Furthermore, there are {\em transition
functions}, which describe how the fiber transforms between two patches $U_{\alpha}$ and $U_{\beta}$
of the base $M$ so that globally it is (generically) {\em not} a trivial product. If the fiber is
a vector space then the bundle is called a {\em vector bundle}. The bundles of interest to us are the {\em tangent
bundle} $TM$, with fiber in a point $p$ the space of tangent vectors $T_p M$, its dual the {\em cotangent
bundle} $T^*M$ with fiber $T_p^*M$, and tensor products of these. A {\em section} $s$ of a bundle $E$ assigns
an element of the fiber to every point of $M$. More formally, a section is a smooth map $s:M \rightarrow E$ such that $\pi(s(p))=p$.
We denote
\eq{
\Gamma(E) : \quad \text{space of sections of } E \, .
}
The sections of the tangent bundle are called {\em vector fields} and of the cotangent bundle {\em one-forms}.

For every vector bundle over $M$ it is possible to define a frame bundle. In particular, for the definition of the
structure group we need the tangent frame bundle, for which we will illustrate some of the concepts we discussed above.
\begin{defi}
The {\em tangent frame bundle} $FM$, associated to the tangent bundle $TM$, is the bundle over
the manifold $M$ with fiber in each point $p \in M$ the set of ordered bases of the
tangent space $T_p M$.
\end{defi}
Locally --- in the language of fiber bundles this means on a patch $U_\alpha$ of $M$ --- the elements
of the bundle just look like a product of base and fiber: $(p,e_a)$ with $p \in U_\alpha$
and $e_a=e^i{}_a \frac{\partial\,\,\,\,}{\partial x^i}$ for $a=1,\ldots,d$ a set of $d$ independent vectors
forming a base of $T_p M$, i.e.\ a local frame. A description like that in each patch is called a {\em local trivialization}.
The $a$-indices are naturally acted upon by
the group GL(d,$\mathbb{R}$), the group of general linear transformations or equivalently
the group of real invertible $d\times d$-matrices.

Let us consider now two different patches $U_\alpha$ and $U_\beta$ with
local trivializations $(p,e_a)$ and $(p,e'_a)$ respectively. On the overlap of these patches
we find the following relation inherited from the associated tangent bundle:
\eq{
e'{}^i{}_a = \frac{\partial x'{}^i}{\partial x^j} e^j{}_a \, ,
}
which can be converted into the action of an element $t_{\beta\alpha}(p) \in$GL(d,$\mathbb{R}$) acting from the right
\eq{
e'{}^i{}_a =  e^i{}_b (t_{\beta\alpha})^b{}_a \, ,
}
The $t_{\beta\alpha}(p)$ are called the {\em transition functions} and contain all the information
about the non-trivial topology of the bundle. They must satisfy the consistency conditions
\eq{
t_{\alpha\beta} t_{\beta\alpha}=1 \, ,
}
and on the triple overlap $U_{\alpha} \cap U_{\beta} \cap U_{\gamma}$,
\eq{
t_{\alpha\beta}t_{\beta\gamma}=t_{\alpha\gamma} \, .
}
This gives the set of transitions functions the properties of a group. The group of transition functions is called the
{\em structure group}, which is in this case GL(d,$\mathbb{R}$). Note that the elements $e^i{}_a$ of the fiber can
themselves be considered as $d\times d$-matrices and thus elements of GL(d,$\mathbb{R}$). The frame bundle is thus
an example of a principal bundle, where the fiber is the structure group. Furthermore, while we have here constructed
the frame bundle from the tangent bundle, we can also take the other way round, i.e.\ given the frame bundle with
local trivializations $(p,e_a)$ in the different patches, we can construct an associated vector bundle with local trivializations $(p,v^a)$
and transition functions such that $v=e_a v^a$ is invariant
\eq{
v'^{a} = (t_{\beta\alpha})^{-1}{}^a{}_b v^b \, .
}

We will be interested in cases where by an appropriate choice of local frame in the different patches
we can introduce a {\em reduced} tangent frame bundle such that the structure group reduces to a proper subgroup $G \subset$GL(d,$\mathbb{R}$).
Whether this is possible depends on the topological properties of the
manifold $M$. So after this extended introduction we are finally ready to give the definition of a $G$-structure.
\begin{defi}
A manifold $M$ has a $G$-structure with $G \subset$GL(d,$\mathbb{R}$) if it is possible to reduce
the tangent frame bundle such that it has structure group $G$.
\end{defi}
An extreme example is the case
where one is able to find a global section of the tangent frame bundle. By appropriate changes of local frames, we can then choose
the $e_a$ so that this section takes everywhere the same form $s^a{}_b \in$GL($d$,$\mathbb{R}$).
The only transition function that preserves this form is the identity so
that the structure group is the trivial group consisting of only the identity element. The manifold $M$
is then a {\em parallelizable} manifold. Prominent examples of parallelizable manifolds are Lie-groups.
Another example is the seven-sphere $S^7$.
\begin{figure}[t]
\centering
\subfigure[Structure group O($d$)]{
\psfrag{Ua}{$U_\alpha$}
\psfrag{Ub}{$U_\beta$}
\psfrag{ea}{$e_a$}
\psfrag{eb}{$e'_a$}
\psfrag{Sod}{O($d$)}
\includegraphics[width=6cm]{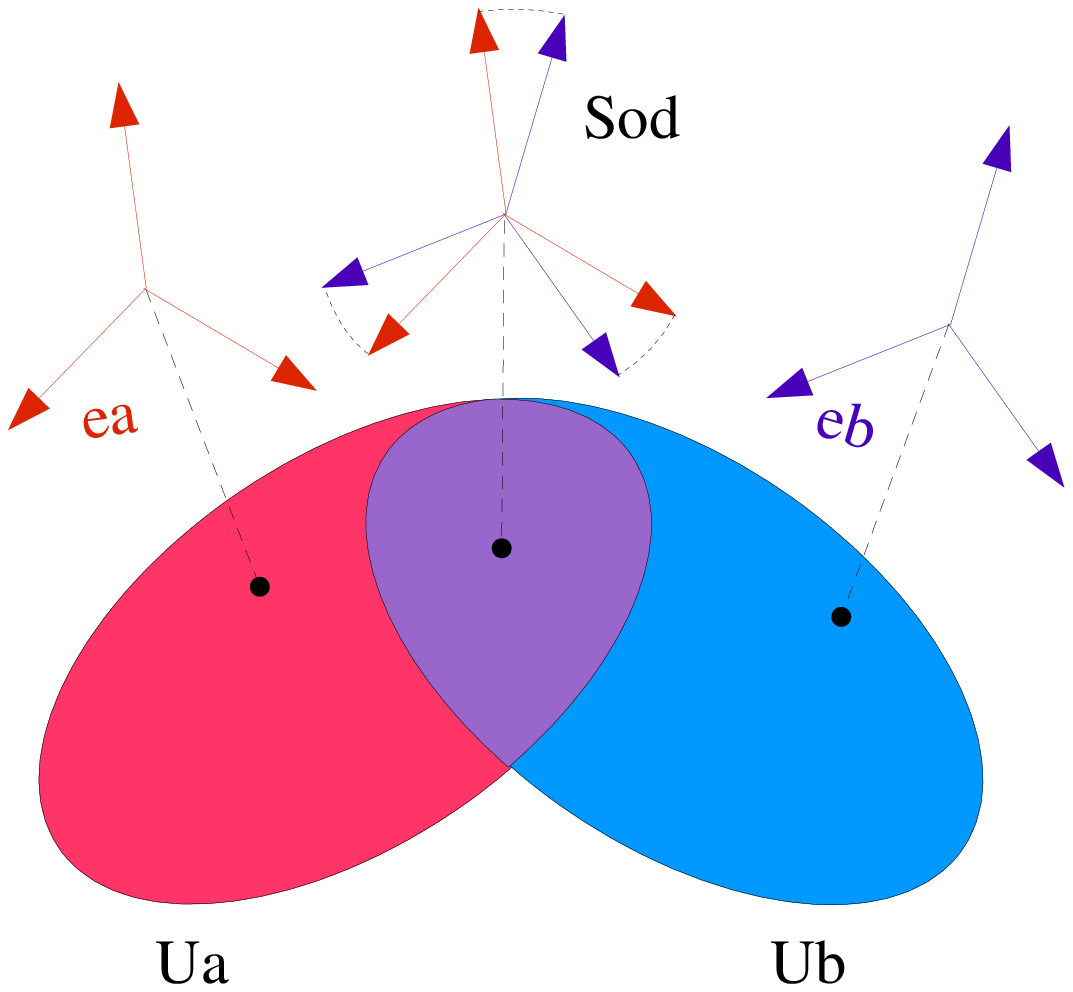}
}\hspace{0.2cm}
\subfigure[A globally defined vector reduces the structure to O($d-1$)]{
\psfrag{z}{$v$}
\psfrag{Ua}{$U_\alpha$}
\psfrag{Ub}{$U_\beta$}
\psfrag{ea}{$e_a$}
\psfrag{eb}{$e'_a$}
\psfrag{Sod}{O($d-1$)}
\includegraphics[width=6cm]{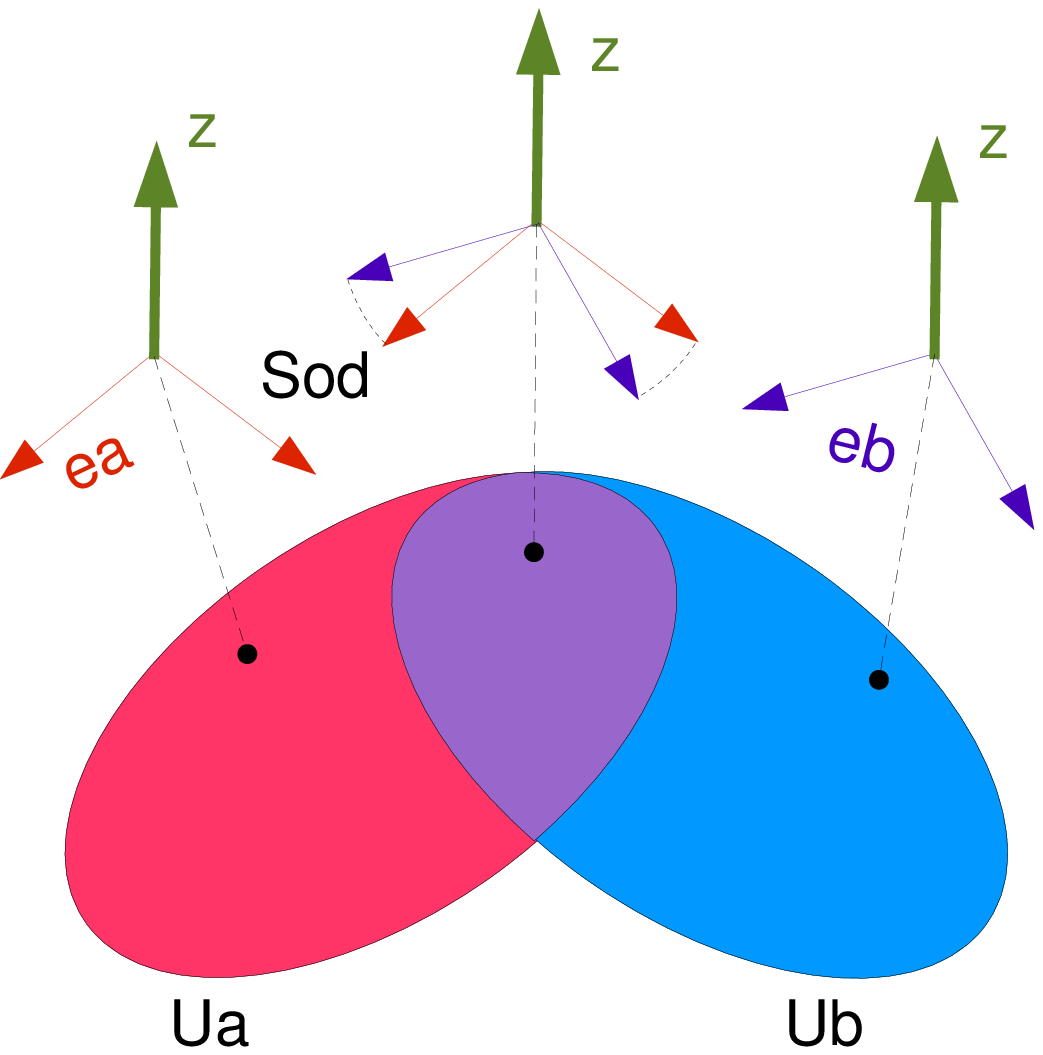}
}
\caption{A set of non-degenerate tensors describes a $G$-structure. On the left: in the special case of the figure we
assume that the structure group is already reduced to O($d$) (see example \ref{metricex}).
On the right: an everywhere non-vanishing vector field $v$ is introduced. Because of the existence of this vector field it is possible to construct a reduced frame bundle, where
on the overlap between the patches only the rotations that leave the vector invariant are
allowed as transition functions, i.e.\ (proper and improper) rotations in a plane orthogonal to the $v$-axis, making up O($d-1$).
The figure is inspired by a similar one from a talk by Davide Cassani.}
\label{structurefig}
\end{figure}

A convenient way to describe a $G$-structure, used a lot by physicists, is via one or
more $G$-invariant tensors --- or spinors as we will see later --- that
are globally defined on $M$ and non-degenerate. Indeed, since these objects are globally defined it is possible to choose
frames $e_a$ in each patch so that they take exactly the same form in all patches. It follows that only those transition functions that
leave these objects invariant are allowed and the structure group reduces to $G$ or a subgroup thereof, see figure \ref{structurefig}.

Note that, typically, such a set of $G$-invariant tensors is not unique, so that there are several descriptions
of the same $G$-structure. Furthermore, it is possible that these tensors are actually invariant under a larger group $G'$, in which
case one can add more tensors to more accurately describe the $G$-structure. The $G$-invariant tensors can be found in a systematic way
using representation theory. Indeed, one should decompose the different representations of GL($d$,$\mathbb{R}$), in which a tensor on $M$
transforms, into irreducible representations of $G$ and scan for invariants. These invariants will then correspond to non-degenerate $G$-invariant
tensors.

If the $G$-structure is already reduced to SO($d$) (see example \ref{metricex}) and the manifold is spin, which means
one can lift the SO($d$) in the transition functions to its double cover Spin($d$) in a globally consistent way, we can also consider spinor bundles.
We will especially be interested in invariant spinors since they are needed to construct the generators of unbroken
supersymmetry.

\subsubsection*{Examples}

After these abstract definitions let us give the main examples of interest:
\begin{ex}[Metric and orientation]
\label{metricex}
If there is a globally defined symmetric two-tensor $g \in \Gamma(S^2(T^* M))$ that
is positive-definitive --- which implies in particular that it is non-degenerate --- the structure group
reduces to $G=$O(d,$\mathbb{R}$), the group of orthogonal $d\times d$-matrices. $g$ is then a {\em metric} and the manifold is called
a {\em Riemannian} manifold. If there is furthermore a globally defined volume-form vol$_{d}$ associated to the metric,
the manifold is orientable and the structure group further reduces to SO(d,$\mathbb{R}$).
\end{ex}
\begin{ex}[Almost complex structure]
Suppose we have a globally defined tensor $J \in \Gamma(T M \otimes T^* M)$
or equivalently a map
\eq{
J : T M \rightarrow T M \, ,
}
which respects the bundle structure, i.e.\ $\pi(Jv)=\pi(v)$ for all $v \in T M$.
Furthermore it must be such that
\eq{
\label{prodstruc}
J^2 = - \bbone \, ,
}
which implies that it is non-degenerate. Then the structure group
reduces to GL($d/2$,$\mathbb{C}$), the group of complex $d/2\times d/2$-matrices, and $J$ is called an {\em almost
complex structure}. It is easy to show that such an almost complex
structure can only exist if $d$ is even. From $J^2=-\bbone$, it follows that the action of $J$ on $T_p M$ has eigenvalues $+i$ and $-i$.
A subtlety is that to allow for complex eigenvalues and eigenvectors we have to consider the complexification\footnote{The complexification $V \otimes \mathbb{C}$ of a vector space $V$ is obtained by extending
the scalar multiplication, i.e.\ number times vector, from the real to the complex numbers. For a bundle, one complexifies the fiber in each point.} of the tangent bundle, namely $T M \otimes \mathbb{C}$.
We denote the subbundles with fibers respectively the $(+i)$- and $(-i)$-eigenspaces in each point
with
\eq{
L,\bar{L} \subset T M \otimes \mathbb{C} \, .
}
Since $J$ is smooth, one can in each patch define two bases of vector fields, spanning respectively $L$ and $\bar{L}$.
Note that these bases are generically not globally defined. Indeed, between patches there are still transition functions
mixing the individual basis vector fields, but, since $G$ preserves $J$, they will preserve
the decomposition in $L$ and $\bar{L}$. Such subbundles of the tangent bundle that are locally spanned by smooth vector fields
are called {\em distributions}.
Since $J$ is real, it follows that if $v \in L$ then $\bar{v} \in \bar{L}$, justifying the notation $\bar{L}$. This implies in particular that $L$ and $\bar{L}$ are subbundles of $T M \otimes \mathbb{C}$ of equal rank\footnote{The rank of a bundle is the dimension of the fiber.}.
\end{ex}
\begin{ex}[Pre-symplectic structure]
If there is a globally defined non-degenerate two-form $\omega \in \Gamma(\Lambda^2 T^*M)$, the structure
group reduces to Sp($d$,$\mathbb{R}$). Note that $\omega$ is non-degenerate if and only if (iff)
\eq{
\omega^{d/2} \neq 0 \, .
}
\end{ex}
\begin{ex}[Hermitian metric]
A metric $g$ and an almost complex structure $J$ satisfying the compatibility condition $J^i{}_k g_{ij} J^j{}_{l}=g_{kl}$
--- also called {\em hermiticity} --- reduce the structure group to U($d/2$), the group of unitary $d/2 \times d/2$-matrices. In fact, it follows from the hermiticity that
\eq{
\label{sympfromg}
\omega_{ij}= g_{ik}J^k{}_j
}
is anti-symmetric so that a pre-symplectic structure is automatically present.

Conversely, an almost
complex structure $J$ and pre-symplectic structure $\omega$ satisfying the compatibility condition
\eq{
\label{hermiticity}
J^i{}_k \omega_{ij} J^j{}_l = \omega_{kl} \, ,
}
define a metric
\eq{
\label{metricJom}
g_{ij}=-\omega_{ik} J^k{}_j \, .
}

Finally, a pre-symplectic structure and a metric also define an almost complex structure. We conclude that two out of the three structures imply the third.
\end{ex}
\begin{ex}[Almost product structure]
\label{prodex}
This one is very similar to an almost complex structure in that it is also a globally defined non-degenerate tensor $R \in \Gamma(T M \otimes T^* M)$, but this time one that satisfies
\eq{
R^2=\bbone.
}
An almost product structure has $(+1)$ and $(-1)$-eigenvalues. We denote the corresponding subbundles with
\eq{
T, N \subset T M
}
respectively.
If the dimension of the $(+1)$-eigenspace is $l$, the structure
group reduces to GL($l$,$\mathbb{R}$)$\times$Gl($d-l$,$\mathbb{R}$), and the ranks of $T$ and $N$, this time generically
not equal, are $l$ and $d-l$ respectively. If the almost product structure satisfies in addition the following orthogonality
condition with respect to a metric $g$,
\eq{
R^i{}_k g_{ij} R^j{}_l = g_{kl}\, ,
}
the structure group further reduces to O($l$)$\times$O($d-l$). Product structures and their generalization will
be useful in studying D-branes.
\end{ex}

\subsection{Lie bracket and integrability}

So far we have discussed the {\em topology} of the tangent bundle and the associated frame bundle. We have argued that the existence of
certain invariant and non-degenerate objects implied a reduction of the structure group. Conversely, a reduced structure group $G$,
typically, also implies that there are $G$-invariant objects (which objects exactly depends on the structure group $G$
as we saw in the examples). Let us now investigate which kind of differential conditions one can impose on these objects.
One natural differential condition is {\em integrability}, which we define presently.
Later on we will express the supersymmetry conditions of supergravity in terms of these differential conditions.

First we point out that for vector fields a Lie-bracket can be defined using the natural differential action
of these vector fields on functions. Indeed, such a vector field $X=X^i \frac{\partial\,\,\,\,}{\partial x^i}$
acts as a differential operator on a function $f$ as follows
\eq{
\mathcal{L}_X(f) = X(f)= X^i \frac{\partial f}{\partial x^i} \, .
}
This is the {\em Lie derivative} on functions.

\begin{defi}
The {\em Lie-bracket} of two vector fields is a new vector
field that acts on functions as the commutator of
the differential actions of the two vector fields:
\eq{
\mathcal{L}_{[X,Y]}f = \mathcal{L}_X \mathcal{L}_Y f - \mathcal{L}_Y \mathcal{L}_X f \, ,
}
for any function $f$.
\end{defi}
As usual, if we define a bracket as a commutator it automatically satisfies
the two requirements for making it a Lie-bracket: {\em antisymmetry} and the {\em Jacobi-identity}.

In coordinates, for two such vector fields
$X=X^i \frac{\partial\,\,\,\,}{\partial x^i},Y=Y^i \frac{\partial\,\,\,\,}{\partial x^i}$
the bracket is given by
\eq{
[X,Y] = \left(X^j \frac{\partial Y^i}{\partial x^j} - Y^j \frac{\partial X^i}{\partial x^j} \right) \frac{\partial}{\partial x^i} \, .
}
Furthermore the Lie-derivative acting {\em on a vector field} is simply defined as $\mathcal{L}_X Y =[X,Y]$.

Now we can check whether a distribution $L$ is closed under the action of the Lie-bracket. Recall that a distribution
is a subbundle, locally spanned by smooth vector fields.
\begin{defi}
A distribution $L$ is {\em involutive}
if for any two vector fields $X,Y$
\eq{
X,Y \in \Gamma(L) \Rightarrow [X,Y] \in \Gamma(L) \, .
}
\end{defi}

The next element is the Frobenius theorem, which provides the necessary and sufficient conditions
for the existence of a solution $x^i(\sigma^1,\ldots,\sigma^{\text{rank}(L)};x_0)$ through every point
$p(x_0) \in M$ to the set of partial differential equations
\eq{
\label{setdiffeq}
\frac{\partial x^i}{\partial \sigma^a} = X^i_a \, ,
}
where the $X_a$, $a=1,\ldots,\text{dim}(L)$, locally span $L$.

\begin{defi}
A distribution $L$ is {\em integrable} if through every point $p \in M$
there exists a solution to eq.~\eqref{setdiffeq} in a neighbourhood of $p$.
\end{defi}

One can see that solving eq.~\eqref{setdiffeq} is equivalent to finding a coordinate transformation $x'=x'(x)$
such that the distribution $L \subset T M$
is locally spanned by vector fields of the simple form $\{ \frac{\partial\,\,\,\,}{\partial {x'}^{i}} \, | \, i=1,\ldots,\text{rank}(L) \}$.
The new coordinates $x'$ are called {\em adapted coordinates}. Indeed, once we have found adapted coordinates the solution
to eq.~\eqref{setdiffeq} through $p(x'_0)$ is given by $x'{}^i(\sigma)=x'_0 + \sigma^1 + \cdots \sigma^{\text{rank}(L)}$.

Now the Frobenius theorem states the following:
\begin{theor}[Frobenius]
\label{theor:frob}
$L$ is integrable if and only if it is involutive.
\end{theor}
Because of the Frobenius theorem involutivity and integrability are often used interchangeably, even in definitions.

\begin{ex}[Complex structure]
\label{complintegr}
We can now apply the theorem to the distribution $L$, corresponding to the $(+i)$-eigenspace of an almost
complex structure $J$. If this bundle is integrable, one can introduce adapted coordinates $z^a$ such that $L$ is spanned
by $\{ \frac{\partial\,\,\,\,}{\partial z^a} \, | \, a=1,\ldots,d/2 \}$. These are the {\em holomorphic coordinates} associated
to $J$, which is then, dropping the ``almost'', a {\em complex structure}. The global existence of $J$ ensures that the transition
functions between patches are restricted to GL($d/2$,$\mathbb{C}$), or in other words, the coordinate transformations relating the adapted coordinates between patches are holomorphic.
Note that the integrability of $L$
implies the integrability of $\bar{L}$ so that we might just as well talk about the integrability of $J$.

According to the Frobenius theorem the necessary
and sufficient condition for integrability of $L$ is that for all vector fields $X$ and $Y$,
\eq{
J X = i X \;\; \text{and} \;\; J Y = i Y \quad \Rightarrow \quad J[X,Y]=i[X,Y] \, ,
}
while the complex conjugate is equivalent the integrability of $\bar{L}$.
This can be conveniently formulated in terms of the {\em Nijenhuis tensor}, $N_J \in \Gamma(TM \oplus \Lambda^2 T^*M)$, defined as follows
\eq{\spl{
\label{nijenhuis}
N_J(X,Y) & =  J[JX,Y] + J[X,JY] - [JX,JY] - J^2[X,Y] \\
& = J[JX,Y] + J[X,JY] - [JX,JY] +[X,Y]\, .
}}
We find then that $J$ is integrable iff
\eq{
N_J(X,Y) = 0 \, ,
}
for all vector fields $X$ and $Y$.
\end{ex}

\begin{ex}[Product structure]
A similar story applies to the case of an almost product structure $R$.
Then the Nijenhuis tensor becomes
\eq{\spl{
N_R(X,Y) & =  R[RX,Y] + R[X,RY] - [RX,RY] - R^2[X,Y] \\
& = R[RX,Y] + R[X,RY] - [RX,RY] - [X,Y] \, .
}}
This time integrability of $T$ does not imply integrability of $N$ nor vice-versa.
One finds
\eq{\spl{
T \; \text{integrable} & \Longleftrightarrow N_R(X,Y)|_N = 0 \, , \\
N \; \text{integrable} & \Longleftrightarrow N_R(X,Y)|_T = 0 \, ,
}}
where $|_N$ and $|_T$ denote the projections to $N$ and $T$ respectively.
If $T$ is integrable and of rank $l$, it defines an $l$-dimensional submanifold through every point $p$ of $M$,
such that the tangent bundle of this submanifold in that point is $T_p$. The solution to eq.~\eqref{setdiffeq}
forms a parametrization $x^i(\sigma)$ of the submanifold through $p$. These submanifolds form
the {\em leaves} of a {\em foliation}.

When both $T$ and $N$ are integrable one can foliate the
manifold in two ways: either with leaves along $T$ or with leaves along $N$ (see figure \ref{foliationfig}) and, dropping again the ``almost'',
$R$ becomes a {\em product structure}. For our purposes, however, it will be sufficient that $T$ is integrable.
In fact, using the language of currents, in chapter \ref{branesection} we will only require $T$ to be defined on
one submanifold. Extending to Generalized Complex Geometry, this generalized submanifold corresponds to a D-brane.
\begin{figure}[t]
\centering
\psfrag{T}{$T$}
\psfrag{N}{\color{red}$N$}
\includegraphics[width=7cm]{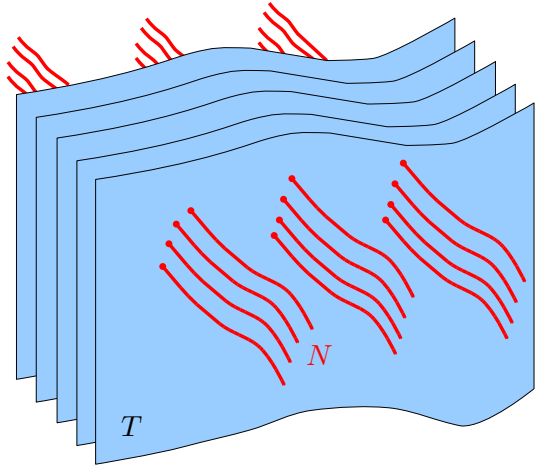}
\caption{When both $T$ and $N$ are integrable one can foliate the
manifold in two ways: either with leaves along $T$ or with leaves along $N$.}
\label{foliationfig}
\end{figure}
\end{ex}

\begin{ex}[Symplectic structure]
\label{sympintegr}
A pre-symplectic structure $\omega$ is integrable if
\eq{
\d \omega = 0 \, ,
}
and is then called, dropping the ``pre'', a {\em symplectic structure}.
A symplectic structure is a bit different since we have not provided a description
in terms of a subbundle of the tangent bundle. In chapter \ref{gengeomfund} we will show
that in Generalized Complex Geometry a symplectic structure can be treated on the same footing
as a complex structure and a description in terms of a subbundle of the generalized tangent bundle can
be given. For now, we note that for an integrable symplectic structure adapted coordinates $(q^a,p_a)$, called
{\em Darboux coordinates}, can be introduced so that $\omega$ takes the standard form
\eq{
\omega = \sum_a \d p_a \wedge \d q^a \, .
}
\end{ex}

\subsection{The language of forms}

We have already remarked that the same $G$-structure can typically be described by different sets
of invariant tensors, leading to different viewpoints. In this section we describe the second point of view often
used in Generalized Complex Geometry. Indeed, all key structures can be reformulated in terms of forms, which are sections of antisymmetric products of the cotangent bundle. For the space of real/complex $l$-forms
we will use the following shorthand notation
\eq{
\Gamma(\Lambda^l T^* M) \equiv \Omega^l(M,\mathbb{R}) \, , \qquad
\Gamma(\Lambda^l T^* M) \otimes \mathbb{C} \equiv \Omega^l(M,\mathbb{C}) \, .
}

\subsubsection*{$G$-structures and forms}

\begin{ex}[Almost complex structure]
So let us start by constructing the invariant form that is associated to an almost complex structure.
We have already seen that such an almost complex structure decomposes the tangent bundle
into two subbundles $L$ and $\bar{L}$. Likewise the cotangent bundle splits as
follows into two bundles of rank $d/2$
\eq{
\Lambda^1 T^* M \otimes \mathbb{C}= \Lambda^{1,0} T^* M \oplus \Lambda^{0,1} T^* M \, ,
}
where $\theta \in \Lambda^{1,0} T^* M$ iff $\theta(X)=0$ for all $X \in \bar{L}$ and
analogously for $\Lambda^{0,1} T^* M$. This induces a decomposition of higher forms as follows
\eq{
\label{formdecomp}
\Lambda^l T^*M = \bigoplus_{0\le p \le l} \, (\Lambda^p T^{*(1,0)}M  \otimes \Lambda^{l-p} T^{*(0,1)}M )\,
=  \bigoplus_{0 \le p \le l}  \Lambda^{p,l-p} T^* M \, .
}
We will introduce the following shorthand notation for the sections:
\eq{
\Omega^{p,q}(M)=\Gamma(\Lambda^{p,q} T^* M) \, .
}
We can now define a local frame of $d/2$
independent $(1,0)$-forms $\theta^{a}$ and a corresponding local section $\Omega$ of
the bundle $\Lambda^{d/2,0} T^*M$ --- called the {\em canonical bundle} ---
\eq{
\label{omsimple}
\Omega = \theta^1 \wedge \cdots \wedge \theta^{d/2} \, .
}
So we used the almost complex structure $J$ to construct $\Omega$. Conversely, from $\Omega$
one can construct $\bar{L}$ (and thus $L$ and finally $J$) as follows:
\eq{
\bar{L} = \{ v \in TM \, | \, \iota_v \Omega =0 \} \, ,
}
where the interior product $\iota_v$ is defined in a moment (definition \ref{interiorderdef}).
It is important to realize that an almost complex structure only determines the
$\theta^{a}$ up to a GL($d/2$,$\mathbb{C}$)-transformation, which implies in turn for
$\Omega$ that it is only determined up to an overall complex function. Between patches such
a GL($d/2$,$\mathbb{C}$)-transformation can thus change the overall factor of $\Omega$.
So an almost complex structure does not quite require
the existence of a globally defined $(d/2,0)$-form, since on the overlap of two patches it allows $\Omega$
to change by a complex factor. If however we do require $\Omega$ to be a globally defined
everywhere non-vanishing $(d/2,0)$-form the structure group is further reduced to SL($d/2$,$\mathbb{C}$).

$\Omega$ is not just any form, it has to be a {\em decomposable} or {\em simple} form,
which means it can (locally) be written as the wedge product of one-forms as in eq.~\eqref{omsimple}.
This condition is quite cumbersome to check. For the special case of $d=6$, \cite{hitchinfunc}
provides a way to construct a decomposable three-form from any real stable three-form as follows.
Start from a real three-form $\rho$ and construct:
\eq{
\label{Jfromrho}
\tilde{J}^i{}_j = \varepsilon^{ii_1\dots i_{5}} \rho_{ji_1i_{2}} \rho_{i_{3}i_4i_{5}} \, ,
}
where $\varepsilon^{i_1 \ldots i_d}=\pm 1$ is the totally antisymmetric symbol. Next we calculate
\eq{
q(\rho) = \frac{1}{6} \tr \tilde{J}^2 \, .
}
If $q(\rho) \neq 0$ then $\rho$ is called {\em stable} and can be used to construct a complex or real decomposable
three-form for $q(\rho) < 0$ and $q(\rho) > 0$ respectively. Let us proceed with $q(\rho)<0$.
The almost
complex structure is then (up to a sign) given by
\eq{
J = \pm \tilde{J}/\sqrt{-q(\rho)} \, ,
}
and the complex decomposable three-form by
\eq{
\Re \Omega = \rho \, , \qquad \Im \Omega = \hat{\rho}=\frac{1}{6} J^i{}_j \left( \iota_i \wedge \d x^j - \d x^j \wedge \iota_i\right) \rho \, .
}
$H(\rho) = \sqrt{-q(\rho)}$ is called the {\em Hitchin function}. We conclude that both the almost complex structure $J$
and the imaginary part, $\Im \Omega$, of a complex decomposable three-form are (up to a sign) completely determined by the real part $\Re \Omega=\rho$.
Note that also in dimensions different from 6, the almost complex structure can (up to a sign) be determined
from $\Re \Omega$ alone using
\eq{
\label{JfromReOm}
J^i{}_j = c \, \varepsilon^{ii_1\dots i_{d-1}} \Re \Omega_{ji_1\ldots i_{d/2-1}} \Re \Omega_{i_{d/2}\ldots i_{d-1}} \, ,
}
with $c$ a function which should be determined by appropriately normalizing $J$ such that $J^2 = -\bbone$.
In this case however, not every real form $\rho$ with negative $q(\rho)$ will lead in this way to a proper almost
complex structure.

\end{ex}

\begin{ex}[Hermitian pre-symplectic structure or U($d/2$)-structure]
As a second example let us turn to a pre-symplectic structure $\omega \in \Omega^2(M,\mathbb{R})$.
Although this is already a form, we will argue in section \ref{puresec} that in the Generalized Complex Geometry
formalism the appropriate associated (poly)form is $e^{i \omega}$. For now, let us just consider $\omega$
and furthermore require that it satisfies the hermiticity condition \eqref{hermiticity}.
To express this condition in the form language note that it is actually equivalent to requiring $\omega$ to
be a (1,1)-form:
\eq{
\omega \in \Omega^{1,1}(M) \, .
}
Considering that $\omega$ is real we can reformulate this as follows:
\eq{
\label{hermform}
\omega \wedge \Omega=0 \, .
}
\end{ex}

\begin{ex}[SU($d/2$)-structure]
We already pointed out that an almost complex structure and a compatible pre-symplectic structure
reduce the structure group to U($d/2$). If there exists a globally defined $(d/2,0)$-form $\Omega$ associated
to the almost complex structure, which means in particular we can eliminate the ambiguity in the overall factor,
then the structure further reduces to SU($d/2$). Expressing this entirely in
terms of the form language we have the following statement.
In the presence of a globally defined, decomposable, complex $d/2$-form $\Omega$, which is non-degenerate
everywhere:
\eq{
\Omega \wedge \bar{\Omega} \neq 0 \, ,
}
and a non-degenerate two-form $\omega$, such that the compatibility condition \eqref{hermform} is satisfied
and such that the associated metric (to be constructed from eqs.~\eqref{JfromReOm} and \eqref{metricJom}) is positive-definite, the structure group reduces to SU($d/2$).
The $(d/2,0)$-form $\Omega$ is usually normalized such that
\eq{
\Omega \wedge \bar{\Omega} = (-1)^{\frac{d/2(d/2-1)}{2}} \, \frac{\,(2i \omega)^{d/2}}{(d/2)!}  \, .
}
The prefactors are chosen in such way that we can find a local basis of $(1,0)$-forms $\theta^{a}$
such that $J$ and $\Omega$ take the standard form
\eq{
J = - \frac{i}{2} \sum_a \theta^{a} \wedge \bar{\theta}^{\bar{a}} \, , \qquad
\Omega = \theta^{1} \wedge \cdots \wedge \theta^{d/2} \, .
}
\end{ex}

\begin{ex}[Almost product structure]
\label{almostprodforms}
Suppose we have an almost product structure with (+1)-eigenbundle $T$ of rank $l$.
Let us introduce the annihilator space
\eq{
\text{Ann} \, T = \{ \theta \in T^*M \, | \, \theta(v)=0,\, \forall v \in T\} \, .
}
We can then define a local frame of $d-l$ independent
base forms $\theta^a$ of Ann $T$, this time {\em real}, and a real local section of $\Lambda^{d-l} T^*M$
\eq{
\tau = \theta^1 \wedge \cdots \wedge \theta^{d-l} \, ,
}
which equivalently describes the subbundle $T$ as follows
\eq{
T = \{ v \in TM \, | \, \iota_v \tau = 0 \} \, .
}
Note that under GL($l$,$\mathbb{R}$)$\times$Gl($d-l$,$\mathbb{R}$)-transformations, $\tau$
is also only defined up to an overall factor.
\end{ex}

\subsubsection*{Integrability}

Since we claimed that many statements can be reformulated in the form language, one
might wonder in particular whether it is possible to reexpress the integrability condition
in terms of the associated form. This is indeed possible but first
we have to introduce some extra technology: namely three different derivative operators on forms.

The first one is just the contraction with the first argument,
which decreases the degree of the form by one.
\begin{defi}
\label{interiorderdef}
The {\em interior product} $\iota_X: \Omega^l(M) \rightarrow \Omega^{l-1}(M)$ for $X \in \Gamma(T^* M)$
acts on a form $\phi \in \Omega^{l}(M)$ and produces an $(l-1)$-form $\iota_X \phi$ such that
\eq{
\label{interiorder}
\iota_X \phi (Y_{1},\ldots,Y_{l-1}) = \phi(X,Y_1,\ldots,Y_{l-1}) \,  ,
}
for all $Y_1,\ldots, Y_{l-1} \in \Gamma(T M)$.
\end{defi}
In coordinates:
\eq{
(\iota_X \phi)_{i_1 \ldots i_{l-1}} = X^{i_1} \phi_{i_1i_2\ldots i_l} \, .
}
It satisfies
\eq{
\iota_X \iota_Y \phi = - \iota_Y \iota_X \phi \, .
}
The second one is the exterior derivative, which increases the degree of the form by one.
\begin{defi}
\label{exteriorderdef}
The {\em exterior derivative} $\d: \Omega^l(M) \rightarrow \Omega^{l+1}(M)$ acts on
a form $\phi \in \Omega^{l}(M)$ and produces an $(l+1)$-form $\d \phi$ such that
\eq{\spl{
\label{exteriorder}
\d \phi(Y_0,\ldots,Y_{l}) = & \sum_{0\le a \le l} (-1)^{a} Y_a\left(\phi(Y_0,\ldots,\hat{Y}_a,\ldots,Y_l)\right) \\
& + \sum_{0\le a<b \le l} (-1)^{a+b} \phi([Y_a,Y_b],Y_0,\ldots,\hat{Y}_{a},\ldots,\hat{Y}_b,\ldots,Y_l) \, ,
}}
for all $Y_0,\ldots, Y_{l} \in \Gamma(T M)$ and where $\hat{Y}_a$ indicates that the vector field with index $a$ is missing.
\end{defi}
In coordinates:
\eq{
\label{extdercoord}
(\d \phi)_{i_0i_1\ldots i_{l}} = (l+1) \partial_{[i_0} \phi_{i_1 \ldots i_l]} \, .
}
From the definition we see that the exterior derivative depends on the Lie-bracket on $\Gamma(T M)$.

The last derivative is the Lie-derivative, which we already defined on functions and vector fields, and
which we now extend to forms.
\begin{defi}
The {\em Lie derivative} $\mathcal{L}_X: \Omega^l(M) \rightarrow \Omega^{l}(M)$ is given by
\eq{
\label{lieder}
\mathcal{L}_X = \{\iota_X,\d \} = \iota_X \d + \d \, \iota_X \, .
}
\end{defi}
From the above definitions one can find the following property
\eq{
\label{bracketprop}
\iota_{[X,Y]} = [\mathcal{L}_X,\iota_Y] = [\{\iota_X,\d\}, \iota_Y] \, .
}
Alternatively, instead of defining the exterior derivative in terms of the Lie-bracket
we could introduce the exterior derivative first (for instance through its expression in terms
of coordinates eq.~\eqref{extdercoord}) and define the Lie-bracket using the above formula. A bracket
defined in this way in terms of a derivative operator is called a {\em derived bracket} (see e.g.~\cite{derivedbrackets}).

With this technology we are ready to formulate the condition for integrability in the
form language. Suppose that $X,Y \in \Gamma(\bar{L})$ which implies $\iota_X \Omega=\iota_Y \Omega=0$.
We can then use \eqref{bracketprop} to obtain
\eq{
\iota_{[X,Y]} \Omega = \iota_Y \iota_X \d \Omega \, .
}
It follows that $J$ is integrable iff
\eq{
\iota_Y \iota_X \d \Omega = 0 \, ,
}
for all $X,Y \in \Gamma(\bar{L})$. This means that $\d \Omega$ must be of complex type $(3,1)$,
\eq{
\label{omegaintegr}
\d \Omega \in \Omega^{3,1}(M) \, ,
}
or equivalently
\eq{
\label{intomcond}
\d \Omega = \theta \wedge \Omega \, ,
}
for some one-form $\theta$. Note that one can indeed rescale $\Omega$
with a function $f$ and still satisfy \eqref{intomcond}, so that the integrability condition
does not depend on the overall factor. A similar condition applies for the integrability of
an almost product structure.

\subsubsection*{Decomposition of forms and the exterior derivative}

Let us consider the decomposition \eqref{formdecomp} on an almost complex manifold.
Acting with the exterior derivative on a $(p,q)$-form $\phi^{p,q}$ we find
\eq{
\label{extdecomp}
\d (\phi^{p,q}) \in \Omega^{p+2,q-1}(M) \cup \Omega^{p+1,q}(M) \cup \Omega^{p,q+1}(M) \cup \Omega^{p-1,q+2}(M) \, .
}
That there are no more terms in the decomposition of $\d (\phi^{p,q})$, can be easily shown from eq.~\eqref{exteriorder}
defining the exterior derivative.

If the almost complex structure is integrable, this decomposition further collapses
to just two terms\footnote{This can be shown through an inductive argument based on the one leading to eq.~\eqref{omegaintegr}.}
\eq{
\label{extdecompintegr}
\d (\phi^{p,q}) \in \Omega^{p+1,q}(M) \cup \Omega^{p,q+1}(M) \, ,
}
and the exterior derivative decomposes into the Dolbeault operators $\partial$ and $\bar{\partial}$,
\eq{
\label{ddecomp}
\d = \partial + \bar{\partial} \, ,
}
with
\eq{
\partial: \Omega^{p,q}(M) \rightarrow \Omega^{p+1,q}(M) \, , \qquad
\bar{\partial}: \Omega^{p,q}(M) \rightarrow \Omega^{p,q+1}(M) \, .
}
They satisfy:
\eq{
\partial^2 = 0 \, , \qquad \bar{\partial}^2= 0 \, , \qquad \partial \bar{\partial} + \bar{\partial} \partial = 0 \, .
}
This allows us to define the Dolbeault cohomology groups
\eq{
{\rm H}_{\bar{\partial}}^{p,q}(M) = \frac{ \{\phi \in \Omega^{p,q}(M) \, | \, \bar{\partial} \phi = 0 \}}{ \{ \phi \sim \phi + \bar{\partial} \lambda \, | \,
\lambda \in \Omega^{p,q-1}(M) \}} \, ,
}
with {\em Hodge numbers} $h^{p,q}= \text{dim}({\rm H}_{\bar{\partial}}^{p,q}(M))$.
At this point, it is tempting to assume there exists a nice decomposition of the ordinary cohomology groups,
\eq{
{\rm H}_l(M,\mathbb{C}) = \frac{ \{\phi \in \Omega^{l}(M,\mathbb{C}) \, | \, \d \phi = 0 \}}{ \{ \phi \sim \phi + \d \lambda \, | \,
\lambda \in \Omega^{l-1}(M,\mathbb{C}) \}} \, ,
}
(with {\em Betti numbers} $b^l = \text{dim}({\rm H}_l(M,\mathbb{C}))$) into Dolbeault cohomology groups as follows
\eq{
{\rm H}_l(M,\mathbb{C}) = \bigoplus_{0 \le p \le l} {\rm H}_{\bar{\partial}}^{p,l-p}(M) \, ,
}
where a representative of an ordinary cohomology group can be written as a sum of representatives of Dolbeault
cohomology groups.
This decomposition only works, however, if the {\em $\partial\bar{\partial}$-lemma} holds
\eq{
\text{Im} \, \partial \cap \text{Ker} \, \bar{\partial} =
\text{Im} \, \bar{\partial} \cap \text{Ker} \, \partial=
\text{Im} \, \partial \bar{\partial} \, .
}
This is for instance the case for K\"ahler manifolds (for a definition see section \ref{SU3structuretorsion}).

\subsection{SU(3)-structure and torsion classes}
\label{SU3structuretorsion}

In this section, let us restrict to
$d=6$ and consider a globally defined complex decomposable $(3,0)$-form $\Omega$ and a
pre-symplectic real $(1,1)$-form $\omega$, so that the structure group is reduced to SU(3).
We can decompose $(\d\omega,d\Omega)$ as follows in terms of SU(3)-representations
\eq{\label{torsioncl}\spl{
\d \omega & = -\frac{3}{2} \Im (\bar{\mathcal{W}}_1 \Omega) + \mathcal{W}_4 \wedge \omega + \mathcal{W}_3 \, , \\
\d \Omega & = \mathcal{W}_1 \omega^2 + \mathcal{W}_2 \wedge \omega + \bar{\mathcal{W}}_5 \wedge \Omega \, ,
}}
where the $\mathcal{W}_i$ are the {\em torsion classes} \cite{greyhervella,chiossal} (see \cite{lustcurio,stringswithtorsion} for early applications to string compactifications): 
$\mathcal{W}_1$ is a complex scalar, $\mathcal{W}_2$ is a complex primitive (1,1)-form, $\mathcal{W}_3$ is a real
primitive $(1,2)+(2,1)$-form, $\mathcal{W}_4$ is a real one-form and $\mathcal{W}_5$ a complex (1,0)-form.
To understand how this decomposition comes about the reader should observe the following:
according to the decomposition \eqref{extdecomp}, the real three-form $\d \omega$ consists of a $(3,0)+(0,3)$-part --- which is described
by $\mathcal{W}_1$ and transforms in the $\mathbf{1} \oplus \mathbf{1}$ of SU(3) --- and a $(2,1)+(1,2)$-part.
A (2,1)-form transforms under SU(3) as
\eq{
\mathbf{3} \otimes \mathbf{3} = (\mathbf{3} \otimes \mathbf{3})_S + (\mathbf{3} \otimes \mathbf{3})_A = \mathbf{6} + \mathbf{\bar{3}} \, ,
}
so that in total, the $(2,1)+(1,2)$-part transforms as
\eq{
(\mathbf{3} + \bar{\mathbf{3}}) + (\mathbf{6} + \bar{\mathbf{6}}) \, .
}
The $(\mathbf{3} + \bar{\mathbf{3}})$-part is described by the real one-form $\mathcal{W}_4$
while the $(\mathbf{6} + \bar{\mathbf{6}})$-part is described by $\mathcal{W}_3$, which in order
to remove the $(\mathbf{3} + \bar{\mathbf{3}})$-part must satisfy
the primitivity condition:
\eq{
\mathcal{W}_3 \wedge \omega=0 \, .
}
Furthermore, again according to \eqref{extdecomp}, $\d \Omega$ consists of a $(3,1)$-part and a $(2,2)$-part.
The $(3,1)$-part, transforming as $\mathbf{\bar{3}}$,
is described by $\mathcal{W}_5$, while the $(2,2)$ part transforms as
\eq{
\mathbf{\bar{3}} \otimes \mathbf{3} = \mathbf{8} + \mathbf{1} \, ,
}
described by respectively the primitive $\mathcal{W}_2$,
\eq{
\mathcal{W}_2 \wedge \omega \wedge \omega = 0 \, ,
}
and again $\mathcal{W}_1$. Because the $\mathcal{W}_i$ in $\d \Omega$
are complex these representations count twice.

It follows from \eqref{intomcond} that if $\mathcal{W}_1=\mathcal{W}_2=0$ the almost complex structure
is integrable and the manifold is {\em complex}. On the other hand, if $\mathcal{W}_1=\mathcal{W}_3=\mathcal{W}_4=0$
we find $\d \omega=0$ and the manifold is called {\em symplectic}. If the manifold is both complex and symplectic, then it
is a {\em K\"ahler} manifold and $\omega$ is called the K\"ahler form. In this case the manifold has U(3)-{\em holonomy}. If on top of that also $\mathcal{W}_5=0$
the holonomy reduces to SU(3) and the manifold is {\em Calabi-Yau}. If $(\omega,\Omega)$ define a Calabi-Yau holonomy up to on overall factor $e^A$, i.e.\
($\omega'$,$\Omega'$)=($e^{2A} \omega, e^{3A} \Omega$) is Calabi-Yau then the geometry is called conformal Calabi-Yau (see example \ref{CYex}).
Another interesting case which is relevant for the study of type IIA
AdS$_4$ compactifications \cite{cveticnk1,cveticnk2} is $\mathcal{W}_2=\mathcal{W}_3=\mathcal{W}_4=\mathcal{W}_5$, which is called {\em nearly K\"ahler} (see section \ref{sec:AdS4SU3ex}). See table \ref{torsiontable}
for an overview containing some more cases.
\begin{table}[tp]
\begin{center}
\rowcolors{2}{blue!40}{blue!10}
\begin{tabular}{|c|c|}
\hline
Torsion classes & Name \\
\hline
$\mathcal{W}_1=\mathcal{W}_2=0$ & Complex \\
$\mathcal{W}_1=\mathcal{W}_3=\mathcal{W}_4=0$ & Symplectic \\
$\mathcal{W}_2=\mathcal{W}_3=\mathcal{W}_4=\mathcal{W}_5=0$ & Nearly K\"ahler \\
$\mathcal{W}_1=\mathcal{W}_2=\mathcal{W}_3=\mathcal{W}_4=0$ & K\"ahler \\
$\Im \mathcal{W}_1=\Im \mathcal{W}_2=\mathcal{W}_4=\mathcal{W}_5=0$ & Half-flat \\
$\mathcal{W}_1=\Im \mathcal{W}_2=\mathcal{W}_3=\mathcal{W}_4=\mathcal{W}_5=0$ & Nearly Calabi-Yau \\
$\mathcal{W}_1=\mathcal{W}_2=\mathcal{W}_3=\mathcal{W}_4=\mathcal{W}_5=0$ & Calabi-Yau \\
$\mathcal{W}_1=\mathcal{W}_2=\mathcal{W}_3=0, (1/2) \mathcal{W}_4=(1/3) \mathcal{W}_5=-\d A$ & Conformal Calabi-Yau \\
\hline
\end{tabular}
\caption{Classification of geometries from vanishing SU(3) torsion classes. Adapted from table 3.1 of \cite{fluxrev1}.}
\label{torsiontable}
\end{center}
\end{table}

For Calabi-Yau manifolds there exists the following celebrated theorem.
\begin{theor}[Calabi-Yau]
On a compact K\"ahler manifold $M$ of dimension $d$ with K\"ahler form $\tilde{\omega}$ and complex structure $J$,
for which there exists a globally defined nowhere-vanishing $(d/2,0)$-form $\Omega$, there is a unique metric with K\"ahler form $\omega$
in the same K\"ahler class as $\tilde{\omega}$ (which means $\tilde{\omega}=\omega + \d \alpha$) such that $(f\Omega,\omega)$, with appropriate
normalization function $f$, is Calabi-Yau.
\end{theor}
The requirement that there exists a globally defined nowhere-vanishing $(d/2,0)$-form is often phrased
as the statement that the integral Chern class $c_1(M,\mathbb{Z})$ vanishes or that the canonical line bundle $\Omega^{d/2,0}(M)$ is
trivial. Sometimes the extra requirement that the fundamental group of $M$ be trivial (and thus $b_1=0$) is imposed in order
to exclude ``trivial'' or reducible Calabi-Yau manifolds, like tori or products of tori with lower-dimensional Calabi-Yau manifolds.
Note that although because of the above theorem it is known that there is a unique Calabi-Yau metric, except for tori,
this metric is not analytically known.

\subsection{The language of spinors}

Finally we come to the last way of describing an SU($d/2$)-structure: an invariant spinor and its complex
conjugate. In fact, since GL($d$,$\mathbb{R}$) does not have a spinor representation we must first introduce
a metric and an orientation to reduce the structure group to SO($d$,$\mathbb{R}$). To be precise, SO($d$,$\mathbb{R}$)
does not have a spinor representation either, but its universal (double) cover Spin($d$,$\mathbb{R}$) does.
In order to be able to lift the SO($d$,$\mathbb{R}$)-structure to a Spin($d$,$\mathbb{R}$)-structure in a globally consistent way an extra topological requirement
is imposed on the manifold, namely that it be a {\em spin manifold}. To the metric we then associate a {\em vielbein},
which consists of a basis of orthonormal one-forms $e^{a}$ in every patch\footnote{To make the connection with the explanation on the frame bundle of section \ref{langstruc}, we note that this vielbein is a local trivialization of the coframe bundle,
which is associated to the {\em cotangent} bundle $T^* M$. This is the inverse of the previously
introduced local trivialization of the frame bundle: $e^j{}_a e^a{}_i  = \delta^j{}_i$. Moreover, the structure group
is reduced to O($d$,$\mathbb{R}$) by requiring the metric to take the simple form $\delta_{ab}$ in these {\em flat coordinates}.} so that
the metric takes a simple standard form $\delta_{ab}$ in this basis:
\eq{
g_{ij} = \delta_{ab} e^{a}{}_i e^{b}{}_j \, .
}
A spinor field $\eta(x)$ is then
a field on which the infinitesimal rotations $R \in$so($d$,$\mathbb{R}$) \footnote{We denote the algebra associated to a group with
small letters. The algebra associated to SO($d$,$\mathbb{R}$) and its fundamental cover Spin($d$,$\mathbb{R}$) is of course the same, and
by displaying the infinitesimal version of the transformation we avoid such global subtleties.} act as follows:
\eq{
\delta_R \eta = \frac{1}{4} R^{ab} \gamma_{ab} \eta \, ,
}
where the $\gamma_{a}$ are the gamma-matrices with defining property $\{ \gamma_{a}, \gamma_{b} \} = 2 \, \delta_{ab}$.
Using the vielbein one can introduce the curved gamma-matrices
\eq{
\label{curvedgamma}
\gamma_{i} = e^{a}{}_{i} \gamma_{a} \, .
}
Instead of the Dirac slash notation, which is not really suitable for long expressions, we will
indicate the contraction of the form $\phi$ with gamma-matrices as follows:
\eq{
\slashchar{\phi} = \frac{1}{l!} \, \phi_{i_1 \ldots i_l} \gamma^{i_1 \ldots i_l} =
\frac{1}{l!} \, \phi_{i_1 \ldots i_l} e^{i_1}{}_{a_1} \cdots e^{i_l}{}_{a_l} \gamma^{a_1 \ldots a_l} \, .
}

For $d$ even a globally defined invariant
pure\footnote{\label{purefootnote}See definition \ref{purespinordef} of a pure spinor, i.e.\ a spinor is pure if it is annihilated by half
the gamma-matrices (which is the maximal amount). In $d=6$ we do not have to worry about this subtle extra requirement
since any Weyl spinor is then pure. An alternative definition, which is
more used in the literature on the Berkovits formalism (e.g.\ in \cite{berkovitsformalism,berkovitsformalismreview}) says that all
bilinears of the form $\eta'{}^\dagger \gamma_{m_1 \ldots m_k} \eta$ must vanish for $k <  d/2$, so that
the bilinear defining $\Omega$ in eq.~\eqref{strucspinor} is the first non-zero one. The equivalence between the
two definitions is proven in \cite{chevalley}.} spinor $\eta$ and its complex conjugate $\eta'= C \eta^*$, with $C$ the charge conjugation matrix ($\gamma_a^*=-C^{-1} \gamma_a C$), reduces the structure group
to SU($d/2$).\footnote{In the special case $d=8$, it is possible to have only the linear combination $\eta+\eta'$ (which is {\em not} pure) invariant
so that the structure group is merely reduced to SO(7). Related is the dimension $d=7$, which is the only odd dimension that
allows for an invariant spinor (without also having an invariant vector). The structure group is then reduced to G$_2$.}
From these spinors we can construct $\omega$ and $\Omega$
as follows:
\eq{
\label{strucspinor}
\omega_{ij} = i \eta^\dagger \gamma_{ij} \eta \, , \qquad \Omega_{i_1 \ldots i_{d/2}} = \eta'{}^\dagger \gamma_{i_1\ldots i_{d/2}} \eta \, .
}

Specializing to $d=6$ we find for the chirality operator $\gamma_{(6)}= - C^{-1} \gamma_{(6)} C$ so
that $\eta$ and $\eta'$ have different chirality.
We label $\eta_+=\eta$ and $\eta_-=\eta'$. There is always a generically torsionful connection\footnote{\label{footconnec} Very roughly one can think of a connection as a covariant derivative.
The precise definition is as follows. On a given vector bundle E,  a {\em connection} $\nabla$ is a map
$\nabla \, : \, \Gamma(E) \rightarrow \Gamma( E \otimes T^*M)$
that satisfies the derivative property $\nabla ( f v) = f \nabla v + v  \otimes \d f$ ,
where $v$ is a smooth section of $E$ and $f$ is a smooth function on $M$.
In an explicit coordinate basis the connection has the form $\nabla = \d x^m \otimes \nabla_m$.
$\nabla_m$ is then usually better-known by physicists as the {\em covariant derivative}.} $\nabla_{T}$,
which satisfies
\subeq{\label{torsion}\al{
& \nabla_{T} g = 0 \, , \\
& \nabla_{T} \eta_+ = \nabla_{\text{LC}} \eta_+ + T \eta_+ = 0 \, ,
}}
where $\nabla_{\text{LC}}$ is the Levi-Civita connection and $T$ is called
the {\em contorsion tensor}. The Levi-Civita connection
is the unique connection that is both compatible with the metric ($\nabla_{\text{LC}}=0$) and
torsion-free. Furthermore, in Riemannian geometry there is a one-to-one correspondence between
the contorsion tensor $T$ and the torsion of $\nabla_T$.
The contorsion tensor decomposes as follows in SU(3)-representations
\eq{
T_{ab}{}^{c} \in (\text{su(3)}\oplus \text{su(3)}^\perp) \otimes V \, ,
}
where the antisymmetric indices $ab$ span the space of antisymmetric matrices,
which is isomorphic to so(6) and the upper index transforms
as a vector (indicated by $V$). The adjoint representation so(6) splits as so(6)$=$su(3)$\oplus$su(3)$^\perp$,
where acting on an SU(3)-invariant spinor the su(3)-part drops out. In
the end we find:
\eq{\spl{
 (\text{su(3)}^\perp) \otimes V  & \hspace{0.2em}=  (\mathbf{1}+\mathbf{3}+\mathbf{\bar{3}})\otimes (\mathbf{3} \oplus \mathbf{\bar{3}}) \\
& \begin{array}{ccccccccc}
= & (\mathbf{1} \oplus \mathbf{1}) &\oplus& (\mathbf{8} \oplus \mathbf{8}) &\oplus& (\mathbf{6} \oplus \mathbf{\bar{6}}) &\oplus& 2(\mathbf{3} \oplus \mathbf{\bar{3}}) \, , \\
 & \mathcal{W}_1 &&  \mathcal{W}_2 && \mathcal{W}_3 && \mathcal{W}_4,\mathcal{W}_5
\end{array}
}}
which exactly corresponds to the representations of the torsion classes appearing in the SU(3)-decomposition of $(\d \omega,\d \Omega)$.
Comparing eqs.~\eqref{strucspinor}, \eqref{torsion} and \eqref{torsioncl} it is
possible to find the exact relations between the components of the contorsion and the torsion classes,
but we will not need these here.

\subsection{Deformation theory}

\begin{ex}[Complex deformations]
Suppose we consider an infinitesimal deformation of the integrable complex structure $J$
\eq{
\label{defprodcompl}
J'{}^i{}_k = J^i{}_k + \tau^i{}_k \, .
}
Of course the new complex structure should still satisfy the defining property \eqref{prodstruc}, which leads
to
\eq{
\label{tauprodcond}
\tau^i{}_j J^j{}_k + J^i{}_j \tau^j{}_k = 0 \, .
}
It will be convenient to switch to holomorphic coordinates $z^a$
and their complex conjugates, which is possible if $J$ is integrable. The condition \eqref{tauprodcond}
then simply reads
\eq{
\tau^a{}_b = 0 \, , \qquad \tau^{\bar{a}}{}_{\bar{b}}=0 \, ,
}
so that only the components with mixed indices can be non-zero.
We want to construct the condition for $J'$ to be integrable as well. From
the vanishing of the Nijenhuis tensor \eqref{nijenhuis} for $J'$ we find then:
\eq{
\label{compldefint}
\partial_{[\bar{a}} \tau^c{}_{\bar{b}]} = 0 \, .
}
On the other hand, we find that the deformation is merely a change of coordinates
$z^{a} \rightarrow z^a + i \epsilon v^a(z,\bar{z})$ and thus trivial
if
\eq{
\tau^{a}{}_{\bar{b}} = \epsilon \, \bar{\partial}_{\bar{b}} v^{a} \, .
}
It follows that the inequivalent deformations of the complex structure are classified
by $H^{0,1}(M, T^{1,0} M)$, i.e.\ the cohomology of (0,1)-forms taking values in the
holomorphic tangent bundle $L=T^{1,0} M$.

Alternatively, in the language of forms the deformation acts like
\eq{
\Omega' = \Omega + \xi \, ,
}
with (restricting to $d=6$ in the rest of the subsection)
\eq{
\label{defformstruc}
\xi_{ijl} = -\frac{3i}{2} \, \Omega_{[ij|k|} \tau^k{}_{l]} \, .
}
From \eqref{tauprodcond} follows
that $\xi$ is a $(2,1)$-form.
The integrability of $J'$ amounts to
\eq{
\d \Omega'|_{2,2} = \bar{\partial} \xi = 0 \, ,
}
where we used \eqref{ddecomp} and thus that $J$ itself is integrable. If furthermore $\d \Omega=0$, one can show that complex coordinate
transformations correspond to $\bar{\partial}$-exact $\xi$. So we find that alternatively
the deformations can be classified by $H^{2,1}(M)$ and the equivalence between both
descriptions is given by \eqref{defformstruc}. So the dimension of the moduli space of complex
deformations is given by the Hodge number $h^{2,1}= \text{dim}(H^{2,1}(M))$.

In fact, for deformations of a Calabi-Yau geometry, one would have to restrict to deformations of $\Omega$
that preserve the compatibility condition \eqref{hermform}, which is the case if
\eq{
\xi \wedge \omega = 0 \, ,
}
where we kept the K\"ahler form $\omega$ invariant.
This is where the extra restriction on the Calabi-Yau that its fundamental group be trivial,
mentioned at the end of section \ref{SU3structuretorsion}, comes in. It implies that the first
and the fifth cohomology group is trivial ($b_1=b_5=0$), so that the above condition on the deformation is always
satisfied. In the case of Generalized Complex Geometry however, there is a similar condition that is not always satisfied and
it is not completely clear how to handle it yet.
\end{ex}

\begin{ex}[K\"ahler deformations]
On the other hand one can keep $\Omega$ fixed and study the deformations of the K\"ahler form
that satisfy the compatibility condition. A similar analysis as above gives
that they are classified by $H^{1,1}(M)$ so that the dimension of the moduli space of K\"ahler deformations
is given by the Hodge number $h^{1,1}$.
\end{ex}

The contents of this chapter is summarized in table \ref{strucformspinor}.
\begin{table}[tp]
\begin{center}
\begin{tabular}{|c|c|c|c|}
\hline
& Structures & Forms & Spinors \\
\hline
\rowcolor{blue!40}
Almost compl.\ structure & $J$ & $\Omega$ & NA \\
\rowcolor{blue!10}
integrability & $N_J=0$ & $\d \Omega = \theta \wedge \Omega$ & NA \\
\rowcolor{blue!40}
Pre-sympl.\ structure & $\omega$ & $c \, e^{i \omega}$ & NA \\
\rowcolor{blue!10}
integrability & $\d\omega=0$ & $\d (c \, e^{i\omega}) = \theta \wedge c e^{i\omega}$ & NA \\
\rowcolor{blue!40}
(S)U($d/2$)-structure & $J,\omega$ & $\Omega, e^{i\omega}$ & $\eta$ \\
\rowcolor{blue!10}
compatibility       & $J^i{}_k \omega_{ij} J^j{}_l = \omega_{kl}$ & $\Omega \wedge \omega = 0$ & automatic \\
\rowcolor{blue!20}
integr.\ (K\"ahler) & $N_J=\d\omega=0$ & $\begin{array}{c} \d \Omega = \theta \wedge \Omega \\ \d (c e^{i\omega}) = \theta' \wedge c e^{i\omega}\end{array}$ & NA \\
\rowcolor{blue!10}
Calabi-Yau          & NA & $\d \Omega=\d e^{i\omega}=0$ & $\nabla_i \eta = 0$  \\
\hline
\end{tabular}
\caption{The languages of respectively structures, forms and spinors. ``NA'' indicates the description
does not exist. For instance, the spinor language can only be used if there is both an almost complex structure
and an pre-symplectic structure leading to SU($d/2$).}
\label{strucformspinor}
\end{center}
\end{table}

\subsection{Exercises}

\begin{exc}[easy]
Show that an almost complex structure can only exist on an even-dimensional manifold and that
it reduces the structure group to GL($d/2$,$\mathbb{C}$). Also show that if there is a Hermitian metric
($J^i{}_k g_{ij} J^j_{l}=g_{kl}$) the structure group further
reduces to U($d/2$).
\end{exc}
\begin{exc}[easy]
Show that a globally defined everywhere non-vanishing $(d/2,0)$-form $\Omega$ further reduces the structure
group to SL($d/2$,$\mathbb{C}$) or in the case of a Hermitian metric to SU($d/2$).
\end{exc}
\begin{exc}[easy]
Write \eqref{lieder} in index notation.
\end{exc}
\begin{exc}[easy]
Show \eqref{bracketprop} from \eqref{interiorder} and \eqref{exteriorder}.
\end{exc}
\begin{exc}[easy]
Show the decomposition \eqref{extdecomp}.
\end{exc}
\begin{exc}[intermediate]
Show the decomposition \eqref{extdecompintegr}.
\end{exc}
\begin{exc}[intermediate]
If a (1,0)-form transform as the $\mathbf{3}$ of SU(3), show that a (2,1)-form transforms
as $\mathbf{3}\otimes\mathbf{3} = \mathbf{\bar{3}} \oplus \mathbf{6}$.  Show that the $\mathbf{1}+\mathbf{1}$-part in $\d \omega$ and $\d \Omega$ are described
by the same torsion class $\mathcal{W}_1$.
\end{exc}

\clearpage

\thispagestyle{beginsection}
\section{Fundamentals of Generalized Complex Geometry}
\label{gengeomfund}

In  this chapter we introduce the formalism of Generalized Complex Geometry.
Most of the concepts explained below were first introduced in Hitchin's paper \cite{hitchinGCY}
or in Gualtieri's PhD thesis \cite{gualtieri} (see also \cite{gualtierinew}).

\subsection{The generalized tangent bundle and the Courant bracket}

The key point of Generalized Complex Geometry is that one replaces the tangent bundle $T M$ by the
sum of the tangent and the cotangent bundle $TM \oplus T^* M$ --- called the {\em generalized tangent bundle} --- and
then extends the concepts  introduced in the previous chapter.

On the generalized tangent bundle, as opposed to the ordinary tangent bundle, already lives a canonical
metric $\mathcal{I}$, defined in the following way: for two generalized tangent vectors $\mathbb{X}= X + \xi$,
$\mathbb{Y}= Y + \eta$, where $X,Y \in TM$ are the vector parts and $\xi,\eta \in T^* M$
the one-form parts, we have
\eq{
\mathcal{I}(\mathbb{X},\mathbb{Y}) = \frac{1}{2} \left( \xi(Y) + \eta(X) \right) \, .
}
This metric is maximally indefinite, i.e.\ it has signature $(d,d)$. It should therefore not be confused with an ordinary
metric, for which, as we will see, we should introduce extra structure.  All this means that the structure
group is not completely generic, but reduces instead to O($d$,$d$). In fact, a natural volume-form $\text{vol}_{\mathcal{I}} \in \Gamma(\Lambda^{2d}(TM \oplus T^* M))$
can be associated to $\mathcal{I}$ ordering all vectors before all one-forms:
\eq{
\text{vol}_{\mathcal{I}} = \frac{1}{(d!)^2} \, \epsilon^{i_1 \ldots i_d} \frac{\partial}{\partial x^{i_1}} \wedge \cdots \wedge \frac{\partial}{\partial x^{i_d}} \wedge
\epsilon_{i_1 \ldots i_d} \d x^{i_1} \wedge \cdots \wedge \d x^{i_d} \, .
}
Despite the appearance of the $\epsilon$-tensor, since it actually appears twice, vol$_{\mathcal{I}}$ does not depend
on a choice of orientation on $M$. As a result, the structure group further reduces to SO($d,d$).
It is generated by elements of the form
\eq{
\left(\begin{array}{cc} A & \mathbf{0} \\ \mathbf{0} & (A^T)^{-1} \end{array}\right) \, , \quad
e^{B}=\left(\begin{array}{cc} \bbone & \mathbf{0} \\ B & \bbone \end{array}\right) \, , \quad
e^{\beta}=\left(\begin{array}{cc} \bbone & \beta \\ \mathbf{0} & \bbone \end{array}\right) \, ,
}
where $A \in$GL($d,\mathbb{R}$), the by now familiar structure group of the tangent bundle. In addition,
there are the {\em $B$-transforms} and {\em $\beta$-transforms}. A $B$-transform is described by a two-form $B$ and we
can rewrite the action of the above matrix on a generalized tangent vector conveniently as follows
\eq{
\label{Btransform}
e^B: X+\xi \rightarrow X+ (\xi - \iota_X B) \, ,
}
Analogously the $\beta$-transform is generated by an antisymmetric two-vector $\beta$ and sends:
\eq{
e^\beta: X+\xi \rightarrow (X - \iota_\xi \beta) + \xi \, ,
}

Now that we have discussed the structure of the generalized tangent bundle, we
should introduce an analogue of the Lie-bracket. This is the Courant bracket.
\begin{defi}
For two generalized vector fields $\mathbb{X}=X+\xi$, $\mathbb{Y}=Y+\eta \in \Gamma(T M \oplus T^* M)$
the {\em Courant bracket} is given by:
\eq{
\label{courant}
[X+\xi,Y+\eta]_C=[X,Y]+ \mathcal{L}_X \eta - \mathcal{L}_Y \xi - \frac{1}{2} d(\iota_X \eta - \iota_Y \xi) \, .
}
\end{defi}
With this definition the projection $\pi_{TM}: TM \oplus T^* M \rightarrow T M$, defined as $\pi_{TM}(X+\xi)=X$, satisfies
\eq{
\label{anchor}
[\pi_{TM}(\mathbb{X}),\pi_{TM}(\mathbb{Y})]=\pi_{TM}([\mathbb{X},\mathbb{Y}]_C) \, .
}
Let us note that the Courant bracket is {\em not} a Lie-bracket since it does not satisfy the Jacobi identity.
The Courant bracket is invariant under diffeomorphisms (the action of $A$ above on the generalized tangent bundle) just
as the Lie bracket, and in addition under $B$-transforms with $\d B=0$. In fact, to take into account $B$-transforms
for which $\d B \neq 0$, we can readily introduce an $H$-twisted
version of the Courant bracket, where $H$ is a closed three-form:
\eq{
\label{couranttwisted}
[\mathbb{X},\mathbb{Y}]_H=[\mathbb{X},\mathbb{Y}]_C + \iota_X \iota_Y H \, .
}
This satisfies the following property under a general $B$-transform
\eq{
\label{propcourant}
[e^B \mathbb{X}, e^B \mathbb{Y}]_{H-dB} = e^B [\mathbb{X},\mathbb{Y}]_H \, .
}
We will show that the three-form $H$ can be related to the NSNS three-form of type II
supergravity.

Suppose that we want to study the $H$-twisted Courant bracket. We want to investigate now whether it possible
to equivalently study the untwisted Courant bracket by making a $B$-transform
of the vector fields with $H = \d B$ so that the $H$-field vanishes on the left-hand side of eq.~\eqref{propcourant}.
Indeed, one can describe an $H$-field  by providing a two-form $B_{\alpha}$
in every patch $U_{\alpha}$ such that $H=\d B_{\alpha}$.
Of course, since $B$ is only a potential we are free to shift it by
a closed form, since that does not change the physical field $H$. As we saw this is a symmetry
of the Courant bracket just like a diffeomorphism. One might try to proceed and apply
a $B$-transform over $B_{\alpha}$ in every patch in order to obtain a description with
an untwisted Courant bracket. One has to be careful though, since when $H$ is non-trivial, $B$
is not a globally defined two-form. In fact, mathematicians would say it is a
curving of a connection on a {\em gerbe}, see e.g.~\cite{hitchingerbe}.
This means that on the overlap between two patches $U_\alpha \cap U_\beta$ one must allow
for a gauge transformation
\eq{
B_{\alpha} - B_\beta = d \Lambda_{\alpha\beta} \, .
}
These gauge transformations must satisfy the consistency condition that
on the triple overlap $U_\alpha \cap U_\beta \cap U_\gamma$,
\eq{
\Lambda_{\alpha\beta} + \Lambda_{\beta\gamma} + \Lambda_{\gamma_\alpha} = \d \Lambda_{\alpha\beta\gamma} \, .
}
If the flux is quantized $H \in H^3(M,\mathbb{Z})$ then $t_{\alpha\beta\gamma}=e^{i\Lambda_{\alpha\beta\gamma}}$
are the gerbe transition functions and satisfy the following cocycle condition on $U_{\alpha} \cap U_{\beta} \cap U_\gamma \cap U_\delta$:
\eq{
t_{\beta\gamma\delta} t_{\alpha\gamma\delta}^{-1} t_{\alpha\beta\delta} t_{\alpha\beta\gamma}^{-1} = 1 \, .
}
Note that these conditions are very similar to the ones for a U(1) gauge bundle with
connection $A$, except that for a gauge bundle there is one step less in the above procedure.
If we now perform
a $B$-transform in every patch to eliminate the $H$-field,
the gerbe structure carries over to the generalized tangent bundle \cite{hitchinbracket}, which we now call
$E$. On the overlap of two patches $U_\alpha \cap U_\beta$ we find
\eq{
\mathbb{X}_\alpha = e^{B_\beta-B_\alpha} \mathbb{X}_\beta = e^{-\d \Lambda_{\alpha\beta}} \mathbb{X}_\beta \, .
}

We will see later that Generalized Complex Geometry allows in this way to describe both the metric $g$ and the $B$-field
on the same level. In this review we will usually not perform this $B$-transform, but instead continue
to work with the $H$-twisted Courant bracket. However, if one opts to make the $B$-transform, one should
remember to replace the generalized tangent bundle $TM \oplus T^* M$ by the twisted bundle $E$ in every definition
in the following.

After we tried to snow the reader under with details on the $B$-transform, he/she might still wonder
what the meaning of the $\beta$-transform is, which is {\em not} a symmetry of the Courant bracket. In fact,
from the above construction follows that the structure group of $E$ is not the whole SO($d,d$)-group of symmetries
of the canonical metric, but rather only the symmetries of the Courant bracket, let us call it the {\em generalized diffeomorphism group},
\eq{
G_{\text{gendiff}} = \text{GL}(d,\mathbb{R}) \rtimes G_{B,\text{closed}} \, ,
}
the semi-direct product of diffeomorphisms and $B$-transforms with closed $B$.
In \cite{grangebeta,gmpw,fluxtwist} it was suggested that the $\beta$-transform is related to two T-dualities and might be used for
the construction of non-geometry in the style of the T-fold approach of \cite{hullTfold1,hullTfold2}.
For alternative attempts to give this a rigourous meaning see \cite{ellwoodnongeom,haganongeom}.

\subsection{Generalized complex structures}
\label{gcssec}

We can now straightforwardly extend the concept of (almost) complex structures to Generalized Complex Geometry.
\begin{defi}
A {\em generalized almost complex structure} is a map
\eq{
\mathcal{J}: TM \oplus T^* M \rightarrow T M \oplus T^* M \, ,
}
respecting the bundle structure: $\pi(\mathcal{J} \mathbb{X})=\pi(\mathbb{X})$,
and which satisfies two further conditions:
it squares to minus one
\eq{
\label{squareone}
\mathcal{J}^2 = - \bbone \, ,
}
and the canonical metric is Hermitian
\eq{
\label{herm}
\mathcal{I}(\mathcal{J} \mathbb{X}, \mathcal{J} \mathbb{Y})  = \mathcal{I}(\mathbb{X},\mathbb{Y}) \, .
}
\end{defi}
A generalized almost complex structure reduces the structure group from SO($d$,$d$) to U($d/2$,$d/2$). Associated to $\mathcal{J}$ are two subbundles
$L_{\mathcal{J}},\bar{L}_{\mathcal{J}} \subset (T M \oplus T^* M) \otimes \mathbb{C}$ with fibers respectively
the $(\pm i)$-eigenspaces of the action of $\mathcal{J}$ on the fiber over each point. In fact, the introduction of a generalized
complex structure $\mathcal{J}$ is equivalent to the specification of a maximally isotropic subbundle $L$ satisfying $L \cap \bar{L}=\{0\}$. The correspondence
is of course given by $L_{\mathcal{J}}=L$. We used here the following definition:
\begin{defi}
\label{isotropic}
A subbundle $L$ is {\em isotropic} if
\eq{
\mathcal{I}(\mathbb{X},\mathbb{Y}) = 0 \, , \quad \text{for all} \,\, \mathbb{X},\mathbb{Y} \in L \, .
}
Moreover, it is {\em maximally isotropic} if its rank is half of the rank of $TM \oplus T^* M$ (namely $d$),
which is the maximal rank for an isotropic subbundle.
\end{defi}

To extend the concept of integrability it is natural to replace the Lie bracket
by the (twisted) Courant bracket.

\begin{defi}
$\mathcal{J}$ is $H$-integrable, making it a generalized complex structure,
if $L_{\mathcal{J}}$ is involutive under the $H$-twisted Courant bracket:
\eq{
[\mathbb{X},\mathbb{Y}]_H \in \Gamma(L_{\mathcal{J}}) \, , \quad \text{for all} \,\, \mathbb{X},\mathbb{Y} \in \Gamma(L_{\mathcal{J}}) \, .
}
\end{defi}
Note that before we had a different definition of integrability and used the Frobenius theorem (theorem \ref{theor:frob}) to equate
integrability with involutivity. Here, it is convenient to directly define the integrability as
involutivity of the Courant bracket. Nevertheless, there is a similar theorem, which
states that if a generalized complex structure is involutive, adapted coordinates can be introduced.
We will however not use this any further.
\begin{theor}[Generalized Darboux theorem, see Theorem 4.35 (and generalization in section 4.9) of \cite{gualtieri}]
Any regular point in a manifold with a $H$-twisted integrable generalized complex structure has a neighbourhood that is equivalent, via a
diffeomorphism {\em and a $B$-transform} with $\d B=H$, to the product of an open set in $\mathbb{C}^k$ (described by complex coordinates)
and an open set in the standard symplectic space $(\mathbb{R}^{d-2k},\omega_0)$ (described by Darboux coordinates).
Here $k$ is the type of the associated pure spinor, which we will define in definition \ref{typedef}.
\end{theor}

From the Generalized Darboux theorem we find that in adapted coordinates a generalized complex structure
interpolates between a complex structure (example \ref{complintegr}) and a symplectic structure (example \ref{sympintegr}).
In fact, an ordinary complex structure and a symplectic structure are two special cases of a generalized complex structure
as the following two examples show:
\begin{ex}[Almost complex structure]
\label{complex}
  From an ordinary complex structure $J$ we can construct a generalized complex structure as follows:
\eq{
\label{complexpr}
\mathcal{J}_J = \left(\begin{array}{cc} -J & \mathbf{0} \\ \mathbf{0} &  J^T\end{array} \right) \, .
}
It is easy to show that $\mathcal{J}_J$ is $H$-integrable iff $J$ is integrable and $H$ is of type $(2,1)\oplus (1,2)$.
\end{ex}
\begin{ex}[Symplectic structure]
\label{symplex}
From a symplectic structure $\omega$ on the other hand we can construct the following
generalized complex structure
\eq{
\label{symplexpr}
\mathcal{J}_{\omega} = \left(\begin{array}{cc} \mathbf{0} &  \omega^{-1} \\  -\omega & \mathbf{0} \end{array} \right) \, .
}
$\mathcal{J}_{\omega}$ is $H$-integrable iff $\d \omega=0$ and $H$=0.
\end{ex}

It follows that both essential parts of an SU(3)-structure, namely the almost complex structure and the pre-symplectic
structure are described in a completely uniform way. This also suggests that Generalized Complex Geometry can be used
to describe mirror symmetry, which interchanges the complex and symplectic structure. We will not go into details, but
refer to \cite{louismicu2,mirror1,mirror2,mirror3,mirror4,mirror5,granaN2part2}.

\begin{ex}[Holomorphic Poisson structure]
See exercise \ref{excpoisson}.
\end{ex}

\subsection{Polyforms and pure spinors}
\label{puresec}

\subsubsection*{Algebraic properties of polyforms and pure spinors}

Let us now see how we can equivalently describe a generalized almost complex structure in the language of forms.
First, introduce a polyform $\phi \in \Omega^\bullet(M)$, this is a sum of forms of different dimensions.
On such a polyform a section $\mathbb{X}=(X,\xi)$ of the generalized tangent bundle acts in a natural way as follows
\eq{
\mathbb{X} \cdot \phi = \iota_X \phi + \xi \wedge \phi \, .
}
With this definition the elements of the generalized tangent bundle act as a Clifford algebra, i.e.\ as gamma-matrices associated
to the SO($d$,$d$)-metric:
\eq{
\{ \mathbb{X}, \mathbb{Y} \} \cdot \phi = \left( \mathbb{X}\cdot \mathbb{Y} + \mathbb{Y} \cdot \mathbb{X} \right) \cdot \phi = 2 \, \mathcal{I}(\mathbb{X},\mathbb{Y}) \phi \, .
}
Compare this with the behaviour of gamma-matrices associated to a metric $g_{ij}$, for which we find $\{ v^i \gamma_i, v^j \gamma_j \} = 2 v^i v^j g_{ij}$.
Since they are acted upon by gamma-matrices, one might conclude that polyforms transform in the spin representation of Spin($d$,$d$). In more detail, the irreducible representations of Spin($d$,$d$)
are Majorana-Weyl. Imposing the Majorana condition would then amount to restricting to real forms, while the Weyl condition
leads to restricting to polyforms with all dimensions even (positive chirality) or odd (negative chirality). In fact, it is almost true that real polyforms of definite chirality transform as
Majorana-Weyl spinors: considering carefully how the spinors transform under
the GL(d,$\mathbb{R}$)-subgroup \cite[Example 2.12]{gualtieri} one finds for the positive and negative chirality spin bundle $S^\pm$
\eq{
S^{\pm} \simeq \Lambda^{E/O} T^* M \otimes |\det T^* M|^{-1/2} \, .
}
This means that the isomorphism between spinors and polyforms depends on the choice of a volume-form $\epsilon \in \Gamma(\det T^*M)$:
\eq{
\label{isospinform}
\phi_s \rightarrow \phi = \epsilon^{1/2} \, \phi_s \, .
}
We called $\epsilon$ a volume-form since a section of $\det T^*M$ transforms in the same way as a section of $\Lambda^d T^*M$.

Another way to see this is as follows: instead of the
usual bilinear form that one can define on spinors using
the charge conjugation matrix $C$,
\eq{
\phi_{1s}{}\!^T C \phi_{2s} \, ,
}
and which produces a scalar, one has on polyforms a natural bilinear form taking values in $\Lambda^d T^* M$, which
is called the Mukai pairing.
\begin{defi}
The {\em Mukai pairing} of two polyforms $\phi_1,\phi_2 \in \Lambda^\bullet T^* M$ is given by
\eq{
\langle \phi_1, \phi_2 \rangle = \phi_1 \wedge \sigma(\phi_2)|_{\text{top}} \, ,
}
where $\sigma$ is the operator reversing all the indices of a polyform as in \eqref{reversal}, and $|_\text{top}$ indicates
we project on the top-form part, i.e.\ the part with dimension $d$.
\end{defi}
The Mukai pairing is the
map of the bilinear form on the spinors under the isomorphism~\eqref{isospinform}. One can show that
the Mukai pairing defined in this way indeed transforms as a top-form under the elements
of Spin($d$,$d$) (while the original bilinear form on spinors is invariant). In particular,
it is easy to show that the Mukai-pairing
is invariant under the $B$-transform:
\eq{
\langle e^B \phi_1, e^B \phi_2 \rangle = \langle \phi_1,\phi_2 \rangle \, .
}
Furthermore we have the following symmetry property:
\eq{
C = (-1)^{d(d-1)/2} C^T \Longleftrightarrow \langle \phi_1, \phi_2 \rangle = (-1)^{d(d-1)/2} \langle \phi_2,\phi_1 \rangle \, .
}
So for $d=6$, which is the case of interest for compactifications from ten to four dimensions,
the Mukai pairing is antisymmetric. Two further properties of the Mukai pairing are:
\subeq{\al{
\label{mukaiprop1}
\langle \mathbb{X}\cdot\phi_1 , \phi_2 \rangle=- (-1)^d \langle \phi_1 , \mathbb{X}\cdot\phi_2 \rangle\, , \\
\label{mukaiprop2}
\int_M \langle d_H\phi_1 , \phi_2 \rangle =(-1)^d \int_M \langle \phi_1 , d_H\phi_2 \rangle\ .
}}
As of now we will only work with polyforms and call them spinors of
Spin($d$,$d$), implicitly assuming the isomorphism \eqref{isospinform}.

Next, we define the null space of a spinor $\phi$.
\begin{defi}
The null space of $\phi$ is the subbundle $L_\phi$
consisting of all the annihilators of $\phi$ i.e.\
\eq{
L_{\phi}= \{ \mathbb{X} \in TM \oplus T^* M \, | \, \mathbb{X} \cdot \phi =0\} \, .
}
\end{defi}
$L_\phi$ is isotropic, since
\eq{
2 \, \mathcal{I}(\mathbb{X},\mathbb{Y}) \phi = \left( \mathbb{X} \cdot \mathbb{Y} + \mathbb{Y} \cdot \mathbb{X} \right) \cdot \phi = 0  \Rightarrow \mathcal{I}(\mathbb{X},\mathbb{Y}) = 0 \, ,
}
for all $\mathbb{X},\mathbb{Y} \in L_\phi$. We will be interested in spinors with a null space of maximal rank, since it turns out
that one can associate pure spinors of this type to a generalized almost complex structure.
\begin{defi}
\label{purespinordef}
$\phi$ is a {\em pure spinor} if its null space $L_\phi$ is maximally isotropic, i.e.\ if its rank is $d$, half the rank of $TM \oplus T^* M$.
\end{defi}
We remark here that the concept of pure spinors can also be introduced for the more familiar
spinors of Spin($d$) with $d$ even in exactly the same way, i.e.\ a spinor is pure if the number of independent gamma-matrices that
annihilate the spinor is maximal and thus $d/2$ (see also footnote \ref{purefootnote}). It turns out that for $d=6$, every Weyl spinor is pure, while for higher dimensions the pure spinor
constraints are non-trivial. Pure spinors for $d=10$ play an important role in the covariant quantization of the superstring \cite{berkovitsformalism} (for
a review see \cite{berkovitsformalismreview}).

One can now introduce a complex pure spinor for which the null space is an isotropic subbundle of the complexification
$(T M \oplus T^* M) \otimes \mathbb{C}$ of the generalized tangent bundle. Then, one can associate to every generalized almost complex
structure $\mathcal{J}$ a pure spinor $\phi_{\mathcal{J}}$ such that its null space is the (+i)-eigenbundle of $\mathcal{J}$
\eq{
L_{\phi_{\mathcal{J}}} = L_{\mathcal{J}} \, .
}
Note that this determines the pure spinor only up to an overall factor. Generically, we also allow the overall factor to change between
patches. Conversely, using the above formula one can construct a generalized almost complex structure from every complex non-degenerate pure spinor ($\phi \wedge \bar{\phi} \neq 0$).
There is thus a one-to-one correspondence between generalized almost complex structures and complex non-degenerate pure spinors modulo an overall
factor.

\subsubsection*{Integrability}

To express the integrability of $\mathcal{J}$ in terms of $\phi_{\mathcal{J}}$ we are looking for
a generalization of eq.~\eqref{bracketprop}, i.e.\ we would like to express the Courant bracket as a derived bracket. Introducing a twisted exterior derivative $\d_H$ acting on a
polyform $\phi$ as
\eq{
\d_H \phi = \d \phi + H \wedge \phi \, ,
}
which satisfied $\d_H^2=0$ since $H$ is closed, we can easily show that
\eq{
\label{courantbracketprop}
[\mathbb{X},\mathbb{Y}]_H \cdot \phi = [\{\mathbb{X},\d_H \},\mathbb{Y}] \cdot \phi - \d\left(\mathcal{I}(\mathbb{X},\mathbb{Y})\right) \wedge \phi \, ,
}
for all $\phi \in \Omega^{\bullet}(M)$, and for all sections $\mathbb{X},\mathbb{Y}$ of the generalized tangent bundle.
Restricting to the isotropic subbundle $L_\phi$, the last term drops out and we find the relation we were looking for.
We can think of $\mathbb{X} \cdot$ as the generalization of $\iota_X$. Moreover, as shown in \cite[Proposition 3.16]{gualtieri},
on an integrable and isotropic subbundle the Courant bracket satisfies the Jacobi identity, turning $(L_\phi,[\cdot,\cdot]_H)$ into a Lie algebra.
In fact, from \eqref{anchor} follows that $\pi_{TM}$ is
a Lie-algebra homomorphism $\pi_{TM}:L_\phi \rightarrow
T M$. A Lie algebra equipped with such a homomorphism onto the Lie algebra $T M$ is called a {\em Lie algebroid}. Therefore on the forms of $L_\phi$
a Lie algebroid derivative $\d_{L_\phi}$ can be defined
in a similar way as in definition~\ref{exteriorderdef}:
\begin{defi}
\label{liealgebroidderdef}
Given a Lie algebroid $(L,\pi_{TM},[\cdot,\cdot]_H)$, the Lie algebroid derivative $\d_L : \Gamma(\Lambda^k L^*) \rightarrow \Gamma(\Lambda^{k+1} L^*)$
acts on a form $\alpha \in \Gamma(\Lambda^k L^*)$ and produces an $(k+1)$-form $\d_L \alpha$ such that
\eq{\spl{
\label{liealgebroidder}
\d_L \alpha(\mathbb{Y}_0,\ldots,\mathbb{Y}_{l}) = & \sum_{0\le a \le l} (-1)^{a} \pi_{TM}(\mathbb{Y}_a)\left(\alpha(\mathbb{Y}_0,\ldots,\hat{\mathbb{Y}}_a,\ldots,\mathbb{Y}_l)\right) \\
& + \sum_{0\le a<b \le l} (-1)^{a+b} \alpha([\mathbb{Y}_a,\mathbb{Y}_b]_H,\mathbb{Y}_0,\ldots,\hat{\mathbb{Y}}_{a},\ldots,\hat{\mathbb{Y}}_b,\ldots,\mathbb{Y}_l) \, ,
}}
for all $\mathbb{Y}_0,\ldots,\mathbb{Y}_l \in \Gamma(L)$.
\end{defi}
We will turn to the relation
between $\d_H$ and $\d_L$ in a moment. For now let us continue our discussion of integrability in terms of $\d_H$.
We find from eq.~\eqref{courantbracketprop} that
\eq{
[\mathbb{X},\mathbb{Y}]_H \cdot \phi = \mathbb{X} \cdot \mathbb{Y} \cdot \d_H \phi \, ,
}
for all $\mathbb{X},\mathbb{Y} \in \Gamma(L_\phi)$. So $L_{\phi}$ is involutive if
the right-hand side vanishes, which means $\d_H \phi$ can be written as $\mathbb{X}\cdot \phi$ for some $\mathbb{X}$.
In terms of the decomposition we will define in eq.~\eqref{decomp} it belongs to $\Gamma(U_{d/2-1})$. It follows that:
\begin{theor}
$\mathcal{J}$ is integrable iff
\eq{
\label{spinint}
\d_H \phi_{\mathcal{J}} = \mathbb{X} \cdot \phi_{\mathcal{J}} \, ,
}
for some $\mathbb{X}$.
\end{theor}
Note that if $\phi_{\mathcal{J}}$ satisfies the above condition, it is also
satisfied for $\phi'_{\mathcal{J}}=f \phi_{\mathcal{J}}$ with $f$ an arbitrary function if
we take $\mathbb{X}' = \mathbb{X} + \d f$. It follows that the integrability condition does not depend
on the undetermined overall factor.

If, however, we find some way to fix the overall factor, and there exists a globally defined
complex pure spinor, the structure group of the generalized tangent bundle further reduces to SU($d/2$,$d/2$).
It is then natural to impose the stronger constraint that $\mathbb{X}=0$.
\begin{defi}
\label{genCYhitchin}
The complex pure spinor $\phi$ is {\em generalized Calabi-Yau \`a la Hitchin}
if it satisfies
\eq{
\d_H \phi = 0 \, .
}
\end{defi}
This terminology is a bit confusing since this is not really a suitable generalization of a Calabi-Yau geometry. An ordinary Calabi-Yau is
namely a special case of a construction with {\em two} generalized complex structures (corresponding
to the ordinary complex structure and the symplectic structure). We will give a more appropriate generalization
of a Calabi-Yau geometry in definition \ref{genCYgualtieri}.

\subsubsection*{Decomposition of polyforms and the twisted exterior derivative}

A complex pure spinor $\phi$ and an associated isotropic null space $L=L_\phi$ induces
a decomposition of the space of polyforms as follows
\eq{
\label{decomp}
\Lambda^\bullet T^* M\otimes\mathbb{C}=\bigoplus_{-d/2\le k \le d/2} U_k\ ,
}
where
\eq{
U_k=\Lambda^{d/2-k}\bar L\cdot \phi \, ,
}
i.e.\ $U_k$ is the subbundle of polyforms that one gets by acting with an antisymmetric
product of $d/2-k$ generalized vectors of $\bar{L}$ on $\phi$. So $\phi$ provides an isomorphism between
\eq{
\label{isoL}
\Gamma(\Lambda^{d/2-k}\bar{L}) \rightarrow U_k : \alpha \rightarrow \alpha\cdot \phi \, .
}
Such a decomposition is called a {\em filtration}.
One can look at this decomposition as building up the spinor
representation by acting with anti-holomorphic gamma-matrices --- taking the role of creation operators ---
on the ``null state'' (see e.g.~\cite[Appendix B]{polchinskiII} for exactly the same construction for
spinors of Spin($d$)). One can give an alternative definition of $U_k$ as the $ik$-eigenbundle of $\mathcal{J}$, acting
in the spinor representation on forms.  To be specific, given a local frame $\mathbb{X}_a$, $a=1,\ldots, 2d$, of the generalized
tangent bundle the action of $\mathcal{J}$ in the spinor representation on an arbitrary $\phi'$ is given by:
\eq{
\label{Jspinorrep}
\mathcal{J}\cdot \phi' = \frac{1}{2} \, \cali^{ab} \mathcal{J}(\mathbb{X}_a) \cdot  \mathbb{X}_b \cdot \phi' \, ,
}
where $\mathcal{I}^{ab}$ is the inverse matrix of $\mathcal{I}_{ab}=\cali(\mathbb{X}_a,\mathbb{X}_b)$.
Note that for the original pure spinor $\phi$ we have
\eq{
\phi\in\Gamma(U_{d/2}) \, .
}
Furthermore, for any polyform $\phi' \in \Gamma(U_k)$ we have $\bar\phi'\in\Gamma(U_{-k})$. The decomposition is also compatible with the
Mukai pairing:
for $\phi_1\in \Gamma(U_k)$
\eq{
\langle \phi_1,\phi_2\rangle=0\quad {\rm if} \quad \phi_2|_{U_{-k}}=0\ ,
\label{mukcomp}
}
where $\phi_2|_{U_{-k}}$ denotes the projection on $U_{-k}$.

For the exterior derivative one finds through an argument by induction that
\eq{
\label{maxder}
\d_H : \Gamma(U_k) \rightarrow \Gamma(U_{k-3})\oplus\Gamma(U_{k-1})\oplus \Gamma(U_{k+1}) \oplus \Gamma(U_{k+3}) \, .
}
Furthermore one can show that if $L$ is integrable the terms in $\Gamma(U_{k \pm 3})$ become zero, and
$\d_H$ splits as follows
\eq{
\d_H = \partial_H + \bar{\partial}_H \, ,
}
where
\eq{
\partial_H : \Gamma(U_k) \rightarrow \Gamma(U_{k+1}) \, , \qquad
\bar{\partial}_H : \Gamma(U_k) \rightarrow \Gamma(U_{k-1}) \, ,
}
satisfying
\eq{
\partial_H^2 =0 \, , \qquad \bar{\partial}_H^2 = 0 \, , \qquad \partial_H \bar{\partial}_H + \bar{\partial}_H \partial_H = 0 \, .
}
This allows us to define the {\em generalized Dolbeault cohomology} groups
\eq{
{\rm H}_{\bar{\partial}_H}^{k}(M) = \frac{ \{\phi \in \Gamma(U_{k}) \, | \, \bar{\partial}_H \phi = 0 \}}{ \{ \phi \sim \phi + \bar{\partial}_H \lambda \, | \,
\lambda \in \Gamma(U_{k+1}) \}} \, .
}
We remark that generalized cohomology, given by
\eq{
{\rm H}_{H}^{\bullet}(M) = \frac{ \{\phi \in \Omega^\bullet(M,\mathbb{C}) \, | \, d_H \phi = 0 \}}{ \{ \phi \sim \phi + d_H \lambda \, | \,
\lambda \in \Omega^{\bullet}(M,\mathbb{C}) \}} \, ,
}
splits nicely into generalized Dolbeault cohomology groups in that every representative of a generalized cohomology group can
be written as a sum of representatives of generalized Dolbeault cohomology groups, iff the generalized
$\partial_H \bar{\partial}_H$-lemma holds (for more details see \cite{caval}).

Finally, the isomorphism \eqref{isoL} maps $\d_L$ into $\bar{\partial}_H$:
\eq{
(\d_L \alpha) \cdot \phi = \bar{\partial}_H (\alpha \cdot \phi) \, ,
}
which gives us the promised relation between $\d_L$ and $d_H$.

\subsubsection{The general form of pure spinors}

Let us now dwell for a while on the general form of pure spinors and how one can construct them without using
the cumbersome definition based on the rank of the null space.
Gualtieri shows in \cite[Theorem 4.8]{gualtieri} that every non-degenerate complex pure spinor can be written as
\eq{
\label{purespinorform}
\phi = \Omega_k \wedge e^{i\omega + B} \, ,
}
with $\omega,B$ real two-forms and $\Omega_k$ a complex {\em decomposable} $k$-form
such that
\eq{
\langle \phi , \bar{\phi} \rangle = (-1)^{k(k-1)/2} \frac{2^{d/2-k}}{(d/2-k)!} \Omega_k \wedge \bar{\Omega}_k \wedge \omega^{d/2-k}\neq 0 \, .
}
$k$ is called the {\em type} of the pure spinor.
\begin{defi}
\label{typedef}
The type of a pure spinor is the lowest appearing form-dimension.
\end{defi}
We can consider pureness as the generalization to Generalized Complex Geometry of decomposability.

\begin{ex}
The pure spinor associated to $\mathcal{J}_J$ of example \ref{complex} is
\eq{
c \, \Omega \, ,
} for some nowhere-vanishing function $c$ and $\Omega$ the complex decomposable three-form associated to $J$. Indeed,
the null space is $\bar{L} \oplus \Lambda^{(1,0)} T^*M$ which is also the $(+i)$-eigenspace of $\mathcal{J}_J$ in eq.~\eqref{complexpr}.
The type is 3. Eq.~\eqref{spinint} leads to
\eq{
\d \Omega = \bar{\mathcal{W}}_5 \wedge \Omega \, , \qquad H \wedge \Omega  = 0 \, .
}
\end{ex}
\begin{ex}
The pure spinor associated to $\mathcal{J}_{\omega}$ of example \ref{symplex} is $c e^{i\omega}$ for some
function $c$. Indeed, the null space is given by
\eq{
\{ X+\xi \in TM \oplus T^*M \, | \, \xi = -i \, \iota_X \omega \} \, ,
}
which is also the $(+i)$-eigenspace of $\mathcal{J}_\omega$ in eq.~\eqref{symplexpr}.
The type is 0. Eq.~\eqref{spinint} leads to
\eq{
\d \omega = 0 \, , \qquad H=0 \, .
}
\end{ex}
\begin{ex}
Consider a polyform
\eq{
\phi = \phi_1 + \Omega \, ,
}
with $\Omega$ a complex decomposable three-form, associated to an ordinary almost complex structure $J$, and
$\phi_1$ a one-form which is not everywhere non-zero. For this to be a pure spinor, $\phi_1$ must be
of type $(1,0)$ with respect to $J$. This pure spinor exhibits the so-called {\em type jumping phenomenon}. Generically
the type is 1, except on the locus where $\phi_1$ is zero, where the type jumps to 3, and the structure
becomes locally a complex structure. Suppose we impose the Calabi-Yau condition \`a la Hitchin.
We find $\d \Omega=0$, which implies that $J$ is integrable, and $\d \phi_1=0$. Locally, we
can then write $\phi_1 = \partial w$. In \cite{lucasuppot,deforms} it was shown
that $w$ can be associated to the superpotential of a D3-brane (see also eq.~\eqref{Dbranesuppot} for the general expression for the superpotential of a D-brane).
\end{ex}

\subsubsection*{Constructing a pure spinor using Hitchin's theorem in $d=6$}

In fact, the characterization of a pure spinor as a spinor of the form \eqref{purespinorform}
is still quite cumbersome for $k>1$ because of the condition that $\Omega_k$ should
be decomposable. We will describe now Hitchin's approach \cite{hitchinGCY}, which
for $d=6$ allows to construct a pure spinor from any stable real spinor through the
Hitchin function. First we define a $\Lambda^6 T^* M$-valued two-form
$\mathcal{K} \in \Gamma(\Lambda^2(T M \oplus T^* M)\otimes \Lambda^6 T^* M)$
associated to a real polyform $\rho \in \Omega^\bullet(M) $ as follows:
\eq{
\mathcal{K}(\mathbb{X},\mathbb{Y}) = \langle \rho, \frac{1}{2} \, \left(\mathbb{X} \cdot \mathbb{Y}- \mathbb{Y}\cdot\mathbb{X}\right)\cdot \rho \rangle \, ,
}
for any two generalized tangent vector fields $\mathbb{X},\mathbb{Y}$. We can raise the indices with the metric $\mathcal{I}$.
Let us calculate the following quartic pseudo-scalar:
\eq{
q(\rho) = \frac{1}{12} \tr (\mathcal{I}^{-1} \mathcal{K})^2 \, .
}
$q(\rho)$ is not really a scalar since it transforms as $(\Lambda^6 T^* M)^2$ under diffeomorphisms.
Under such diffeomorphisms the sign does not change, so that one can assign a definite sign to $q(\rho)$.
In practice, choose any volume-form $\epsilon$ and read off the sign of the scalar $q(\rho)/\epsilon^2$, which does
not depend on the choice of $\epsilon$.
\begin{defi}
A real form $\rho$ is stable if $q(\rho) \neq 0$ everywhere.
\end{defi}
Hitchin showed the following theorem, which can be used to construct pure spinors:
\begin{theor}
\label{hitchintheorem}
For a real stable polyform $\rho$
\begin{enumerate}
\item $q(\rho)>0$ everywhere iff there exist $\alpha,\beta$ real pure spinors such that $\rho=\alpha+\beta$
and $\langle \alpha, \beta \rangle \neq 0$. The pure spinors $\alpha$ and $\beta$ are unique up to ordering.
\item $q(\rho)<0$ everywhere iff there exists a pure spinor $\phi$ such that $\rho=\Re \phi$ and
$\langle \phi,\bar{\phi}\rangle\neq 0$. $\phi$ is unique up to complex conjugation. We can then define the {\em Hitchin
function} (actually a top-form) as follows:
\eq{
\label{hitchinfunc}
H(\rho) = \sqrt{-q(\rho)} \, .
}
\end{enumerate}
\end{theor}
The real pure spinors of the first case are
associated to real subbundles of the generalized tangent bundle corresponding to a generalized almost
product structure, which we will discuss in section \ref{genprodstruc}.
As we will see in section \ref{gensubman} they are useful for describing generalized submanifolds corresponding to D-branes.

The second type is of interest here and provides
a way to construct a complex pure spinor $\phi$ from every real stable spinor $\rho$ that satisfies $q(\rho)<0$.
Indeed, given $\rho$ we take $\Re \phi=\rho$. Furthermore, we can construct the generalized almost complex structure
as
\eq{
\mathcal{J} = \pm \frac{\mathcal{I}^{-1} \mathcal{K}}{\sqrt{-\frac{1}{12}\tr (\mathcal{I}^{-1} \mathcal{K})^2}} \, ,
}
and finally the imaginary part of $\phi$ as
\eq{
\label{impart}
\Im \phi = \hat{\rho} = \mp \frac{1}{3} \mathcal{J} \cdot \rho \, ,
}
where $\mathcal{J}$ acts in the spinor representation (see eq.~\eqref{Jspinorrep}).
In particular, we find that
the imaginary part of a pure spinor is (up to a sign) completely determined in terms of the real part through \eqref{impart}.
This will be useful in constructing the effective theory, where the moduli will come from the deformations of the real
part only. The Hitchin function itself is related to the K\"ahler potential on the moduli space as we will discuss in section \ref{sec:effective}.

\subsection{SU($d/2$)$\times$SU($d/2$)-structure}
\label{sudsud}

We have seen that a globally defined invariant spinor, a prerequisite for
supersymmetry, puts an SU($d/2$)-structure on the manifold. Moreover, such an SU($d/2$)-structure corresponds
to {\em both} a pre-symplectic and an almost complex structure, satisfying some compatibility conditions.
Then, in section \ref{gcssec} we have seen that these are both special cases of a generalized almost complex
structure. In this section, we will introduce the generalization of an SU(3)-structure, which then not surprisingly
consists of {\em two} generalized almost complex structures.

\subsubsection*{Structures}

\begin{defi}\label{J1J2def}
A U($d/2$)$\times$U($d$/2)-structure consists of two generalized almost complex structures $\mathcal{J}_1$
and $\mathcal{J}_2$ such that they commute
\eq{
\label{J1J2comm}
[\mathcal{J}_1,\mathcal{J}_2]=0\, ,
}
and such that the generalized
metric $G=-\mathcal{I} \mathcal{J}_1\mathcal{J}_2$ is positive-definite.
\end{defi}

\begin{defi}
A U($d/2$)$\times$U($d/2$)-structure is an $H$-twisted {\em generalized K\"ahler structure} if both $\mathcal{J}_1$
and $\mathcal{J}_2$ are $H$-integrable.
\end{defi}

\begin{ex}[K\"ahler structure]
\label{genkahl}
An obvious example is an ordinary U($d/2$)-structure, which consists of an almost complex
structure $J$ and a symplectic structure $\omega$, from which we make two generalized
almost complex structures as in examples \ref{complex} and \ref{symplex}. The compatibility
condition $[\mathcal{J}_1,\mathcal{J}_2]=0$ indeed amounts to \eqref{hermiticity} and
the generalized metric is
\eq{
\label{genkahlermetric}
G = \left(\begin{array}{cc} - \omega J & \mathbf{0} \\ \mathbf{0} & J \omega^{-1} \end{array}\right) =
\left(\begin{array}{cc} g & \mathbf{0} \\ \mathbf{0} & g^{-1} \end{array} \right) \, ,
}
where $g$ is the ordinary metric given by \eqref{metricJom}.
\end{ex}

Let us discuss in some more detail a metric $G$ on the generalized tangent bundle
$T M \oplus T^* M$. In contrast to the canonical metric $\mathcal{I}$ this metric
is positive-definite and reduces the structure from O($d$,$d$) to O($d$)$\times$O($d$).
It satisfies $(\mathcal{I}^{-1}\mathcal{G})^2=\bbone$.
Equivalently, one can specify $G$ by giving a subbundle $C_+$ such that the canonical
metric $\mathcal{I}$ reduced to $C_+$ is positive-definite. Indeed, if $C_-$ is the orthogonal
complement to $C_+$ we can recover $G$ as follows
\eq{
G(\cdot,\cdot) = \mathcal{I}(\cdot,\cdot)|_{C_+} - \mathcal{I}(\cdot,\cdot)|_{C_-} \, .
}
Conversely, given $G$, $C_+$ is the bundle with fibers the $+1$-eigenspaces of $\mathcal{I}^{-1} \mathcal{G}$.
Adding now a generalized almost complex structure $\mathcal{J}_1$ that commutes with $\mathcal{I}^{-1} \mathcal{G}$
the structure is further reduced to U($d/2$)$\times$U($d/2$).

Since for a U($d/2$)$\times$U($d$/2)-structure the $\mathcal{J}_1$ and $\mathcal{J}_2$ commute
we can simultaneously diagonalize them and define:
\eq{\spl{
L_1^+ & = L_1 \cap L_2 \, , \\
L_1^- & = L_1 \cap \overline{{L}_2} \, .
}}
where $L_{1}=L_{\mathcal{J}_1}$ and $L_2=L_{\mathcal{J}_2}$ are the isotropic subbundles associated to $\mathcal{J}_1$ and
$\mathcal{J}_2$ respectively. From the positive-definiteness of $G$ follows that both
$L_1^+$ and $L_1^-$ have rank $d/2$. It is furthermore easy to see that $C_+$ has as fibers
the generalized vectors with equal eigenvalues under $\mathcal{J}_1$ and $\mathcal{J}_2$ while
for $C_-$ the eigenvalues are opposite. We can write this as
\eq{\spl{
C_+ \otimes \mathbb{C} & = L_1^+ \oplus \overline{L_1^+} \, , \\
C_- \otimes \mathbb{C} & = L_1^- \oplus \overline{L_1^-} \, .
}}

One can show that generically $\mathcal{I}^{-1} G$ takes the form of a $B$-transform of
the metric \eqref{genkahlermetric} of example \ref{genkahl}:
\eq{
\label{genmetric}
\mathcal{I}^{-1} G = e^{B} \left(\begin{array}{cc} \mathbf{0} & g^{-1} \\ g & \mathbf{0} \end{array} \right) e^{-B} \, ,
}
so that it contains both an ordinary metric $g$ as well as a two-form $B$. There is an easier way to extract $g$ and $B$
than explicitly writing $\mathcal{I}^{-1} G$ in the form \eqref{genmetric}. Indeed, using the fact that the elements
$\mathbb{X}_\pm=X +\xi \in C_{\pm}$ correspond to the $(\pm 1)$-eigenvectors of $\mathcal{I}^{-1} G$ respectively, one finds that the general
form of these elements is:
\eq{
\mathbb{X}_\pm=(X,\xi)=e^B(X,\pm g X)=(X,(\pm g + B) X) \, .
}
Taking $\mathbb{X}_\pm,\mathbb{Y}_\pm\in C_\pm$ we can then read off the metric $g$ and the $B$-field
as follows:
\eq{\spl{
\mathcal{I}(\mathbb{X}_\pm,\mathbb{Y}_\pm) & = \pm g(X,Y) \, , \\
\mathcal{A}(\mathbb{X}_\pm,\mathbb{Y}_\pm) & = B(X,Y) \, ,
}}
where $\mathcal{A}(\mathbb{X},\mathbb{Y}) = \frac{1}{2} \left( \eta(X) - \xi(Y) \right)$ is the
canonical antisymmetric bilinear product.

\subsubsection*{Polyforms}

We can also describe all this in the language of forms:
\begin{theor}\label{Ud2Ud2}The generalized tangent bundle has a U($d/2$)$\times$U($d/2$)-structure iff there
exist pure spinor line bundles $\Psi_1$ and $\Psi_2$ (which means they are only
defined up to an overall scalar function) that satisfy the compatibility
condition
\eq{
\label{spinorcomp}
\Psi_2 \in \Gamma(U_0) \, ,
}
where $U_i$ is the filtration \eqref{decomp} associated to $\Psi_1$, and that are such that the
metric associated to their generalized almost complex structures is positive-definite.
\end{theor}
Note that eq.~\eqref{spinorcomp} can be equivalently expressed as
\eq{
\label{compcondalt}
\Psi_1 \in \Gamma(V_0) \, ,
}
where $V_i$ is the filtration associated to $\Psi_2$. It
expresses the fact that $\mathcal{J}_1$ and $\mathcal{J}_2$ commute in terms of polyforms.
For $d=6$ it can be formulated as:
\eq{
\langle \Psi_1 , \mathbb{X} \cdot \Psi_2 \rangle = \langle \Psi_1 , \mathbb{X} \cdot \bar{\Psi}_2 \rangle = 0 \, .
}
This is the generalization of the ordinary compatibility condition in terms of forms, eq.~\eqref{hermform}.

If we can remove the ambiguity of the overall factor so that
we can construct globally defined non-degenerate pure spinors, the structure of the generalized tangent
bundle further reduces to SU($d/2$)$\times$SU($d/2$). We can then normalize the pure spinors.
\begin{defi}
The generalized tangent bundle has an SU($d/2$)$\times$SU($d/2$)-structure if there
exist globally defined pure spinors $\Psi_1$ and $\Psi_2$ that satisfy the normalization
condition
\eq{
\label{normcond}
\langle \Psi_1,\bar{\Psi}_1 \rangle = \langle \Psi_2,\bar{\Psi}_2 \rangle \neq 0 \, ,
}
and the compatibility condition
\eq{
\label{compcond}
\Psi_2 \in \Gamma(U_0) \, ,
}
where $U_i$ is the filtration \eqref{decomp} associated to $\Psi_1$, and that are such that the metric
associated to their generalized almost complex structures is positive-definite.
\end{defi}

Using the U($d/2$)$\times$U($d/2$)-structure we can define a {\em generalized Hodge decomposition} of the forms
as in eq.~\eqref{decomp}, but now in terms of the eigenvalues of both $\mathcal{J}_1$ and $\mathcal{J}_2$:
\eq{
\label{decomp2}
\Lambda^\bullet T^* M\otimes\mathbb{C}=\bigoplus_{\stackrel{|k|+|l| \le d/2}{k+l-d/2 \, \text{even}}} U_{k,l} \, ,
}
where
\eq{
U_{k,l}=\Lambda^{\frac{d/2-(k+l)}{2}}\overline{L_1^+} \cdot \Lambda^{\frac{d/2-(k-l)}{2}}\overline{L_1^-} \cdot \Psi_1 =
\Lambda^{\frac{d/2-(k+l)}{2}}\overline{L_1^+} \cdot \Lambda^{\frac{d/2-(l-k)}{2}} L_1^- \cdot \Psi_2 \, .
}
We can represent this decomposition in the form of a {\em generalized Hodge diamond}:
\begin{equation*}\label{genhodgediamond}
\centerline{
\xymatrix{
&&&  U_{0,d/2}\ar@{->}[ld]<0.5ex>^{L_1^-}\ar@{->}[rd]<0.5ex>^{\overline{L_1^+}} &&& \\
&&\cdots\ar@{->}[ru]<0.5ex>^{\overline{L_1^-}}\ar@{->}[ld]<0.5ex>^{L_1^-} &&\cdots\ar@{->}[rd]<0.5ex>^{\overline{L_1^+}}\ar@{->}[lu]<0.5ex>^{L_1^+}&&\\
&U_{d/2-1,1}\ar@{->}[rd]<0.5ex>^{\overline{L_1^+}}\ar@{->}[ld]<0.5ex>^{L_1^-}\ar@{->}[ru]<0.5ex>^{\overline{L_1^-}}\ar@{->}[rd]<0.5ex>^{\overline{L_1^+}}&& &&U_{-d/2+1,1}\ar@{->}[lu]<0.5ex>^{L_1^+}\ar@{->}[ld]<0.5ex>^{L_1^-}\ar@{->}[rd]<0.5ex>^{\overline{L_1^+}}&\\
U_{d/2,0}\ar@{->}[rd]<0.5ex>^{\overline{L_1^+}}\ar@{->}[ru]<0.5ex>^{\overline{L_1^-}}&&\ar@{->}[lu]<0.5ex>^{L_1^+}\ar@{->}[ld]<0.5ex>^{L_1^-}   &\cdots &\ar@{->}[rd]<0.5ex>^{\overline{L_1^+}}\ar@{->}[ru]<0.5ex>^{\overline{L_1^-}}&&U_{-d/2,0}\ar@{->}[lu]<0.5ex>^{L_1^+}\ar@{->}[ld]<0.5ex>^{L_1^-}\\
&U_{d/2-1,-1}\ar@{->}[lu]<0.5ex>^{L_1^+}\ar@{->}[rd]<0.5ex>^{\overline{L_1^+}}\ar@{->}[ru]<0.5ex>^{\overline{L_1^-}}&&
&&U_{-d/2+1,-1}\ar@{->}[ru]<0.5ex>^{\overline{L_1^-}}\ar@{->}[lu]<0.5ex>^{L_1^+}\ar@{->}[ld]<0.5ex>^{L_1^-}&\\
&&\ldots\ar@{->}[lu]<0.5ex>^{L_1^+}\ar@{->}[rd]<0.5ex>^{\overline{L_1^+}}&&\ldots\ar@{->}[ru]<0.5ex>^{\overline{L_1^-}}\ar@{->}[ld]<0.5ex>^{L_1^-}&&\\
&&&  U_{0,-d/2}\ar@{->}[ru]<0.5ex>^{\overline{L_1^-}}\ar@{->}[lu]<0.5ex>^{L_1^+}  &&&
}}
\end{equation*}
Note that $\Psi_1 \in \Gamma(U_{d/2,0})$ and $\Psi_2 \in \Gamma(U_{0,d/2})$. We have for $\phi \in \Gamma(U_{k,l})$
\eq{
\mathcal{J}_1 \cdot \phi = i k \phi \, , \qquad \mathcal{J}_2 \cdot \phi = i l \phi \, .
}
In the case of an ordinary K\"ahler structure (example \ref{genkahl}) this decomposition does {\em not}
correspond to the Dolbeault decomposition of eq.~\eqref{formdecomp}, rather it is the so-called {\em Clifford decomposition} \cite{michelsohn}.

The decomposition is compatible with the Mukai pairing:
for any $\phi_{k,l} \in \Gamma(U_{k,l})$
\eq{
\langle \phi_{k,l},\phi'\rangle=0\quad {\rm if} \quad \phi'|_{U_{-k,-l}}=0\ ,
\label{mukcomp2}
}
where $\phi'|_{U_{-k,-l}}$ denotes the projection of $\phi'$ on $\Gamma(U_{-k,-l})$. Moreover, the subbundles $U_{k,l}$
are invariant under the combination of the action of the reversal operator of the indices, $\sigma$, and of the $B$-twisted Hodge duality, defined as
\eq{
*_B = e^{B} *_d e^{-B} \, ,
}
with $*_d$ the Hodge dual \eqref{hodgestar}. This can be shown by first proving
\eq{
\label{actionhodge}
*_B \sigma(\mathbb{X_\pm} \cdot \phi) = \mp \mathbb{X_\pm} \cdot *_B \sigma(\phi) \, ,
}
for all $\mathbb{X}_\pm \in \Gamma(C_\pm)$. From this formula we learn in particular that the annihilator
space does not change under the action of $*_B \sigma$. It follows that the pure spinor line bundles
are invariant, i.e.\ $*_B \sigma(\Psi_1) \propto \Psi_1$, $*_B \sigma(\Psi_2) \propto \Psi_2$. To fix the
sign of the proportionality factor we must make a choice of orientation for the volume-form $\text{vol}_d$ (since such
a choice appears in the definition of the Hodge duality \eqref{hodgestar}).
Let us choose:
\eq{
\langle \Psi_1 , \bar{\Psi}_1 \rangle = \langle \Psi_2 , \bar{\Psi}_2 \rangle = (2i)^{d/2} \text{vol}_d \, .
}
Note that we have normalized the overall factor in $\Psi_1$ and $\Psi_2$, but that this normalization does not
affect the sign on the right-hand side. Our convention is chosen such that it is consistent with $\Psi_1 = e^{i\omega}$
and $\text{vol}_d = \frac{1}{(d/2)!} \, \omega^{d/2}$. We find then
\eq{
\label{hodgepure}
*_B \sigma(\Psi_{1,2}) = (-i)^{d/2} \Psi_{1,2} \, .
}
It follows that for every $\phi_{k,l} \in \Gamma(U_{k,l})$ we have
\eq{
\label{actionhodge2}
*_B \sigma(\phi_{k,l}) = (-1)^{\frac{d/2-(k+l)}{2}} (-i)^{d/2} \phi_{k,l} \, .
}

When $\mathcal{J}_1$ and $\mathcal{J}_2$ are $H$-integrable, i.e.\ in the generalized K\"ahler case,
there is a corresponding split of the exterior derivative \cite{gualtierihodge}:
\eq{
\d_H = \partial_{H+} + \partial_{H-} + \bar{\partial}_{H+} + \bar{\partial}_{H-} \, ,
}
where
\eq{\spl{
& \partial_{H+} : \Gamma(U_{k,l}) \rightarrow \Gamma(U_{k+1,l+1}) \, , \qquad
\bar{\partial}_{H+} : \Gamma(U_{k,l}) \rightarrow \Gamma(U_{k-1,l-1}) \, , \\
& \partial_{H-} : \Gamma(U_{k,l}) \rightarrow \Gamma(U_{k+1,l-1}) \, , \qquad
\bar{\partial}_{H-} : \Gamma(U_{k,l}) \rightarrow \Gamma(U_{k-1,l+1}) \, ,
}}
and we can define corresponding cohomology groups.

A stronger condition than $H$-integrability of both $\mathcal{J}_1$ and $\mathcal{J}_2$, which applies
to the SU($d/2$)$\times$SU($d/2$) case, is the following:
\begin{defi}
\label{genCYgualtieri}
A generalized Calabi-Yau geometry \`a la Gualtieri \cite[Definition 6.40]{gualtieri} is an SU($d/2$)$\times$SU($d/2$)-structure
such that
\eq{
\d_H \Psi_1 = 0 \, , \qquad  \d_H \Psi_2=0 \, .
}
\end{defi}
Building on example \ref{genkahl}, we see that an ordinary Calabi-Yau geometry is an example of a generalized
Calabi-Yau geometry.

\subsubsection*{Spinor bilinears}

Finally, we want to discuss the relation between the pure spinors $\Psi_1$ and $\Psi_2$
of Spin($d$,$d$) and the ordinary spinors of Spin($d$). In fact, the generalized
metric $G=(g,B)$, given by eq.~\eqref{genmetric}, allows to define an isomorphism between polyforms and operators acting on
spinors through the Clifford map:
\eq{
\label{clifford}
\phi'=e^B \wedge \phi \longleftrightarrow \slashchar{\phi} = \sum_l \frac{1}{l!} \, \phi_{i_1 \ldots i_l} \gamma^{i_1\ldots i_l} \, ,
}
where the gamma-matrices are defined as in \eqref{curvedgamma} using the vielbein associated to $g$. In what follows
we will put $B=0$ and work with the untwisted $\phi$. The relations below are easily generalized to $B \neq 0$.
It turns out that an SU($d/2$)$\times$SU($d/2$)-structure is associated to two (not necessarily everywhere independent)
spinors $\eta^{(1)}$ and $\eta^{(2)}$, normalized such that $\eta^{(1)^\dagger}\eta^{(1)}=\eta^{(2)^\dagger}\eta^{(2)}=1$, defining two (again not necessarily everywhere independent)
SU($d/2$)-structures. Indeed, using the isomorphism \eqref{clifford} we can associate the polyforms $\Psi_1$ and $\Psi_2$
to the following spinor bilinears:
\eq{
\label{purespinorsclifford}
\slashchar{\Psi_1} = (\text{dim}(S)) \; \eta^{(1)} \, \eta^{(2)\dagger} \, , \qquad
\slashchar{\Psi_2} = (\text{dim}(S))\; \eta^{(1)} \, \eta'{}^{(2)\dagger} \, ,
}
where $\eta'{}^{(2)}$ is the complex conjugate of $\eta^{(2)}$, i.e.\ $\eta'{}^{(2)}=C \eta^{(2)*}$, and $\text{dim}(S)=2^{d/2}$ is the dimension
of the spinor representation. Fierzing we find then:
\subeq{\al{
&\sigma(\Psi)_{1\, i_1 \ldots i_l} =  \eta^{(2)\dagger} \gamma_{i_1\ldots i_l}\eta^{(1)} \, , \\
&\sigma(\Psi)_{2\, i_1 \ldots i_l} = \eta'{}^{(2)\dagger} \gamma_{i_1 \ldots i_l}\eta^{(1)}  \, .
}}
\begin{ex}
Let us take $d=6$ so that $\eta^{(1,2)}=\eta^{(1,2)}_+$ has positive chirality
and $\eta'{}^{(1,2)}=\eta^{(1,2)}_-$ has negative chirality. The most general relation between
$\eta^{(1)}$ and $\eta^{(2)}$ is
\eq{
\eta^{(2)}_+ = c \, \eta^{(1)}_+ + \frac{1}{2}\,V^i \gamma_i \eta^{(1)}_- \, .
}
In order to construct the pure spinors $\Psi_1$ and $\Psi_2$ it will be convenient to slightly
rewrite this as follows
\eq{
\eta^{(1)}_+ = e^{i \vartheta/2} \eta_+ \, , \qquad \eta^{(2)}_+ = e^{-i \vartheta/2} \left( \cos \varphi \, \eta_+ + \sin \varphi \, \chi_+ \right) \, ,
}
where we take $0 \le \varphi \le \pi/2$ and we introduced mutually orthogonal unit spinors $\eta_+$ and $\chi_+ = \frac{1}{2} v^i \gamma_i \eta_-$ (with $|v|^2=2$), i.e.\
they satisfy $\eta_+^\dagger \eta_+=\chi_+^\dagger \chi_+=1$ and $\chi_+^\dagger \eta_+=0$. $\varphi$  indicates
the angle between $\eta^{(1)}$ and $\eta^{(2)}$, which may vary over the manifold $M$.
The relation between both descriptions is
\eq{
c = e^{-i \vartheta} \cos \varphi \, , \qquad V^i = v^i \sin \varphi \, .
}
Note that at points where $\sin \varphi=0$, $\chi_+$ does not need to be defined.
In all other points, the orthogonal spinors $\eta_+$ and $\chi_+$ define a {\em local SU(2)-structure},
which is described by the following forms
\subeq{\al{
& v^i = \eta^\dagger_- \gamma^i \chi_+ \, ,\\
& \omega_{ij} =  (i/2) \eta^{\dagger}_+ \gamma_{ij} \eta_+ - (i/2)\chi^{\dagger}_+ \gamma_{ij} \chi_+ \, , \\
& \Omega_{ij} =  \chi^{\dagger}_+ \gamma_{ij} \eta_+  \, ,
}}
satisfying
\subeq{\al{
& \Omega \wedge \omega = \Omega \wedge \Omega = 0 \, ,\\
& \iota_v \omega = \iota_v \Omega = \iota_v \bar{\Omega} = 0  \, , \\
& \Omega \wedge \bar{\Omega} = 2 \, \omega^2 \, .
}}
The pure spinors are then given by
\subeq{\al{
& \Psi_1 = e^{i\vartheta} \, e^{\frac{1}{2} v \wedge \bar{v}} \left[ \cos \varphi \, e^{i\omega} - \sin \varphi \, \Omega \right] \, , \\
& \Psi_2 = - v \wedge\left(\cos \varphi \, \Omega + \sin \varphi \, e^{i \omega} \right) \, .
}}

The following terminology is used \cite{granaN1,andriot}:
\begin{itemize}
\item Strict SU(3)-structure: $\sin \varphi=0$ everywhere. The types of the pure spinors $(\Psi_1,\Psi_2)$ are $(0,3)$.
\item Static SU(2)-structure: $\cos \varphi=0$ everywhere. The types of the pure spinors $(\Psi_1,\Psi_2)$ are $(2,1)$.
\item Intermediate SU(2)-structure: in generic points both $\cos \varphi$ and $\sin \varphi$ are non-vanishing, and the types of the pure
spinors $(\Psi_1,\Psi_2)$ are $(0,1)$. If the angle $\varphi$ changes over the manifold $M$ the intermediate SU(2)-structure is called dynamic.
A dynamic SU(2)-structure can be type-changing, i.e.\ on a specific locus one could have $\cos \varphi=0$,
so that the type jumps to $(2,1)$, or $\sin \varphi=0$, so that the type jumps to $(0,3)$.
\end{itemize}
Since for static and intermediate SU(2)-structure that does not change type to $(0,3)$,
$\sin \varphi$ is everywhere non-vanishing and the $\chi_+$ is everywhere defined,
the local SU(2)-structure turns into a global SU(2)-structure.
The structure of the tangent bundle then reduces to SU(2).
In the other cases no extra constraints beyond SU(3)-structure are imposed on the topology of the tangent bundle
as the two internal spinors $\eta^{(1)}$ and $\eta^{(2)}$ are not everywhere independent.
\end{ex}

Using the fact that
\eq{
\ul{\text{vol}_d} \,\, \ul{\phi} = \ul{* \sigma(\phi)} \, ,
}
we find that choosing $\eta^{(1)}$ of positive chirality and defining
the chirality operator $\gamma_{(d)}$ (with defining properties $\gamma_{(d)} \gamma_i = - \gamma_i \gamma_{(d)}$,
$\gamma_{(d)}^2=\bbone$) as follows:
\eq{
\gamma_{(d)} = (-i)^{d/2} \ul{\text{vol}_d} \, ,
}
is compatible with the sign convention in eq.~\eqref{hodgepure}.

To conclude the discussion on spinor bilinears, consider the action of elements $\mathbb{X}_\pm \in \Gamma(C_\pm)$ on polyforms. Taking into account that
they take the form $\mathbb{X}_\pm = (X,\pm g X)$, we find that
\eq{
\slashchar{\mathbb{X}_+ \cdot \phi} = X^i \gamma_i \, \slashchar{\phi} \, , \qquad
\slashchar{\mathbb{X}_- \cdot \phi} =  -(-1)^{\text{deg}(\phi)} \, \slashchar{\phi} \, \gamma_i X^i \, ,
}
where $(-1)^{\text{deg}(\phi)}$ is the parity of $\phi$. So we see that the elements of $C_+$
act as gamma-matrices from the left, and the elements of $C_-$ as gamma-matrices from the right.
It follows immediately that if $\eta^{(1)}$ and $\eta^{(2)}$ are pure spinors of
Spin($d$) (which is trivial for $d=6$), $\Psi_1$ and $\Psi_2$ defined in this way are pure spinors of Spin($d,d$). Indeed,
the null space of $\Psi_1$ in every point has dimension $d$: it consists of $d/2$ elements
of $C_+$ constructed from annihilators of the pure spinor $\eta^{(1)}$ and $d/2$ elements of $C_-$
constructed from annihilators of the pure spinor $\eta^{(2)}$. Analogously for $\Psi_2$.
Furthermore the conditions \eqref{normcond} and \eqref{compcond} are also automatically
fulfilled.

\subsubsection*{Bihermitian geometry}

Let us consider $J_+$ and $J_-$, the almost complex structures associated to respectively $\eta^{(1)}$
and $\eta^{(2)}$  through eqs.~\eqref{strucspinor} and the metric $g$. One finds
\eq{\label{jpjm}\spl{
& -J_+ = \pi_{TM} \circ \mathcal{J}_1 \circ (\pi_{C_+})^{-1} = \pi_{TM} \circ \mathcal{J}_2 \circ (\pi_{C_+})^{-1} \, , \\
& -J_- = \pi_{TM} \circ \mathcal{J}_1 \circ (\pi_{C_-})^{-1} = -\pi_{TM} \circ \mathcal{J}_2 \circ (\pi_{C_-})^{-1} \, , \\
}}
with $\pi_{C_\pm}=\pi_{TM}|_{C_\pm}: (X,\pm g X) \rightarrow X$ the isomorphism between $C_\pm$ and $TM$. Using these formulae, one can
write $\mathcal{J}_{1,2}$ explicitly in terms of $J_\pm$:
\eq{
\label{gencompexpl}
\mathcal{J}_{1,2} = \frac{1}{2} \left(\begin{array}{cc} -(J_+ \pm J_-) & \omega_+^{-1}\mp \omega_+^{-1} \\ -(\omega_+ \mp \omega_+) & (J_+ \pm J_-)^T  \end{array}\right)\, ,
}
where $\omega_\pm = g J_\pm$.

The action of a string in a curved background with non-zero NSNS three-form $H$
in the RNS formalism is given by a supersymmetric non-linear sigma model with $\mathcal{N}=(1,1)$
world-sheet supersymmetry. A second left- and right-moving supersymmetry on the world-sheet
is a necessary condition for supersymmetry on the target space. It was found in \cite{ghr} that the world-sheet
theory has this $\mathcal{N}=(2,2)$ supersymmetry iff the target space is a bihermitian geometry.

\begin{defi}
A {\em bihermitian geometry} consists of a metric $g$, a closed three-form $H$ and two
complex structures $J_+,J_-$ with respect to which the metric is Hermitian. Moreover they must satisfy:
\eq{
\label{bihermcond}
\nabla^{\pm} J_\pm =  0 \, ,
}
where $\nabla^\pm$ are the {\em Bismut
connections}, which are related to the Levi-Civita connection as follows:
\eq{
\nabla^{\pm} = \nabla_{\text{LC}} \pm \frac{1}{2} g^{-1} H \, .
}
In index notation we have
\eq{
\nabla^{\pm}_i J^j{}_k = \nabla_{\text{LC} \, i} J^j{}_k \pm (1/2) \, g^{jp} H_{pli} J^l{}_k \mp (1/2) \, g^{lp} H_{pki} J^j{}_l = 0 \, .
}
\end{defi}

In \cite[Section 6.4]{gualtieri} the following was shown:
\begin{theor}
A bihermitian geometry is equivalent to a generalized K\"ahler structure.
\end{theor}

Since having a bihermitian geometry is a necessary condition for space-time supersymmetry in the presence of an $H$-field
this is the first example of this review where a supersymmetry condition can be concisely reformulated in terms of Generalized Complex Geometry.
The study of these supersymmetry conditions in the presence of both NSNS and RR-fluxes will be the topic of chapter \ref{sugrasusycond}.

As opposed to the generalized complex structures, $J_+$ and $J_-$ generically do not commute.
From $[J_+,J_-]=(J_+ - J_-)(J_+ + J_-)$ we find the following decomposition of the tangent space:
\eq{
\label{decompJ+J-}
T M = \text{ker}(J_+ - J_-) \oplus \text{ker}(J_+ + J_-) \oplus \text{coim}([J_+,J_-]) \, .
}
The first two subspaces can also be related to the kernels of the Poisson structures
\eq{
\omega_+^{-1} \mp \omega_-^{-1} \, ,
}
which we found in eq.~\eqref{gencompexpl} to appear in the upper-right block of the generalized complex structures $\mathcal{J}_{1,2}$.
The third subspace corresponds to the {\em symplectic leaf} (for a definition see exercise \ref{excpoisson}) of a third Poisson structure
\eq{
[J_+,J_-] g^{-1} \, .
}
One can show that (see exercise \ref{excpoisson} and \cite{tomasiellogenpot})
\eq{
\text{dim}(\text{ker}(J_+ \mp J_-)) = 2\,k_{1,2} \, ,
}
where $k_{1,2}$ are the types of the pure spinors $\Psi_{1,2}$ respectively. \cite{lindstromproof,lindstromgenpot} shows that there is
an $\mathcal{N}=(2,2)$ superspace description of the sigma-model in terms of chiral, twisted-chiral and semichiral multiplets associated
to the decomposition \eqref{decompJ+J-}. In terms of the types of the pure spinors we find
\eq{\spl{
k_1: \quad & \# \, \text{chiral multiplets}\, , \\
k_2: \quad & \# \, \text{twisted-chiral multiplets} \, , \\
(d-k_1-k_2)/2 : \quad & \# \, \text{semichiral multiplets} \, .
}}

\begin{ex}
Every even-dimensional compact semisimple Lie-group admits a bihermitian geometry/generalized K\"ahler structure.
Indeed, one can introduce a complex structure at one point of the manifold, say the unit element of the group,
and then transport it over the whole manifold using either the left or the right action of the group.
In this way one obtains left- and right-invariant complex structures $J_L$ and $J_R$. If the group is semisimple, they
can be chosen so that the Cartan-Killing metric is Hermitian with respect to both. Moreover it can be shown that they obey \eqref{bihermcond}
for $H$ given by
\eq{
H(X,Y,Z) = g([X,Y],Z) \, ,
}
for arbitrary vector fields $X,Y,Z$ and where $g$ is the Cartan-Killing metric. In this way, one obtains a bihermitian geometry.
\end{ex}

A related example is the Hopf surface $S^3 \times S^1$, which can be seen as the Lie-group SU(2)$\times$U(1) \cite{sevrinWZW}.
Further examples have been constructed in \cite{hitchinpoisson,gualtierireduction,gualtierigenkahl,hitchindelpezzo}

\subsection{Generalized product structures}
\label{genprodstruc}

\subsubsection*{Definitions}

A generalized almost product structure is closely related to a generalized almost complex structure, and many of
the concepts reviewed in this chapter carry over. Since it will be useful to describe the embedding of D-branes (see e.g.~\cite{zabzinebranes}
for early work) in chapter \ref{branesection} we will briefly define it here and give some examples.

\begin{defi}
A {\em generalized almost product structure} is a map
\eq{
\mathcal{R}: TM \oplus T^* M \rightarrow T M \oplus T^* M \, ,
}
respecting the bundle structure: $\pi(\mathcal{R} \mathbb{X})=\pi(\mathbb{X})$,
and which satisfies two further conditions:
it squares to one
\eq{
\label{Rsquareone}
\mathcal{R}^2 = \bbone \, ,
}
and it is compatible with the canonical metric
\eq{
\label{Rantiherm}
\mathcal{I}(\mathcal{R} \mathbb{X}, \mathcal{R} \mathbb{Y})  = - \mathcal{I}(\mathbb{X},\mathbb{Y}) \, .
}
\end{defi}
Associated to $\mathcal{R}$ are two subbundles
$T_{\mathcal{R}},N_{\mathcal{R}} \subset (T M \oplus T^* M)$ with fibers respectively
the $(\pm 1)$-eigenspaces of the action of $\mathcal{R}$ on the fiber over each point.  The sign in eq.~\eqref{Rantiherm}
is important as it makes both $T_{\mathcal{R}}$ and $N_{\mathcal{R}}$ into {\em isotropic} subbundles (see definition \ref{isotropic}).
In fact, since their sum must make up the full generalized tangent space, they must both be {\em maximally isotropic}.
In contrast to an ordinary almost product structure the subbundles $T_{\mathcal{R}}$ and $N_{\mathcal{R}}$ have thus equal rank $d$, equal to half
the rank of $T M \oplus T^* M$.

For integrability we introduce the following obvious definition:
\begin{defi}
A generalized almost product structure is $H$-integrable, making it into a generalized product structure, if both $T_{\mathcal{R}}$
and $N_{\mathcal{R}}$ are involutive under the $H$-twisted Courant bracket.
\end{defi}
We will see in section \ref{gensubman} that for the description of generalized submanifolds one only needs to impose
the involutivity of $T_{\mathcal{R}}$.

We can associate a real pure spinor line bundle, such that its null space is equal to $T_{\mathcal{R}}$.
In six dimensions such pure spinors are of the type with $q(\rho)>0$ discussed in theorem \ref{hitchintheorem}.

Finally, we can also impose compatibility with the positive-definite metric $G$ introduced in definition \ref{J1J2def}
\eq{
\mathcal{R}^T G \mathcal{R} = \mathcal{R} \, .
}

\subsubsection*{Examples}

\begin{ex}[Foliation of generalized submanifolds]
Consider an almost product structure of the form
\eq{
\mathcal{R}_{(R,F)} = e^{-F} \mathcal{R}_R e^{F} =
\left(\begin{array}{cc} R & \mathbf{0} \\ -F R - R^T F &  - R^T\end{array} \right) \, ,
}
where $R$ is an ordinary almost product structure and $F$ is a two-form. It easy to verify that $\mathcal{R}_{(R,F)}$
satisfies both properties (\ref{Rsquareone}-\ref{Rantiherm}) of the definition. Using
the property \eqref{propcourant} of the Courant bracket, it is easy to show that $\mathcal{R}_{(R,F)}$
is $H$-integrable iff $R$ is integrable and $\d F = H$.

The (+1)-eigenbundle $T_{(R,F)}$ takes the form
\eq{
\label{genprodtan}
T_{(R,F)} = \{ X + \xi \in T \oplus T^*M \,\,\, \big| \,\,\, \xi|_T = \iota_X F|_T \} \, ,
}
where $T \subset TM$ is the (+1)-eigenbundle associated to the product structure $R$, and $|_T$
indicates the restriction to $T$. The associated pure spinor line bundle is
\eq{
\label{pureprod}
\tau_{(T,F)} = e^{-F} \wedge \theta^1 \wedge \cdots \wedge \theta^{d-l} \, ,
}
where the $\theta^a$ span Ann $T$ as in example \ref{almostprodforms}.

Suppose we restrict $F$ to $T$, i.e. $F|_N = 0$. Requiring only the integrability of $T_{(R,F)}$
(and not necessarily of $N_{(R,F)}$) amounts to the integrability of $T$ and $H|_T = \d F$. We then have a foliation such that through every point
of $M$ there is a submanifold $\Sigma$, such that $T \Sigma = T$, and carrying a two-form $F$ satisfying $\d F = H|_\Sigma$.
\end{ex}

\begin{ex}[Poisson structure]
Consider an almost product structure of the form
\eq{
\mathcal{R}_{P} = \left(\begin{array}{cc} -\bbone & P \\ \mathbf{0} &  \bbone\end{array} \right) \, ,
}
where $P$ is an antisymmetric two-vector. The integrability of $\mathcal{R}_{P}$ amounts
to
\eq{
\label{poissonintcond}
P^{[i|l} \partial_l P^{|jk]} =0 \, ,
}
which makes $P$ into a Poisson structure (see also exercise \ref{excpoisson}). A Poisson structure can be used to define a Poisson bracket as follows
\eq{
\{ f, g \} = P^{ij} \partial_i f \partial_j g \, ,
}
for two functions $f,g \in C^{\infty}(M)$. The Jacobi identity of the bracket is then ensured by eq.~\eqref{poissonintcond}.
\end{ex}

\subsection{Deformation theory}
\label{deformgencomp}

In order to construct the 4D low-energy effective theory around supersymmetric backgrounds,
which will be introduced in chapter \ref{sugrasusycond}, we have to figure out what the moduli are.
Indeed, if the solutions to the supersymmetry
equations \eqref{susyconds} or \eqref{susycondsAdS} have many arbitrary parameters,
these parameters will show up as massless fields in the low-energy theory. So it is useful to study deformations of the solutions
to the equations \eqref{susyconds} or \eqref{susycondsAdS} that are themselves solutions. The analysis is complicated by the fact that the pure spinor
$\Psi_1$ is not integrable, and the general solution is still unknown. Assuming the $\partial_H\bar{\partial}_H$-lemma holds, this
problem was studied in detail in \cite{martuccibulkdef}. Here we will restrict ourselves to
presenting the deformations of only one {\em integrable} pure spinor (like $e^{3A-\Phi} \Psi_2$) and only briefly comment
on the deformation of a generalized K\"ahler structure.

A generalized complex structure $\mathcal{J}$ is completely determined by the isotropic subbundle $L=L_{\mathcal{J}}$. To describe the
deformations of $\mathcal{J}$ one can thus equivalently describe how $L$ varies. This variation is determined by
the map $\epsilon: L \rightarrow \bar{L}$ which describes how each $\mathbb{X} \in L$ deforms:
\eq{
\label{genJdeform}
\mathbb{X} \rightarrow \mathbb{X} + \epsilon(\mathbb{X}) \, .
}
Of course the deformed $L$ should still be isotropic so that we must require
\eq{
\mathcal{I}(\epsilon(\mathbb{X}),\mathbb{Y})+\mathcal{I}(\mathbb{X},\epsilon(\mathbb{Y})) = 0 \, ,
}
for all $\mathbb{X},\mathbb{Y} \in L$. Using the metric $\mathcal{I}$, which provides an isomorphism $\bar{L}\simeq L^*$,
one can turn $\epsilon$ into a two-form $\epsilon \in \Gamma(\Lambda^2 L^*)$:
\eq{
\epsilon(\mathbb{X},\mathbb{Y}) = \mathcal{I}(\epsilon(\mathbb{X}),\mathbb{Y}) = - \mathcal{I}(\mathbb{X},\epsilon(\mathbb{Y})) = -\epsilon(\mathbb{Y},\mathbb{X}) \, .
}
In \cite[Section 5.1 and theorem 3.37]{gualtieri} it is shown that the deformed complex structure is integrable iff
\eq{
\label{Lint}
\d_L \epsilon + \frac{1}{2} [\epsilon,\epsilon]_H = 0 \, ,
}
where $\d_L$ is the Lie algebroid derivative (definition~\ref{liealgebroidderdef}) and $[\cdot,\cdot]_H$ is the
Schouten bracket.
\begin{defi}
The {\em Schouten bracket} is a bilinear bracket extending the $H$-twisted Courant bracket to antisymmetric products of vector fields.
It acts on $\mathbb{X}_1 \wedge \cdots \wedge \mathbb{X}_k \in \Gamma(\Lambda^k (T M \oplus T^* M))$
and $\mathbb{Y}_1 \wedge \cdots \wedge \mathbb{Y}_l \in \Gamma(\Lambda^l (T M \oplus T^* M))$ as follows
\begin{multline}
[\mathbb{X}_1 \wedge \cdots \wedge \mathbb{X}_k,\mathbb{Y}_1 \wedge \cdots \wedge \mathbb{Y}_l]_H = \\ \sum_{i,j} (-1)^{i+j}  [\mathbb{X}_i,\mathbb{Y}_j]
\wedge \mathbb{X}_1 \wedge \cdots \wedge \mathbb{X}_i \wedge \cdots \wedge \mathbb{X}_k \wedge \mathbb{Y}_1 \wedge \cdots \wedge \mathbb{Y}_j \wedge \cdots \wedge \mathbb{Y}_l \, .
\end{multline}
\end{defi}

So, infinitesimally we find that the deformations preserving the integrability of a generalized complex
structure are in the kernel of $\d_L : \Gamma(\Lambda^2 L^*) \rightarrow \Gamma(\Lambda^3 L^*)$.
Now, two deformations are considered equivalent if they are related by a diffeomorphism and
an exact $B$-transform. It turns out that such a trivial deformation can be described by a
generalized vector $\mathbb{X}=(X,\xi)$ acting on the pure spinor as
\eq{
\mathcal{L}_{\mathbb{X}} \phi = \d_H (\mathbb{X} \cdot \phi) = (\d_L \mathbb{X}) \cdot \phi \, ,
}
where we used $\d_H \phi=0$, which --- strictly speaking --- is, as we saw in \eqref{spinint}, stronger than
just the integrability of $\mathcal{J}$. It follows that the inequivalent deformations are classified by the
second cohomology of $\d_L$: $H^2(L)$.

We can now turn to the question of deforming a generalized K\"ahler structure, which consists
of {\em two} generalized complex structures $\mathcal{J}_1$ and $\mathcal{J}_2$. The extra complication
is that the deformation $\epsilon$ must preserve the compatibility condition \eqref{spinorcomp}.
It is immediate that deformations of the type
\eq{\spl{
& \epsilon_1 : L_1^+ \rightarrow \overline{L_1^-} \, , \\
& \epsilon_2 : L_1^+ \rightarrow L_1^- \, ,
}}
leave $\mathcal{J}_2$ (with null space $L_2=L_1^+ \cup \overline{L_1^-}$) respectively $\mathcal{J}_1$ (with null space $L_1=L_1^+ \cup L_1^-$) invariant.
Also $\epsilon_1$ will leave the $\mathcal{J}_2$-eigenvalue of $\Psi_1$ invariant, while $\epsilon_2$ leaves
the $\mathcal{J}_1$-eigenvalue invariant, so that the compatibility condition \eqref{spinorcomp}
is still satisfied. Generically, there are however also deformations that change both $\mathcal{J}_1$ and $\mathcal{J}_2$
at the same time. It turns out that these deformations leave the metric and the $B$-field invariant, so that they should
probably not be considered as physical moduli.

The contents of this chapter is summarized in table \ref{strucformspinorgen}.
\begin{table}[tp]
\begin{center}
\begin{tabular}{|c|c|c|c|}
\hline
& Structures & Polyforms & Spinors \\
\hline
\rowcolor{blue!40}
Almost gen.\ compl.\ structure & $\mathcal{J}$ & $\Psi$ & NA \\
\rowcolor{blue!10}
integrability & $L_{\mathcal{J}}$ involutive & $\d \Psi = \mathbb{X} \cdot \Psi$ & NA \\
\rowcolor{blue!40}
(S)U($d/2$)$\times$(S)U($d/2$)-structure & $\mathcal{J}_1,\mathcal{J}_2$ & $\Psi_1,\Psi_2$ & $\eta^{(1)},\eta^{(2)}$ \\
\rowcolor{blue!10}
compatibility       & $[\mathcal{J}_1,\mathcal{J}_2]=0$ & $\Psi_2 \in \Gamma(U_0)$ & automatic \\
\rowcolor{blue!20}
integr.\ (gen.\ K\"ahler) & $L_1,L_2$ involutive & $\begin{array}{c} \d \Psi_1 = \mathbb{X} \cdot \Psi_1 \\ \d \Psi_2 = \mathbb{Y} \cdot \Psi_2\end{array}$ & NA \\
\rowcolor{blue!10}
gen.\ Calabi-Yau          & NA & $\d \Psi_1=\d \Psi_2=0$ & $\begin{array}{c} \nabla^+_i \eta^{(1)} = 0 \\ \nabla^-_i \eta^{(2)}=0\end{array}$  \\
\hline
\end{tabular}
\caption{The languages of respectively structures, forms and spinors in Generalized Complex Geometry.}
\label{strucformspinorgen}
\end{center}
\end{table}

\subsection{Exercises}

\begin{exc}[intermediate]
Show property \eqref{propcourant} of the $H$-twisted Courant bracket. The solution may be found
in proposition 3.23 and 3.42 of \cite{gualtieri}.
\end{exc}
\begin{exc}[easy]
Show that $L$ and $\bar{L}$ are maximally isotropic subbundles.
\end{exc}
\begin{exc}[hard]
Show that the Courant bracket restricted to an isotropic integrable subbundle satisfies the Jacobi identity.
The solution may be found by using proposition 3.16 of \cite{gualtieri}.
\end{exc}
\begin{exc}[easy]
Work out example \ref{complex}. Show that $\mathcal{J}$ is indeed a generalized almost complex structure.
Construct $L_{\mathcal{J}}$ and $\bar{L}_{\mathcal{J}}$. Check the condition for integrability.
\end{exc}
\begin{exc}[easy]
Same as above for example \ref{symplex}.
\end{exc}
\begin{exc}[intermediate]
Show \eqref{courantbracketprop}, the expression of the Courant bracket as a derived bracket.
\end{exc}
\begin{exc}[intermediate]
Show that if $L$ is integrable $\d_H$ splits into
$\partial_H$ and $\bar{\partial}_H$.
Hint: apply \eqref{courantbracketprop} in an inductive argument. For the solution
see theorem 4.23 of \cite{gualtieri}.
\end{exc}
\begin{exc}[intermediate]
\label{excpoisson}
In general $\mathcal{J}$ takes the following form
\eq{
\mathcal{J}= \left(\begin{array}{cc} -J & P \\ L & J^T \end{array}\right) \, ,
}
where from eq.~\eqref{herm} follows that $P$ is an antisymmetric bivector and $L$ a two-form.
Show that the integrability of $\mathcal{J}$ with $H=0$
implies that $P$ is a Poisson structure i.e.\ it satisfies
\eq{
\label{poissonint}
P^{[i|l} \partial_l P^{|jk]} = 0 \, .
}
A Poisson structure can be used to define a Poisson bracket as follows
\eq{
\{ f, g \} = P^{ij} \partial_i f \partial_j g \, ,
}
for two functions $f,g \in C^{\infty}(M)$. Check that the Jacobi identity of the Poisson bracket
is ensured by eq.~\eqref{poissonint}. Note that a Poisson structure does not need to be invertible.
In fact, since in \eqref{purespinorform} $\Omega_k$ is decomposable we can write it
locally as $\Omega_k = \theta^1 \wedge \cdots \wedge \theta^k$. Show that $\text{ker} \, P$ is
spanned by $\{ \theta^1, \ldots, \theta^k, \bar{\theta}^{\bar{1}},\ldots, \bar{\theta}^{\bar{k}} \}$. Show that if
$\mathcal{J}$ is integrable, then $Q=\text{Ann}(\text{ker}\, P)=\{ X \in T M \, | \, \xi(X)=0 \, , \forall \, \xi \in \ker P \}$ is
integrable, and gives rises to a foliation. $\omega=P^{-1}|_Q$ is a symplectic structure so that the leaves are called
{\em symplectic leaves}. Poisson structures play an important role in the study of the $\mathcal{N}=(2,2)$ supersymmetry
of the world-sheet sigma-model with $H\neq 0$ \cite{lindstromWS1,lindstromproof}.
Suppose we put further $L=0$, then we find from \eqref{squareone} that $J$ is an ordinary almost complex structure
and $JP - J^T P=0$ so that  $P$ is a $(2,0)+(0,2)$-form. $P|_{(2,0)}$ is a {\em holomorphic Poisson structure}.
Show that this is the $\beta$-transform of an ordinary complex structure.
\end{exc}
\begin{exc}[intermediate]
\label{stabex}
Show that if the isotropic subbundle $L_2$ is stable under the generalized almost complex
structure $\mathcal{J}_1$, i.e.\ for all $\mathbb{X} \in L_2$ it follows $\mathcal{J}_1 \mathbb{X} \in L_2$,
then $\Psi_2 \in \Gamma(U_0)$, where $\Psi_2$ is the pure spinor associated to $L_2$ and $U_i$
is the filtration associated to $\mathcal{J}_1$. The solution can be found in \cite[Lemma 2.1]{caval}.
Show then the equivalence between $[\mathcal{J}_1,\mathcal{J}_2]=0$ (eq.~\eqref{J1J2comm}) and $\Psi_2 \in \Gamma(U_0)$ (eq.~\eqref{spinorcomp}).
\end{exc}
\begin{exc}[intermediate]
Show the property~\eqref{actionhodge} of the $B$-twisted Hodge duality. Use this to show the transformation
property~\eqref{actionhodge2} of the sections of $U_{k,l}$ under $*_B \sigma$.
\end{exc}
\begin{exc}[easy]
Show eq.~\eqref{jpjm}.
\end{exc}
\begin{exc}[intermediate]
Show the following relation, which can be found in lemma 2 of \cite{li}:
\begin{multline}
\d_H(\alpha \cdot \beta\cdot \phi) = (-1)^{(p-1)q} \beta \cdot d_H(\alpha \cdot \phi) + (-1)^{p} \alpha \cdot d_H(\beta \cdot \phi)
+(-1)^{p-1} [\alpha,\beta]_H \cdot \phi \\ + (-1)^{p+q+1} \alpha \cdot \beta \cdot d_H\phi \, ,
\label{liid}
\end{multline}
for $\alpha \in \Gamma(\Lambda^p \bar{L}), \beta \in \Gamma(\Lambda^q \bar{L})$ and $\phi$ a polyform.
Show furthermore that the deformation $\epsilon$ of eq.~\eqref{genJdeform} has the following effect on the pure spinor $\phi_\epsilon$
associated to the deformed $\mathcal{J}_\epsilon$:
\eq{
\label{puredeform}
\phi_\epsilon = e^{-\epsilon} \cdot \phi \, .
}
Use this equation and eq.~\eqref{liid} to show  that the integrability condition for the
deformation of $L$ is indeed given by \eqref{Lint}.
\end{exc}

\clearpage

\thispagestyle{beginsection}
\section[Supersymmetry conditions of a SUGRA background]{Supersymmetry conditions of a supergravity background}
\label{sugrasusycond}

In the introduction we have explained that the reason we introduced Generalized
Complex Geometry is its potential to describe the supersymmetry conditions of string
theory and supergravity. At the end of section \ref{sudsud} we have already seen
the first application: it was reviewed there that the requirement of $N=(2,2)$ supersymmetry of a non-linear sigma
model --- corresponding to a string world-sheet action --- in the presence of a curved target-space metric
and an NSNS three-form $H$, is exactly that the target space should have a generalized K\"ahler structure.
These conditions should be completed by the requirement that the sigma model be conformally invariant.
Using the results of \cite{sevrinbeta} it was shown in \cite{tomasiellogenpot} that at one-loop, this
extra condition leads exactly to the requirement of generalized Calabi-Yau geometry. When introducing RR-fluxes,
the RNS world-sheet action is not valid anymore and we should use another formalism like the
Green-Schwarz string or the Berkovits formalism \cite{berkovitsformalism}. Instead, in
this chapter we study the supersymmetry conditions of the low-energy limit of type II string theory,
namely type II supergravity. For conventions and more information on type II supergravity, and in particular
the use of the democratic formalism \cite{democratic} to describe the RR-fluxes, see appendix \ref{sugra}.

\subsection{Minkowski compactifications}
\label{sec:minsusy}

\subsubsection*{Compactification ansatz and SU(3)$\times$SU(3)-structure}

First, let us construct a vacuum solution by compactifying to a flat 4D Minkowski space, which means
we consider type II supergravity on 10D space-times of (warped) factorized form $\mathbb{R}^{3,1} \times_w M$
where $M$ is the 6D internal space and $\mathbb{R}^{3,1}$ is flat Minkowski space.
The 10D metric then takes the form
\eq{
\label{metricansatz}
\d s^2_{(10)}=e^{2A(y)}\d s^2_{(4)}+g_{mn}(y)\d y^m\d y^n\ ,
}
where $\d s^2_{(4)}$ is the flat Minkowski metric, and $A$ is the warp factor.
All the background fluxes should preserve the Poincar\'e symmetry of $\mathbb{R}^{3,1}$, which implies they can only depend on the
internal coordinates $y^m$. It follows furthermore that the NSNS $H$-field can only have internal indices and
that the polyform $F=\sum_n F_n$ of RR fields splits as follows
\eq{
\label{formansatz}
F=\text{vol}_4\wedge e^{4A}\tilde F+\hat F\ \, ,
}
where $\tilde{F}$ and $\hat{F}$ are respectively the ``electric'' and ``magnetic'' components.
Here $\text{vol}_4$ is the (unwarped) flat Minkowski volume-form and $\tilde F$ and $\hat F$ have only internal indices.
The Hodge duality \eqref{Fduality2} implies the following relation between the electric and magnetic components of the RR fields
\eq{
\label{dualin6}
\tilde F= *_6 \sigma(\hat F) \, .
}

The compactification ansatz for the 10D supersymmetry generators $\epsilon^{1,2}$ is
\eq{\label{adskilling}\spl{
\epsilon^1 & = \zeta_+\otimes \eta^{(1)}_+ +\, (\text{c.c.})\ ,\cr
\epsilon^2 & = \zeta_+\otimes \eta^{(2)}_\mp+\, (\text{c.c.})\ ,
}}
for IIA/IIB\footnote{Here and in the following, the upper signs are for IIA while the lower are for IIB.}.
In the above, $\zeta_+$ is any constant complex Weyl spinor on the flat Minkowski space (four real components),
where each choice describes one of the four 4D supersymmetry generators. We thus end up with minimal $\mathcal{N}=1$
supersymmetry in four dimensions. One could have used different 4D spinors $\zeta_+^{(1,2)}$ in the decomposition of $\epsilon^{1,2}$
respectively, leading to an $\mathcal{N}=2$ effective theory in four dimensions. In the presence of RR-fluxes, the supersymmetry variations \eqref{susyvar}
relate $\epsilon^1$ and $\epsilon^2$ and therefore also $\zeta_+^{(1)}$ and $\zeta_+^{(2)}$. The solutions, as opposed to the effective theory, therefore generically only have $\mathcal{N}=1$.
In fact, we will assume the presence of orientifolds --- the necessity of which we will discuss in a moment --- relating $\zeta_+^{(1)}$ and $\zeta_+^{(2)}$ also off-shell and leading
to an $\mathcal{N}=1$ effective 4D theory.

The two internal chiral spinors $\eta^{(1)}_+$ and $\eta^{(2)}_+$, on the other hand, are {\em fixed} spinors that characterize the background geometry.
Indeed, as explained in the previous chapter they define a reduction of the structure group of the generalized tangent bundle $T M\oplus T M^*$ from SO(6,6) to
SU(3)$\times$SU(3). Supersymmetry therefore imposes first a constraint on the {\em topology} of the generalized tangent bundle. We will now see
how to reformulate the supersymmetry conditions \eqref{susyvar} as a {\em differential} constraint on this SU(3)$\times$SU(3)-structure.

Suppose $\eta^{(1)}_+$ and $\eta^{(2)}_+$ have the same norm $|a|^2=||\eta^{(1)}_+||^2=||\eta^{(2)}_+||^2$, which as we will see in section \ref{sec:gencal} is a necessary condition for the background to allow for supersymmetric D-branes. We define the pure spinors $\Psi_1$ and $\Psi_2$ using the Clifford map as in
eq.~\eqref{purespinorsclifford} (appropriately normalized and with an extra phase factor for convenience):
\eq{
\label{nps}
\slashchar{\Psi_1} = \slashchar{\Psi}^{\mp}=-\frac{8i}{|a|^2} \; \eta^{(1)}_+ \, \eta^{(2)\dagger}_\mp \, , \qquad
\slashchar{\Psi_2} = \slashchar{\Psi}^{\pm}=-\frac{8i}{|a|^2} \; \eta^{(1)}_+ \, \eta^{(2)\dagger}_\pm \, ,
}
for type IIA/IIB. Note that the positive-chirality and the negative-chirality pure spinor are interchanged when going
from type IIA to IIB.

\subsubsection*{Differential conditions for supersymmetry}

In \cite{granaN1} it was found that the supersymmetry conditions \eqref{susyvar} can be rewritten as the following elegant conditions
on the pure spinors
\subeq{\label{susyconds}
\al{
\label{susycond0}
& \d_H \big(e^{4A-\Phi} \Re \Psi_1 \big) =  e^{4A} \tilde{F} \, , \\
& \d_H \big(e^{3A-\Phi} \Psi_2) = 0 \, , \label{susycond1} \\
& \d_H \big(e^{2A-\Phi} \Im \Psi_1 \big) =  0 \, .
}}
\begin{figure}[t]
\centering
\psfrag{J1}{$\mathcal{J}_1/\Psi_1$}
\psfrag{J2}{$\mathcal{J}_2/\Psi_2$}
\psfrag{J1int}{$\mathcal{J}_1$ integr.}
\psfrag{J2int}{$\mathcal{J}_2$ integr.}
\psfrag{SU}[cc]{SU(3)$\times$SU(3)}
\psfrag{genk}{generalized K\"ahler}
\psfrag{Minkowski}{Minkowski}
\psfrag{AdS}{AdS$_4$}
\includegraphics[width=7cm]{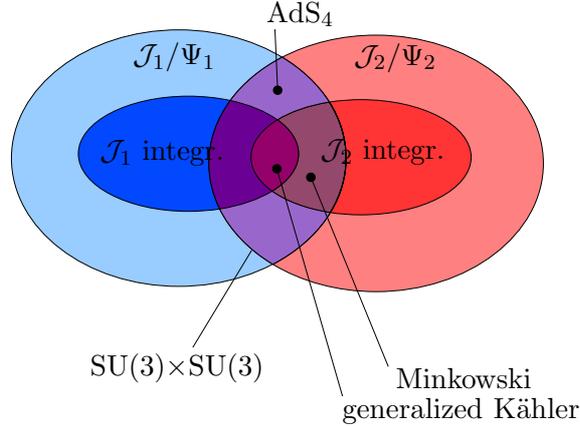}
\caption{Minkowski and AdS$_4$ compactifications indicated in the space
of SU(3)$\times$SU(3)-structures.}
\label{SU3SU3fig}
\end{figure}
The calculation is quite long-winded and technical so we do not repeat it here.
While the original reference does not contain a lot of detail, we refer to \cite[Appendix B]{lucacal} and
\cite[Appendix A]{granascan} for more elaboration. We will given an interpretation of each of these conditions in terms of
a D-brane calibration in section \ref{sec:gencal}.

It follows immediately from \eqref{susycond1} that $e^{3A-\Phi} \Psi_2$ is a generalized
Calabi-Yau structure \`a la Hitchin (definition \ref{genCYhitchin}), and in particular that
the associated generalized complex structure
$\mathcal{J}_2$ is integrable. On the other hand, because of the presence of the
RR-fluxes, the generalized almost complex structure $\mathcal{J}_1$ associated to $\Psi_1$ is {\em not}
integrable. See figure \ref{SU3SU3fig}. This means that although $\Psi_1$ and $\Psi_2$ define an SU(3)$\times$SU(3)-structure they do
not satisfy the appropriate integrability to form a generalized Calabi-Yau geometry \`a la Gualtieri (definition \ref{genCYgualtieri}). Some
proposals for Exceptional Generalized Complex Geometry have been made in \cite{hullexcep,walexcep,granaexcep,granaexcep2}, that
can possibly also include the RR-fields in a more natural way. However, we will see that the present formalism already
does quite a good job in describing supersymmetric backgrounds and D-branes probing them.

\subsubsection*{Supersymmetry together with Bianchi identities implies all EOMs}

We observe that acting with $\d_H$ on \eqref{susycond0} leads to
\eq{
\d_H \left(e^{4A} \tilde{F} \right) = -\sigma\left(\d_{-H} \left( e^{4A} *_6 \hat{F}\right)\right) = 0 \, .
}
Comparing with \eqref{eomB} and plugging in the compactification ansatz \eqref{formansatz}, we see that
these are exactly the equations of motions for the magnetic
part $\hat{F}$ of the RR-fluxes or equivalently, using the Hodge duality, the Bianchi identities for the electric part $\tilde{F}$.
Taking the magnetic part $\hat{F}$ as a reference we refer to them as equations of motion. More importantly,
we see that supersymmetry implies that they should be satisfied {\em without} sources. In fact, these sources
are disallowed anyway since they would correspond to instantons. This is because their current contains
the external volume $\text{vol}_4$, which means they are not extended in any of the four external dimensions, in particular
the time. We will come back to the definition of a current in section \ref{gensubman}, but for now it suffices to know
that a current is a sort of Poincar\'e dual and has indices in directions {\em orthogonal} to the source.
Furthermore, although it is somewhat more complicated to show, the equation of motion for $H$, eq.~\eqref{eomH},
also follows from the supersymmetry conditions \eqref{susyconds} \cite{integr}.

On the other hand the Bianchi identities for $\hat{F}$, which do allow for sources and read
\eq{
\label{bianchicomp}
\d_H \hat{F} = - 2 \kappa_{10}^2 \, j_{\text{total}} \, ,
}
and the Bianchi identity for $H$,
\eq{
\label{bianchiH}
\d H =0 \, ,
}
do {\em not} follow from the supersymmetry and must be imposed separately to obtain a vacuum solution.

In fact, the following can be shown \cite{lt,gauntlettIIB,integr}:
\begin{theor}
\label{theor:integr}
For compactifications of type II supergravity to Minkowski or AdS$_4$, the supersymmetry
equations for the bulk (eqs.~\eqref{susyconds} and \eqref{susycondsAdS} respectively) and for the
sources (see chapter~\ref{branesection}), together with the Bianchi identities for $H$ and $\hat{F}$ imply all the other equations of motion:
the Einstein equation, the dilaton equation and the equations of motion for $H$ and $\hat{F}$.
\end{theor}
In fact, under mild conditions the theorem can be extended to generic type II supergravity solutions, and also to eleven-dimensional
supergravity \cite{gauntlettMkilling}. In chapter \ref{branesection} we will study the supersymmetry conditions for sources and find that they require the sources to be generalized calibrated.

\subsubsection*{Example}

\begin{ex}
\label{CYex}
As an example of a solution to the eqs.~\eqref{susyconds} consider a type IIB SU(3)-structure geometry with
fluxes. In this case
\eq{\label{su3ansatz}
\Psi_1=e^{i\vartheta} e^{i \omega}\quad\text{and}\quad \Psi_2=\Omega\ ,
}
where $e^{i\vartheta}$ is a phase appearing as follows in the relation between the internal spinors:
$\eta^{(2)}_+=-i e^{-i\vartheta} \eta^{(1)}_+$. As we will see later, this phase determines which calibrated
D-brane probes are possible. Let us take $e^{i\vartheta}=1$ which amounts to a vacuum that allows for
supersymmetric D3/D7-branes. Plugging the ansatz~\eqref{su3ansatz} into \eqref{susyconds} we find:
\subeq{\al{
\label{susywarpsol}
& \d (e^{3A-\Phi} \Omega) = 0 \, , \qquad \d (e^{2A-\Phi} \omega) = 0 \, , \\
& H \wedge \Omega = H \wedge \omega = 0 \, , \\
& *_6 G_3 =i G_3 \, , \\
& \bar{\partial} \tau = 0 \, , \\
& 4 \d A - \d \Phi = e^\Phi *_6 \! F_{(5)} \, ,
}}
where $G_{(3)}=F_{(3)}+ie^{-\Phi}H$ and $\tau=C_{(0)}+ie^{-\Phi}$. This is the F-theory solution.
If we further take the dilaton $\Phi$ constant we recover the famous warped Calabi-Yau geometry \cite{gkp}.
\end{ex}

\subsubsection*{No-go theorem}

The no-go theorem of \cite{malnun} (see also the earlier \cite{nogoold}) states the following:
\begin{theor}[Maldacena-N\'u\~{n}ez]
\label{nogomalnun}
In order to obtain a Minkowski solution with non-zero
fluxes or a dS solution of supergravity (without $\alpha'$-corrections) on a compact internal manifold,
one needs to introduce sources with negative tension.
\end{theor}
Luckily, in string theory there are candidates
for such negative-tension sources: the orientifolds. For instance, the example above requires the introduction of orientifold three-planes (03-planes)
in order to satisfy the Bianchi identities. The constraints induced by the integrated version of these Bianchi identities in a compact space are better
known in the literature as the {\em tadpole cancelation} constraints. Finding supergravity backgrounds different from the well-known warped Calabi-Yau/F-theory
background is difficult, partly because of the presence of localized orientifolds, as required by the no-go theorem. One approximation would be
to smear the source terms along some or all directions of the internal manifold. In this way Minkowski compactifications with strict SU(3)-structure
or static SU(2)-structure were found in e.g.~\cite{granascan}, and with intermediate structure in \cite{integr,andriot}. To the best of our knowledge no Minkowski solutions
to the full supergravity equations were so far found that incorporate a dynamic (let alone type-changing) SU(3)$\times$SU(3)-structure. On the other hand, recently AdS$_4$ compactifications (without
sources since they can circumvent the no-go theorem) with such a structure have been constructed \cite{tomasiellomassive2,petriniSU3SU3,lusttsimpisSU3SU32,tomasiellomassive3} as geometric duals
to a Chern-Simons-matter theory with unequal levels, although they are not completely known in analytic form (see section \ref{sec:AdS4SU3ex}). These are certainly
very interesting backgrounds, because their analysis would require the full power of Generalized Complex Geometry.

Since AdS$_4$ compactifications provide a way to circumvent the no-go theorem without the complication of having to introduce sources,
let us take a closer look at them.

\subsection{AdS$_4$ compactifications: supersymmetry conditions}

Let us describe now how the supersymmetry conditions \eqref{susyconds} change
for AdS$_4$ compactifications. In this case we consider a warped compactification as in \eqref{metricansatz}, except that $\d s^2_{(4)}$ is now the AdS$_4$-metric,
 and again
we require the RR and NSNS-fluxes to only depend on the internal coordinates and this time respect the symmetry of AdS$_4$. The RR-fluxes still
take the form \eqref{formansatz} where $\text{vol}_4$ is now the volume-form on AdS$_4$.
The first major change is that the 4D spinors $\zeta_\pm$ can no longer be taken constant. Instead, they must satisfy
the AdS Killing spinor equation:
\eq{
\nabla_\mu \zeta_-=\pm \frac12 W_0 \gamma_\mu \zeta_+ \label{spinori}\ ,
}
for IIA/IIB, where
\eq{
\label{W0val}
W_0=\frac{e^{-i\theta}}{R} \, ,
}
with $R$ the AdS radius. It can be shown \cite{effective} (see also exercise \ref{exc:W0val}) that $W_0$ is proportional to the on-shell value
of the superpotential $\mathcal{W}$ of the
4D effective theory, given by eq.~\eqref{suppot}. Furthermore $|W_0|^2 = - \Lambda/3$ with $\Lambda$ the effective 4D cosmological constant.

The supersymmetry conditions \eqref{susyvar} are now equivalent to the minimal set
of equations \cite{granaN1}:
\subeq{\label{susycondsAdS}
\al{
\label{susycond0AdS}
\d_H \big(e^{4A-\Phi} \Re \Psi_1 \big) & = (3/ R) \, e^{3A-\Phi} \Re (e^{i\theta} \Psi_2) + e^{4A} \tilde{F} \, , \\
\d_H \big(e^{3A-\Phi} e^{i\theta}\Psi_2 \big) & = (2/ R) \, i \, e^{2A-\Phi}\Im \Psi_1 \label{susycond1AdS} \, .
}}
Acting with $\d_H$ on eq.~\eqref{susycond1AdS} we find that
this equation implies as an integrability condition the further equation:
\eq{\label{intcondAdS}
\d_H(e^{2A-\Phi}\Im \Psi_1)=0 \, .
}
This time neither of the generalized almost complex structures is integrable. See figure \ref{SU3SU3fig}.


Remember that these equations must be completed with the Bianchi identity \eqref{bianchicomp} for $\hat{F}$. There is
an alternative formulation for eq.~\eqref{susycond0AdS}, where the Hodge duality (hidden in $\tilde{F}$ given by eq.~\eqref{dualin6})
with its explicit dependence on the metric, does not appear anymore \cite{tomasiellomoduli,tomasiellomassive2}. In order
to derive it one uses the generalized Hodge decomposition \eqref{decomp2} and the property \eqref{actionhodge2} of the Hodge duality.
Taking the Mukai pairing of eq.~\eqref{susycond0AdS}
with $e^{3A - \Phi} e^{i \theta} \Psi_2$ and using \eqref{susycond1AdS} one can derive
\eq{
\label{Faux}
\hat{F}|_{U_{0,-3}} = -\frac{i}{2} e^{-A-\Phi} W_0 \bar{\Psi}_2 \, , \quad {\text{ and c.c.}}
}
Furthermore, taking into account eq.~\eqref{maxder} and using condition \eqref{intcondAdS} we find
\eq{\label{aux2}\spl{
&\d_H \left( e^{2A-\Phi} \Re \Psi_1 \right)|_{U_{2,\pm 1}} = i \, \d_H \left( e^{2A-\Phi} \Im \Psi_1 \right)|_{U_{2,\pm 1}} = 0\\
&\Longrightarrow \d_H \left( e^{4A-\Phi} \Re \Psi_1 \right)|_{U_{2,\pm 1}} = - e^{4A} \d_H \left( e^{-\Phi} \Re \Psi_1 \right)|_{U_{2,\pm 1}} \, ,
}}
and similarly for the $U_{-2,\pm 1}$-part. By means of these two auxiliary equations one finds from the $U_{k,l}$-decomposition
of eq.~\eqref{susycond0AdS}:
\subeq{\label{susycond0AdSdecomp}\al{
& \d_H(e^{-\Phi} \Re \Psi_1)|_{U_{0,-3}} = e^{-A -\Phi} W_0 \bar{\Psi}_2 = 2 i \, \hat{F}|_{U_{0,-3}} \, , \qquad \text{and c.c.} \, , \\
& \d_H(e^{-\Phi} \Re \Psi_1)|_{U_{\pm 2,-1}} = i \hat{F}|_{U_{\pm 2,-1}} \, , \qquad \text{and c.c.} \, , \\
& \d_H(e^{-\Phi} \Re \Psi_1)|_{U_{0,-1}} = i \hat{F}|_{U_{0,-1}} \, , \qquad \text{and c.c.}
}}
Since these expressions do not depend anymore on the eigenvalue of $\mathcal{J}_1$ (the $k$-value in the $U_{k,l}$-decomposition),
we can concisely rewrite this as
\eq{
\label{susycond0AdSalt}
\mathcal{J}_2 \cdot \d_H\left( e^{-\Phi} \Re \Psi_1 \right) = \hat{F} - 5 e^{-A - \Phi} \Im \left(\bar{W}_0 \Psi_2\right) \, ,
}
where $\mathcal{J}_2$ acts in the spinor representation \eqref{Jspinorrep}. This condition can be taken to replace eq.~\eqref{susycond0AdS}.

Before coming to the examples we state the following no-go theorem for the structure of supersymmetric AdS$_4$ compactifications
\cite{effectivecosets}, see also exercise \ref{nogoads}.
\begin{theor}
In classical type IIB supergravity there are no supersymmetric AdS$_4$ compactifications with strict SU(3)-structure.
SU(3)$\times$SU(3)-structure is still possible provided it is type-changing with a locus where the structure is static SU(2).
On the other hand, in classical type IIA supergravity there are no supersymmetric AdS$_4$ compactifications with static SU(2)-structure (see also \cite{ltb}).
SU(3)$\times$SU(3)-structure is still possible provided $e^{3A-\Phi} \Psi_1|_0$ (where $|_0$ projects on the zero-form part)
is non-constant.
\end{theor}

\subsection{AdS$_4$ compactifications: examples}
\label{sec:AdS4SU3ex}

\subsubsection*{L\"ust-Tsimpis AdS$_4$ compactifications with strict SU(3)-structure}

We will now work out the conditions for supersymmetric type IIA AdS$_4$ compactifications with strict SU(3)-structure \cite{lt}.
In this case we find for the pure spinors
\eq{
\Psi_1 = \Psi_- = i \Omega \, , \qquad e^{i\theta}\Psi_2 = e^{i\theta}\Psi_+ = -i e^{i\theta +i\vartheta} e^{i \omega} = -i e^{i \hat{\theta}} e^{i \omega} \, ,
}
where we combined the phase $e^{i \theta}$ of $W_0$ and the phase $e^{i \vartheta}$ (defined as $\eta^{(2)}_+=e^{-i\vartheta} \eta^{(1)}_+$)
into $e^{i\hat{\theta}}$. Plugging this ansatz into the supersymmetry conditions~\eqref{susycondsAdS} we find
after a straightforward calculation
\subeq{\label{geomAdS4}\al{
& \d \omega = 2  \, e^{-A} |W_0| \sin \hat{\theta} \, \Re \Omega \, , \label{geomAdS4a}\\
& \d \Omega = \frac{4}{3} i e^{-A} |W_0| \sin \hat{\theta} \, \omega \wedge \omega + i W_2 \wedge \omega + d A \wedge \Omega \, ,
}}
from which we read off the torsion classes of eq.~\eqref{torsioncl}:
\eq{
\label{geomAdS42}
\mathcal{W}_1 = i W_1 = \frac{4}{3} i e^{-A} |W_0| \sin \hat{\theta} \, , \qquad \mathcal{W}_2 = i W_2 \quad \text{purely imaginary} \, ,
\qquad \mathcal{W}_5 = \partial A \, .
}
Since $\mathcal{W}_1$ and $\mathcal{W}_2$ are purely imaginary, for convenience we introduce the real quantities $W_1, W_2$
denoting their imaginary parts. For the warp factor, the dilaton and the phase factor $\hat{\theta}$
we find
\subeq{\al{
& 3 \, \d A - \d \Phi = 0 \, , \label{dilwarp} \\
& \d \hat{\theta} = 0 \, ,
}}
and finally for the form fields
\subeq{\label{fluxAdS4}\al{
& H = - 2 \, e^{-A} |W_0| \cos \hat{\theta} \, \Re \Omega \, , \label{fluxHAdS4}\\
& \hat{F}_0 = - 5 \, e^{-\Phi-A} |W_0| \cos \hat{\theta} \, , \label{fluxF0AdS4}\\
& \hat{F}_2 = -\frac{1}{3} e^{-\Phi-A} |W_0| \sin \hat{\theta} \, \omega - e^{-\Phi} W_2 - 2 \, e^{-\Phi} \iota_{\d A} \Re \Omega \, , \label{fluxF2AdS4} \\
& \hat{F}_4 = - \frac{3}{2} \, e^{-\Phi - A} |W_0| \cos \hat{\theta} \, \omega \wedge \omega \, , \\
& \hat{F}_6 = 3 \, e^{-\Phi - A} |W_0| \sin \hat{\theta} \, \text{vol}_6 \, ,
}}
where we turned the one-form $\d A$ into a vector by raising the index with the metric $g$ (associated to $(\omega,\Omega)$).
In order to verify these results, the reader has to evaluate the Hodge star in \eqref{susycond0AdS} (or alternatively
start from eq.~\eqref{susycond0AdSalt}), for which it is convenient to use the
following property, which applies to any primitive $(1,1)$-form
\eq{
\label{hodgesimple}
*_6 (W_2 \wedge \omega) = - W_2 \, ,
}
and similar properties for $\omega$ and $\Omega$
\eq{
\label{hodgeomOm}
*_6 \omega = \frac{1}{2} \, \omega \wedge \omega \, , \qquad *_6 \Omega = - i \Omega \, ,
}
which can be found from eq.~\eqref{actionhodge2}.

As follows from theorem \ref{theor:integr}, in order to obtain a solution to the supergravity equations of motion, we must also
impose the Bianchi identities \eqref{bianchicomp} and \eqref{bianchiH}, where in this case we are looking for solutions without source terms ($j_{\text{total}}=0$).
The Bianchi identity for $\hat{F}_0=m$, which just says the Romans mass $m$ should be constant, immediately puts
\eq{
\d A = \d \Phi = \mathcal{W}_5 = 0 \, ,
}
{\em unless} the Romans mass $m$ itself vanishes. For now
we will assume that the warp factor and dilaton are indeed constant. The constant warp factor can then be absorbed into the
definition of the AdS-radius $|W_0|=1/R$. The Bianchi identities
for $H$ and $\hat{F}_4$ are automatically satisfied and the only further condition comes from
the Bianchi identity for $\hat{F}_2$,
\eq{
\d \hat{F}_2 + m H=0\,,
}
which, plugging in eqs.~\eqref{geomAdS4a}, \eqref{geomAdS42}, \eqref{fluxHAdS4}, \eqref{fluxF0AdS4} and \eqref{fluxF2AdS4}, implies
\eq{
\d W_2 = \left[ \frac{2}{5} \, e^{2 \phi} m^2 - \frac{3}{8} \left(W_1\right)^2\right] \Re \Omega \, .
}
At the same time, by taking the exterior derivative of $\Omega \wedge W_2 = 0$ one can derive
\eq{
\d W_2 = - \frac{1}{4} \, \left(W_2 \cdot W_2\right)\, \Re \Omega \, .
}
Combining both, one finds the following bound on the torsion classes:
\eq{
\label{SU3IIAbound}
\frac{16}{5} e^{2 \phi} m^2 = 3 \, \left(W_1\right)^2 - 2 \, W_2 \cdot W_2 \ge 0 \, .
}
This bound can be relaxed by introducing negative-tension sources, like orientifolds (see e.g.~\cite{cosets}).
Since the warp factor is constant, these orientifold sources must be necessarily smeared, so that their proper
string theory interpretation becomes problematic. In this section we will therefore
avoid introducing such sources.

It is interesting to observe that:
\begin{theor}
\label{enhsusy}
{\em Unless} the Romans mass $m$ is zero, strict SU(3)-structure for type IIA AdS$_4$ compactifications
is incompatible with extended supersymmetry $\mathcal{N}>1$ \cite{tomasiellomassive}.
\end{theor}
This comes about as follows: each 4D
supersymmetry is generated by a set of internal spinors $\left(\eta^{(1)}_+,\eta^{(2)}_+\right)$,
inducing a pair of pure spinors $(\Psi_1,\Psi_2)$ as in eq.~\eqref{nps}.
Each of these pure spinor pairs must satisfy the supersymmetry conditions \eqref{susycondsAdS},
which for strict SU(3)-structure, described by $(\omega,\Omega)$, reduce to the above eqs.~(\ref{geomAdS4}-\ref{fluxAdS4}).
In particular, from eq.~\eqref{fluxHAdS4} follows that
$\Re \Omega$ is proportional to the physical field $H$ so that $\Re \Omega$ is the same in every set of SU(3)-structures.
Now, as we discussed below eq.~\eqref{Jfromrho}, the real part of $\Omega$ completely determines the imaginary part as well as the almost
complex structure, so that in fact the
full $\Omega$ and the almost complex structure is the same in every set. Furthermore from eq.~\eqref{sympfromg} we find that if we know the metric, $\omega$ is completely
determined in terms of the almost complex structure. It follows that all the sets $(\omega,\Omega)$
are equivalent and the solution does not have extended supersymmetry at all. The loophole is that the argument does not go through
if the proportionality factor in \eqref{fluxHAdS4} vanishes, so exactly when $m=0$.

Concluding, in order to obtain a supersymmetric $\mathcal{N}=1$ AdS$_4$ compactification of type IIA supergravity without sources and
of strict SU(3)-structure type, the internal manifold $M$ must possess an SU(3)-structure with the only non-vanishing torsion classes
the purely imaginary $\mathcal{W}_1$ and $\mathcal{W}_2$, satisfying the bound \eqref{SU3IIAbound}. This is a special
case of a half-flat geometry (see the entry in table \ref{torsiontable}, which applies after redefining $\Omega$ so that the
phases of $\mathcal{W}_{1,2}$ shift). The
fluxes can then be found from eqs.~\eqref{fluxAdS4} and \eqref{SU3IIAbound}, and the dilaton and warp
factor are constant. In the special case of vanishing Romans mass, it is possible to also have non-constant
warp factor and dilaton, satisfying eq.~\eqref{dilwarp}, and non-zero $\mathcal{W}_5$, but then the condition coming
from the Bianchi identity of $\hat{F}_2$ is more complicated than we discussed above.

So far we have in fact merely simplified
the conditions on the geometry, in the next subsection we will give an overview of actual geometries that satisfy them.

\subsubsection*{Overview of known geometries satisfying the conditions of L\"ust and Tsimpis}

The known examples of geometries satisfying the conditions of L\"ust and Tsimpis roughly belong
to two classes (with overlap): families of solutions containing a nearly K\"ahler geometry and solutions that
can be lifted to M-theory. Interestingly, most of these solutions play a role in the ABJM-duality \cite{abjm} and
can be considered as the geometric dual to a Chern-Simons-matter theory in the regime where the type IIA description
is valid.

Let us start with the 6D {\em nearly K\"ahler manifolds}, for which the only non-vanishing
torsion class is $\mathcal{W}_1$ (for a review see \cite{nkreview}). For $\mathcal{W}_2=0$ the bound \eqref{SU3IIAbound} is obviously satisfied.
Unfortunately, nearly K\"ahler manifolds are rare. In fact, the only known 6D examples are homogeneous and moreover
it has been shown that the only homogeneous examples are the group and coset manifolds \cite{nkreview}
\eq{\label{nkgeom}\spl{
& \mathbb{CP}^3 = \frac{\text{Sp(2)}}{\text{S}(\text{U(2)}\times \text{U(1)})}, \qquad \mathbb{F}(1,2;3) = \frac{\text{SU(3)}}{\text{U(1)}\times \text{U(1)}} \, , \\
& S^3 \times S^3 = \text{SU(2)}\times \text{SU(2)}, \qquad S^6 = \frac{\text{G}_2}{\text{SU(3)}} \, ,
}}
where $\mathbb{F}(1,2;3)$ is the {\em flag manifold} of complex lines and complex planes in $\mathbb{C}^3$ such that the
line belongs to the plane. See also \cite{lustcoset} for early examples of compactifications on these coset manifolds.
These nearly K\"ahler geometries are unique up to an overall scale. It is still an open question whether there are any
other (necessarily non-homogeneous) 6D nearly K\"ahler manifolds. Plugging the nearly K\"ahler geometry into \eqref{SU3IIAbound}
and \eqref{fluxAdS4} we find then the full solution, which in the end has two parameters: the overall scale and the dilaton.

It was discovered in \cite{tomasiellocosets,cosets} that for both geometries in the first line of eq.~\eqref{nkgeom},
namely $\mathbb{CP}^3$ and $\mathbb{F}(1,2;3)$, there exists a family of solutions around the nearly K\"ahler solution, with $\mathcal{W}_2 \neq 0$.
These families have respectively one and two shape parameters. Apart from the shape parameters there are again two more parameters corresponding
to the overall scale and the dilaton. For a proper string theory solution these continuous parameters must take discrete values because of flux
quantization. For the geometries in the second line, if one does not allow for source terms, there is only
the nearly K\"ahler solution. See table \ref{cosetsolutions}.
\begin{table}[tp]
\begin{center}
\rowcolors{2}{blue!40}{blue!10}
\begin{tabular}{|c|c|c|c|c|}
\hline
& SU(2)$\times$SU(2) \rule[1.2em]{0pt}{0pt} & $\frac{\text{G}_2}{\text{SU(3)}}$ & $\frac{\text{Sp(2)}}{\text{S}(\text{U(2)}\times \text{U(1)})}$ & $\frac{\text{SU(3)}}{\text{U(1)}\times \text{U(1)}}$ \\
\hline
\# of shape parameters  & 0 & 0 & 1 & 2 \\
$\mathcal{W}_2\neq 0$ &
No & No & Yes & Yes \\
\hline
\end{tabular}
\caption{6D cosets that satisfy the necessary and sufficient conditions
for a $\mathcal{N}=1$ strict SU(3)-structure compactification to AdS$_4$ in the absence of sources.}
\label{cosetsolutions}
\end{center}
\end{table}

Let us look in some detail at the family of solutions on $\mathbb{CP}^3$. We will just describe
some noteworthy features of these solutions and refer for more details to the original papers \cite{tomasiellocosets,cosets}.
The metric, the SU(3)-structure $(\omega,\Omega)$ and all the form fields are invariant under the action of Sp(2),
the numerator $G$ of the coset description. There is one shape parameter, let us call it $\sigma$, which in the conventions of both papers takes values between
\eq{
\frac{2}{5} \le \sigma \le 2 \, .
}
Roughly speaking, considering $\mathbb{CP}^3$ as an $S^2$ bundle over $S^4$ (this is its description as a {\em twistor space}, see \cite{tomasiellocosets}),
the parameter $\sigma$ describes the relative scale of the $S^2$ with respect to the $S^4$. The nearly K\"ahler point then
corresponds to $\sigma=1$.
At the extreme values $\sigma=2,2/5$ the bound \eqref{SU3IIAbound} is saturated and the Romans mass
$m$ becomes zero. This implies that the solution can be lifted to M-theory, and it turns out it corresponds
to a compactification on the round $S^7$ and the squashed $S^7$ respectively (see table \ref{table:redsols}).
At the special value $\sigma=2$, first investigated in \cite{nilssonpope}, the metric becomes the familiar Fubini-Study metric. The bosonic symmetry
group $G$ of the solution then enhances from Sp(2) to SU(4) and it is possible to describe $\mathbb{CP}^3$ as a coset in a perhaps more familiar way
\eq{
\mathbb{CP}^3 = \frac{\text{SU(4)}}{\text{S}(\text{U(3)}\times \text{U(1)})} \, .
}

At the same time the supersymmetry enhances from $\mathcal{N}=1$ to $\mathcal{N}=6$. As we saw in theorem \ref{enhsusy}
such and extension of the supersymmetry is indeed possible if $m=0$. Associated to the Fubini-Study metric is a U(3)-structure
$(\tilde{\omega},\tilde{J})$, where $\d \tilde{\omega}=0$ and $\tilde{J}$ is an {\em integrable} complex structure, making the U(3)-structure into
a {\em K\"ahler structure} (see table \ref{torsiontable}). There is however no globally defined holomorphic three-form associated to $\tilde{J}$, so that this U(3)-structure is {\em not}
an SU(3)-structure. It is {\em different} from the SU(3)-structure $(\omega,\Omega)$ entering the supersymmetry
conditions. In fact, to each linear combination of the six supersymmetry generators a different SU(3)-structure $(\omega,\Omega)$
is associated, which is invariant under a different Sp(2)-embedding in SU(4). Indeed, there is an $S^5$ worth of
Sp(2)-embeddings in SU(4) corresponding to the choice of (normalized) supersymmetry generator. There is also an $S^5$ worth
of mass deformations, depending on the choice of Sp(2) that preserves the massive solution (they are all equivalent under the action of
an element of SU(4) though). The major disadvantage of the description of the $\sigma=2$ solution in terms of $(\omega,\Omega)$ is that it obscures
the $\mathcal{N}=6$ supersymmetry, making only one supersymmetry manifest. The existence of both the integrable and non-integrable complex structure (associated to the same metric)
is a generic property of twistor spaces \cite{standardJ,nonstandardJ}, which is the description that was used in \cite{tomasiellocosets}
to originally construct this family of solutions.

The lift to M-theory of the solution for $\sigma=2$ is the geometric dual to the original ABJM Chern-Simons matter theory \cite{abjm}.
The dual for the other massless configuration, $\sigma=2/5$, was proposed in \cite{oogurisquashed} and for the massive solutions
with $2/5 < \sigma < 2$ in \cite{tomasiellomassive}, where it was argued that non-zero Romans $m$ corresponds to having $k_1 + k_2 \neq 0$,
where $k_1,k_2$ are the Chern-Simons levels.

For $\mathbb{F}(1,2;3)$ a similar, but somewhat more complicated story --- since there are two shape parameters --- applies.
For a one-dimensional subfamily the Romans mass becomes zero, and the solution lifts to M-theory (see the third and the fourth line of table
\ref{table:redsols}). The CFT-duals are still unclear, but are probably related the proposal of \cite{tomasielloN3}.

We come now to the second class of known solutions (because of overlap with the first class we already saw some examples), namely
the solutions that lift to M-theory. Necessarily, these solutions have $m=0$, since there is so far no known way to lift the Romans mass to
M-theory. Table \ref{table:redsols} gives a list of cases that have been studied in some more detail in the literature. See also \cite[Table 6]{KKreview} for an overview
of homogeneous M-theory solutions.
\begin{table}[tp]
\begin{center}
\rowcolors{2}{blue!40}{blue!10}
\begin{tabular}{|c|c|c|c|}
\hline
7d manifold & \# SUSY 7d & 6d reduction & \# SUSY 6d\\
\hline
Round $S^7$ & 8 & $\mathbb{CP}^3$ & 6 \\
Squashed $S^7$ & 1 & Squashed $\mathbb{CP}^3$ & 1 \\
$N^{1,1}|_I$ & 3 & $\frac{\text{SU(3)}}{\text{U(1)}\times \text{U(1)}}$ & 1 \\
$N^{1,1}|_{II}$, $N^{k,l}|_{I,II}$ & 1 & $\frac{\text{SU(3)}}{\text{U(1)}\times \text{U(1)}}$ & 1 \\
$M^{3,2}$, $Y^{p,q}(\mathbb{CP}^2)$ & 2 & $S^2 \rightarrow \mathbb{CP}^2$ & 2 \\
$Q^{1,1,1}$, $Y^{p,q}(\mathbb{CP}^1\times\mathbb{CP}^1)$ & 2 & $S^2 \rightarrow \mathbb{CP}^1 \times \mathbb{CP}^1$ & 2 \\
\hline
\end{tabular}
\caption{Type IIA AdS$_4$ compactifications obtained from a reduction of an M-theory solution.}
\label{table:redsols}
\end{center}
\end{table}

The M-theory solutions listed in the table are all of the so-called {\em Freund-Rubin} type. This means that
the internal part of the M-theory four-form flux $G_4$ vanishes. The external part is then just proportional
to the 4D volume-form and the proportionality factor is called the {\em Freund-Rubin} parameter.  It
follows from the Einstein equation in M-theory that the internal space $M_7$ should be Einstein with positive Ricci scalar, i.e.\
\eq{
\label{einsteinspc}
R_{mn} = 6 \, |c|^2  g_{mn}\, ,
}
for some constant $c$ related to the Freund-Rubin parameter. Moreover, from the requirement of supersymmetry
follows that there should exist at least one globally defined nowhere-vanishing spinor $\eta$ that satisfies the Killing spinor equation
\eq{
\label{killing7d}
\nabla_i \eta = \frac{i}{2} c \, \gamma_i \eta \, .
}
Note that this Killing spinor equation implies eq.~\eqref{einsteinspc}, but not the other way round. From eq.~\eqref{killing7d} follows that the cone has a covariantly constant
spinor \cite{bar}. Depending on the number of Killing spinors, the manifold is either {\em weak $G_2$} (for just one Killing spinor),
Sasaki-Einstein (for two Killing spinors), tri-Sasakian (for three Killing spinors) or $S^7$ (for eight Killing spinors).
For a review on Sasakian and tri-Sasakian geometry see \cite{sasakireview}.
From the point of view of the AdS$_4$/CFT$_3$ correspondence, Sasakian manifolds have been studied in \cite{martellisparksnotesSE7,martellisparksnotesSE72}, and certain tri-Sasakian manifolds
in \cite{tomasielloN3}.

Note that if one would start with
a supersymmetric M-theory solution {\em with} internal four-form flux (e.g.\ the ``stretched and the warped'' solution of \cite{stretched1,stretched2,stretched3})
and one would be able to apply a reduction to type IIA that preserves supersymmetry, that solution would necessarily have SU(3)$\times$SU(3)-structure.
Indeed, the internal M-theory four-form flux would lead to non-zero $H$ and $\hat{F}_4$ in type IIA. Since the
Romans mass $m$ is zero, that would be incompatible with eqs.~\eqref{fluxAdS4}, which follow
from the strict SU(3)-structure ansatz.

As we mentioned already the coset spaces $N^{k,l}|_{I,II}$ (for their definition including the explicit form of the Einstein metrics see \cite{casnpqr}),
correspond to the lift of the massless solutions on $\mathbb{F}(1,2;3)$. To explain the last two lines in table \ref{table:redsols} we note that
in \cite{martellisparksYpq} a construction of an infinite family of $2n+3$-dimensional Sasaki-Einstein
manifolds with explicit metric, $Y^{p,q}(B_{2n})$, was presented for every $2n$-dimensional K\"ahler-Einstein
manifold $B_{2n}$. The topology of these manifolds is that of a certain $S^3/\mathbb{Z}_p$-bundle over $B_{2n}$.
In the case of interest here, we take $n=2$ in order to obtain a seven-dimensional Sasaki-Einstein space.
The smooth four-dimensional K\"ahler-Einstein manifolds with positive curvature were classified in
\cite{tianyau}: they are $\mathbb{CP}^2$, $\mathbb{CP}^1\times \mathbb{CP}^1$ and $dP_l$ ($l=3,\ldots,8$).
The K\"ahler-Einstein metric on the del Pezzo surfaces $dP_l$ is not explicitly know, so we restrict
here to $\mathbb{CP}^2$ and $\mathbb{CP}^1 \times \mathbb{CP}^1$. Furthermore it turns out that the
homogeneous Sasaki-Einstein manifolds (for an overview see e.g.~\cite{casromoverview})
\eq{
M^{3,2} = \frac{\text{SU(3)}\times\text{SU(2)}}{\text{SU(2)}\times\text{U(1)}} \, , \qquad
Q^{1,1,1} = \frac{\text{SU(2)}^3\times\text{U(1)}}{\text{U(1)}^3} \, ,
}
and their quotients by $\mathbb{Z}^r$ belong to the $Y^{p,q}(\mathbb{CP}^2)$ and $Y^{p,q}(\mathbb{CP}^1 \times \mathbb{CP}^1)$
family respectively:
\eq{
M^{3,2}/\mathbb{Z}_r = Y^{2r,3r}(\mathbb{CP}^2) \, , \qquad Q^{1,1,1}/\mathbb{Z}_r = Y^{r,r}(\mathbb{CP}^1 \times \mathbb{CP}^1) \, .
}
As proposed in \cite{martellisparksnotesSE7} and explicitly worked out in \cite{petriniSU3SU3,lusttsimpisSU3SU32} in all these cases
it is possible to perform a reduction to type IIA along a specific U(1) in the $S^3/\mathbb{Z}_p$ fiber, which preserves $\mathcal{N}=2$ supersymmetry. We stress that the reduction is not
along the so-called {\em Reeb vector} of the Sasaki-Einstein, which would break all the supersymmetry. Moreover, for a generic
$Y^{p,q}$ (so apart from the coset spaces $M^{3,2}$ and $Q^{1,1,1}$) the reduction along the Reeb vector would result in
a singular manifold. The resulting SU(3)-structure type IIA solutions are {\em not} homogeneous and have a non-constant warp factor and dilaton.
Note that the table is not complete as there are other known (even homogeneous) Sasaki-Einstein manifolds, like the Stiefel manifold $V_{5,2}$ (see \cite{martellistieffel}
for a discussion in the AdS$_4$/CFT$_3$-context). It was also argued in \cite{tsimpistoric} that toric geometry might be used in systematically constructing
new examples.

\subsubsection*{Solutions with SU(3)$\times$SU(3)-structure}

Interestingly, from the CFT-side it was argued in \cite{tomasiellomassive} that on $\mathbb{CP}^3$ there are two further families of massive supersymmetric type IIA solutions
with $\mathcal{N}=2$ and $\mathcal{N}=3$ respectively. Because of theorem \ref{enhsusy} they cannot have strict SU(3)-structure, but must necessarily have
dynamic SU(3)$\times$SU(3)-structure. In \cite{tomasiellomassive2} they were then constructed up to first order in the Romans mass. Unfortunately, no full analytic expression
up to all orders is known. In \cite{tomasiellomassive3} the solution was constructed up to three functions obeying a set of three first-order non-linear ordinary differential equations.

Relatedly, it was shown in \cite{petriniSU3SU3,lusttsimpisSU3SU32} that it is possible to deform the type IIA solutions
obtained from the reduction of $Y^{p,q}(\mathbb{CP}^2)$ and $Y^{p,q}(\mathbb{CP}^1\times\mathbb{CP}^1)$ by turning on a Romans mass, and
all the while preserving the $\mathcal{N}=2$ supersymmetry. Again the structure must be SU(3)$\times$SU(3). These solutions can be completely
constructed in terms of two functions satisfying a set of two first-order non-linear ordinary differential equations, for which there is no analytic solution yet (although they can be solved numerically).

\subsection{K\"ahler potential and superpotential of the $\mathcal{N}=1$ effective theory}
\label{sec:effective}

The $\mathcal{N}=1$ 4D effective theory corresponding to compactifications
with fluxes was, for the case of strict SU(3)-structure, systematically studied in \cite{gl1,gl2}.
In there expressions for the K\"ahler potential, the superpotential and the gauge-kinetic function
were proposed, giving a complete description of the $\mathcal{N}=1$ effective theory. For related work
see also \cite{louismicuhet,louismicu1,louismicu2,housepalti,louisSU2}. This was then extended to SU(3)$\times$SU(3)-structures
in \cite{granaN2part1,grimmN1,granaN2part2,effective}, where the K\"ahler potential and superpotential were calculated
in different ways. For more work see also \cite{micupalti,cassanibilal,cassanipot,martuccibulkdef,louisSU2SU2}.

The weak point of \cite{gl1,gl2} is its reduction ansatz. The {\em reduction ansatz} consists of a choice of modes kept
in the effective theory and a prescription of how to expand the 10D fields in terms of them. In \cite{gl1,gl2} the expansion forms are chosen to be harmonic,
just like in the fluxless Calabi-Yau case \cite{candelas}, which is unsuitable in the presence of fluxes. Indeed, the forms $(\omega,\Omega)$  describing
the SU(3)-structure are generically not closed anymore and one would expect at least some expansion forms that are not closed.

In fact, the choice of a suitable set of expansion forms is a difficult problem (see e.g.\ \cite{minpoor,poorNK} for a discussion of
the constraints on such a set). There are two ways to go about it. First one can ask for a {\em consistent truncation}, which means that
every solution of the 4D effective theory should lift to a 10D solution (where the 10D fields are
constructed from the 4D ones using the reduction ansatz). Secondly, one can take a more physical approach and expand in a
set of modes that have a much lower mass with respect to the others, and are thus the only ones to be excited in the low-energy theory.
In practice, identifying such a set is quite difficult. In some cases however, there is a natural set of such expansion forms. For instance, in the case
of coset manifolds $G/H$ it has been shown that expanding in forms that are invariant under the action of the group $G$ leads
to a consistent truncation \cite{cassred}. This has been used to construct the effective theory in \cite{effectivecosets}.

In this subsection, which should be considered more as a foretaste, we will avoid the issue of the reduction altogether.
Rather we will present the K\"ahler potential and superpotential of the $\mathcal{N}=1$ description, but keep all the KK-modes.
So effectively we end up with a ``4D'' theory with an infinite amount of fields. To be really useful this should still
be supplemented with a reduction ansatz, which should be substituted into the expressions~\eqref{kahlpot} and \eqref{suppot}
for the K\"ahler potential and the superpotential respectively. We will not derive these expressions, but provide an a posteriori justification
by deriving the F-terms and comparing with the supersymmetry conditions \eqref{susycondsAdS}.

In an $\mathcal{N}=1$ theory the scalar fields sit in chiral multiplets (and their complex conjugate anti-chiral multiplets).  There is thus a complex structure,
splitting the scalar fields into holomorphic and anti-holomorphic fields. In our case, it turns out \cite{effective} that the holomorphic scalar fields should be found in the expansions
of \eq{
e^{B}\mathcal{Z} = e^{3A - \Phi} e^{B} \Psi_2 \, , \qquad
e^{B}\mathcal{T} =  e^B \left(e^{-\Phi}\Re \Psi_1 - i C\right) \, ,
}
where $C$ are the RR-potentials. First we note that the degrees of freedom in the pure spinor $\Psi_1$ combine
with the degrees of freedom in the RR-potentials into chiral multiplets. Secondly, the fields are most accurately
described in the $B$-twisted description, which makes the degrees of freedom in the $B$-field explicit. It is however also
possible to express the K\"ahler potential and the superpotential in the untwisted picture, which we will do in the following.

The K\"ahler potential and the superpotential of the $\mathcal{N}=1$ 4D effective theory are then given
by
\al{
\label{kahlpot}\mathcal{K} & =  - \ln 4 \int_M e^{-4A} |C|^{-6} H(\Re \mathcal{Z}) - 2 \ln 4 \int_M e^{2A} H(\Re\mathcal{T}) + 3 \ln(8 \kappa_{10}^2)\, , \\
\label{suppot}\mathcal{W} & = \frac{i}{4 \kappa_{10}^2} \int_M\langle C^{-3} \mathcal{Z},\d_H \mathcal{T}\rangle\ .
}
where the Hitchin function $H(\Re\Psi)$ is defined in eq.~\eqref{hitchinfunc}. The Hitchin function can be related to the Mukai
pairing as follows
\eq{
H(\Re \Psi) = \frac{i}{4} \, \langle \Psi, \bar\Psi \rangle \, ,
}
keeping in mind that on the right-hand side, $\Im \Psi$ should be determined from $\Re \Psi$ through eq.~\eqref{impart}.
This leads to an expression of the K\"ahler potential that is more widely used in the literature, albeit less accurate.
The variation of the Hitchin function is given by \cite{hitchinGCY}
\eq{
\delta H(\Re \Psi) = \langle \delta(\Re \Psi) , \Im \Psi \rangle \, .
}

Note that the superpotential only depends on holomorphic fields as it should, while the K\"ahler potential does not, since
it explicitly depends on the real parts $\Re \mathcal{T}, \Re \mathcal{Z}$ separately. Furthermore, there are {\em K\"ahler transformations},
which read
\eq{
\mathcal{W}' = f^3 \mathcal{W} \, , \qquad \mathcal{K} = \mathcal{K} - 3 \, \ln f - 3 \, \ln f^* \, ,
}
for an arbitrary $f$ depending holomorphically on the coordinates on the moduli space (but
{\em not} on the internal coordinates on $M$). Therefore, there is an arbitrary function $C$ of the coordinates on moduli space
in the expressions for the K\"ahler potential and the superpotential. Another viewpoint is that through K\"ahler transformations we
can arbitrarily change the overall factor of $\mathcal{Z}$ appearing in the K\"ahler potential and superpotential.

Let us now construct the F-term equations, which in a standard $\mathcal{N}=1$ supergravity theory are given by
\eq{
\text{F-terms:} \qquad D_a \mathcal{W} = \partial_a \mathcal{W} + \left( \partial_a K\right) \mathcal{W} \, ,
}
where the $\partial_a$ indicate derivatives with respect to the holomorphic coordinates on moduli space. For our case without
reduction, we write:
\eq{
\text{F-terms:} \qquad D W = \delta \mathcal{W} + (\delta K) \mathcal{W} \, ,
}
where $\delta$ is the variation with respect to the holomorphic variables $\mathcal{Z}$ and $\mathcal{T}$.
The variation $\delta Z$ will only take values in $\Gamma(V_{3})$ and $\Gamma(V_{1})$ corresponding to
a deformation of the overall factor and a deformation of the associated generalized complex structure (see the infinitesimal version of eq.~\eqref{puredeform}) respectively.
Here we denoted by
$V_k$ the decomposition \eqref{decomp} associated to $\mathcal{J}_2$. We find then for the variations $\delta \mathcal{Z}$ and $\delta \mathcal{T}$:
\subeq{\al{
\label{varT}\delta \mathcal{T} & : \qquad \d_H \mathcal{Z} - 2 i W_0 \, e^{2A} \Im \mathcal{T} = 0 \, , \\
\label{varZ-3}\delta \mathcal{Z}|_{V_{3}} & : \qquad \left(\d_H \Re \mathcal{T} - i e^{-\Phi} \hat{F}\right)_{V_{-3}} -\frac{1}{2} W_0 \, e^{-4A} \bar{\mathcal{Z}} = 0 \, , \\
\label{varZ-1}\delta \mathcal{Z}|_{V_{1}} & : \qquad \left(\d_H \Re \mathcal{T} - i e^{-\Phi} \hat{F}\right)_{V_{-1}}  = 0 \, ,
}}
where $W_0$ is given by eq.~\eqref{W0val}. One finds immediately that eq.~\eqref{varT} is equivalent
to eq.~\eqref{susycond1AdS}, and using \eqref{Faux} that eqs.~\eqref{varZ-3} and \eqref{varZ-1} are equivalent
to eq.~\eqref{susycond0AdSdecomp} or equivalently eq.~\eqref{susycond0AdSalt}.

We conclude that eq.~\eqref{susycond1AdS} and \eqref{susycond0AdSalt} can be considered as F-term conditions.
In \cite{effective} it was show that eq.~\eqref{intcondAdS} on the other hand is a D-term equation.
In a general $\caln=1$ supergravity with non-zero vacuum expectation value of $\mathcal{W}$ --- here $W_0 \neq 0$ --- the D-term is indeed implied by the F-terms
(see e.g.~\cite{dealwiseff} where this point is particularly stressed).

\begin{ex}[Type IIB with SU(3)-structure]
Plugging in the type IIB SU(3)-ansatz eqs.~\eqref{su3ansatz} we find:
\eq{\spl{
& \mathcal{K} = - \ln \left[-i |C|^{-6} \int_M e^{2A - 2 \Phi} \Omega \wedge \bar{\Omega} \right] - 2 \, \ln \left[ \frac{4}{3} \int_M e^{2A - 2 \Phi }J^3 \right]+ 3 \ln(8 \kappa_{10}^2) \, , \\
& \mathcal{W} = \frac{i e^{i\vartheta}C^{-3}}{4 \kappa_{10}^2} \int_M e^{3A-\Phi} \Omega \wedge (G_{(3)} - \d \omega) \, .
}}
The K\"ahler potential is the one found in \cite{gl1}. In the case of a warped Calabi-Yau, $\d \omega=0$, and the superpotential reduces to the famous
Gukov-Vafa-Witten superpotential \cite{GVW1,GVW2,GVW3}.
\end{ex}
Since the supersymmetry conditions can be derived from the superpotential and the K\"ahler potential as
described above, they should also change upon adding a non-per\-tur\-ba\-tive correction (from instantons or
gaugino condensation) to the superpotential. As was discussed in \cite{effective} this might deform an ordinary
complex structure into a proper generalized complex structure.

\subsection{Exercises}
\begin{exc}[easy]
\label{exc:W0val}
Demonstrate eqs.~\eqref{Faux}, \eqref{aux2} and \eqref{susycond0AdSdecomp}.
Use \eqref{Faux} to calculate the on-shell value of $e^{\mathcal{K}/3} \mathcal{W}$ for a supersymmetric
AdS$_4$ solution and show that, putting $C=1$, it is equal to $W_0$.
\end{exc}
\begin{exc}[difficult]
\label{nogoads}
Use \eqref{susycondsAdS} to show that there are no supersymmetric IIB AdS$_4$ compactifications
with strict SU(3)-structure at the classical supergravity level. Show that, on the other hand, for type IIA there are no static SU(2) AdS$_4$
compactifications. What about compactifications with intermediate structure? See \cite[Appendix B.2]{effectivecosets} for the solution.
\end{exc}
\begin{exc}[intermediate]
Derive the eqs.~(\ref{geomAdS4}-\ref{fluxAdS4}), which are the conditions of L\"ust and Tsimpis, from
the pure spinor conditions \eqref{susycondsAdS}. Use the properties of the Hodge duality \eqref{hodgesimple} and
\eqref{hodgeomOm}. Alternatively, replace eq.~\eqref{susycond0AdS} by eq.~\eqref{susycond0AdSalt} as a starting point.
In this case $\mathcal{J}_2 = \mathcal{J}_{\omega}$ given in eq.~\eqref{symplexpr}.
\end{exc}

\clearpage

\thispagestyle{beginsection}
\section{D-branes}
\label{branesection}

In this chapter we discuss the embedding of supersymmetric D-branes in the backgrounds of the previous chapter.
Just like the backgrounds themselves are  naturally described in terms of Generalized Complex
Geometry we argue that supersymmetric D-branes are described by generalized calibrations.
Moreover, this concept will also provide a nice physical interpretation of the pure spinor supersymmetry
conditions \eqref{susyconds}.

\subsection{Calibrations}

In this section we review the theory of calibrations \cite{harveylawson} before extending it
to Generalized Complex Geometry. Calibrations were first used for the construction of supersymmetric
branes in \cite{beckerstrominger,beckercal2,calgibbons}.

Calibrations provide a mechanism to find minimal-volume surfaces in a curved space.
This means concretely that calibrated submanifolds $\Sigma$ minimize the action
$S = \int_\Sigma \sqrt{g}$. In general, this is a difficult problem that involves
second-order partial differential equations. Now, in certain cases, if the manifold
is equipped with a so-called calibration form the problem can be reduced to solving first-order equations.
This mechanism is analogous to how the first-order self-duality conditions in Yang-Mills theory
provide a solution to the second-order equations of motion. In Yang-Mills theory solutions to
the self-duality equations can be related to supersymmetric or BPS configurations.
Likewise generalized calibrated submanifolds correspond to D-branes that are supersymmetric. In fact, the
theory of generalized calibrations takes into account both the volume of the D-brane and the world-volume
gauge field so that it interpolates between (an extension to higher dimensions \cite{corrigan,thesis} of) the concept of
self-duality in abelian Yang-Mills and the theory of ordinary calibrations.

So let us get down to business and give the definition of a calibration form.
\begin{defi} A {\em calibration form} $\phi$ is an $l$-form on M
that satisfies
\begin{enumerate}
\item an algebraic condition:
in every point $p\in M$ and for every $l$-dimensional oriented subspace
of the tangent space $T$, spanned by an oriented basis $t_1,\ldots,t_l \in T_p M$,
we must have
\eq{
\label{calbound}
\sqrt{\det g|_T} \ge \phi(T) \, ,
}
where we defined $\phi(T)=(\iota_{t_l}\cdots\iota_{t_1}\phi)|_0$, and
$g|_{T\,ab}=g_{ij}t^i{}_a t^j{}_b$. Furthermore, in every point there must exist subspaces $T$ such
that the above bound is saturated. This requires in particular that $\phi$ is
appropriately normalized.
\item a differential condition:
\eq{
\label{caldif}
\d \phi=0 \, .
}
\end{enumerate}
\end{defi}
Note that the tangent space of an oriented submanifold $\Sigma$  provides
such an oriented subspace in every point of $\Sigma$. Indeed, suppose $\Sigma$ is
described by a parametrization $y^i(\sigma)$,
where $\sigma^{1},\ldots,\sigma^{l}$ are the world-volume coordinates, then the $t_a$
are given as follows: $t^i_a= \frac{\partial y^i}{\partial \sigma^a}$. The above
bound then implies:
\eq{
\d^l \sigma \sqrt{\det g|_\Sigma}  \ge \phi|_\Sigma \, ,
}
in every point of $\Sigma$, and where $|_{\Sigma}$ denotes the pull-back, defined in eq.~\eqref{pullback}.

\begin{defi}
A submanifold $\Sigma$ is calibrated if in every point $p \in \Sigma$ the above bound
is saturated:
\eq{
\label{calcond}
\d^l \sigma \sqrt{\det g|_\Sigma}  = \phi|_\Sigma \, .
}
\end{defi}

It follows that a calibrated submanifold is the manifold with the smallest volume within its
homology class. Indeed, suppose there is another manifold $\Sigma'$ within
the same homology class, which implies there is a submanifold $\mathcal{B}$ so that $\partial \mathcal{B}=\Sigma'-\Sigma$. In words:
the manifold $\mathcal{B}$ is such that its boundary is the difference between $\Sigma'$
and $\Sigma$. It follows that
\begin{multline}
\text{Vol}(\Sigma') = \int_{\Sigma'} \d^l \sigma \sqrt{\det g|_{\Sigma'}} \ge \int_{\Sigma'} \phi|_{\Sigma'}= \\
\int_{\Sigma} \phi|_{\Sigma} + \int_{\mathcal{B}} \d \phi|_{\mathcal{B}} = \int_\Sigma \phi|_{\Sigma} = \int_\Sigma \d^l \sigma \sqrt{\det g|_\Sigma}=\text{Vol}(\Sigma) \, ,
\end{multline}
where we used the calibration bound \eqref{calbound} for $\Sigma'$, Stokes' theorem, the differential condition \eqref{caldif},
and the calibration condition \eqref{calcond} for $\Sigma$.

So we have reduced the problem of constructing a minimal-volume submanifold to finding a solution
to the first-order differential equation \eqref{calcond}. It seems though that the difficulty
is merely transferred to finding a calibration form $\phi$ on $M$ with its peculiar properties. Luckily, at least on a Calabi-Yau
manifold there are a number of natural calibration forms

\begin{ex}[Complex submanifold]
\label{compex}
$\frac{1}{l!} \omega^l$. The calibrated manifolds are $2l$-dimensional complex submanifolds.
\end{ex}
\begin{ex}[Special Lagrangian submanifold]
\label{slagex}
$\Re (e^{i\theta} \Omega)$. The calibrated manifolds are special Lagrangian. They
satisfy:
\eq{\spl{
\omega|_\Sigma & = 0 \qquad \text{Lagrangian} \, , \\
\Im (e^{i\theta} \Omega)|_\Sigma & = 0 \qquad \text{special} \, .
}}
\end{ex}
Later on, we will introduce the notion of a generalized complex submanifold (see definition \ref{gencomplsubdef}) with
respect to a generalized complex structure. For $\mathcal{J}_J$ associated to an ordinary complex structure (example \ref{complex})
this means just a complex submanifold, while for $\mathcal{J}_\omega$ associated to a symplectic structure (example \ref{symplex}),
this means Lagrangian. Observe the peculiar property that a submanifold that is calibrated with respect to forms in one pure spinor
of the Calabi-Yau manifold, i.e.\ $e^{i\omega}$ or $e^{i\theta} \Omega$, is generalized complex
with respect to the {\em other} pure spinor (in this case respectively complex and Lagrangian).
We find that this continues to hold in the generalized case.

Before we can generalize the concept of calibrations, we need first to discuss generalized submanifolds.

\subsection{Generalized submanifolds}
\label{gensubman}

Generalized submanifolds correspond to D-branes in that they consist of a submanifold $\Sigma$ and a
world-volume two-form field $\mathcal{F}$. $\mathcal{F}$ cannot be just the pull-back of $B$, since
$B$ is only defined up to a gauge transformation. This can be remedied by introducing an extra ingredient
carried by the D-brane, namely a world-volume
gauge field $F$ defined only on $\Sigma$, which satisfies $\d F=0$.\footnote{For $N$ coinciding D-branes
the world-volume field $F$ becomes a non-abelian U($N$) gauge field. In fact, also the D-brane coordinates describing
the embedding become matrix-valued. See e.g.~\cite{myersreview,taylorreview} for reviews. It seems there should still be
a supersymmetry condition, and thus calibration condition, although the analysis becomes very complicated because
of ordering ambiguities and derivative corrections. For the special case of D9-branes see for instance \cite{koerberalpha3,koerberalpha4}, where
the analysis is performed order by order in $\alpha'$ and the calibration condition (actually the D-flatness condition) is an $\alpha'$-correction of the Donaldson-Uhlenbeck-Yau
condition of Yang-Mills \cite{DUY1,DUY2}.} We can define a gauge potential $A$ so
that $F=\d A$. Then $\mathcal{F}=B|_\Sigma + 2 \pi \alpha' F$ can be made invariant under the gauge
transformations of $B$ if we introduce a suitable transformation of $A$:
\eq{\spl{
& B \rightarrow B + \d \Lambda \, , \\
& A \rightarrow A - \frac{\Lambda|_\Sigma}{2 \pi \alpha'} \, .
}}
In what follows we will only work with this gauge-invariant combination, which satisfies
\eq{
\label{fdh}
\d \mathcal{F} = H|_\Sigma \, .
}
Associated to a generalized submanifold consisting of the pair $(\Sigma,\mathcal{F})$
is a subbundle of the generalized tangent bundle of $M$.

\begin{defi}
A generalized tangent bundle $T_{(\Sigma,\mathcal{F})}$ of a generalized submanifold is defined as follows
\eq{\label{gtb}
T_{(\Sigma,\calf)} = \{X+ \xi \in T \Sigma \oplus T^* M|_\Sigma \,\,\, \big| \,\,\, \xi|_\Sigma = \iota_X \calf \}\ .
}
\end{defi}
$T_{(\Sigma,\calf)}$ is a real maximally isotropic subbundle of the restricted bundle $T M\oplus T^* M|_\Sigma$,
which is $H$-integrable iff \eqref{fdh} is satisfied. It is related to the maximally isotropic subbundle $T_{(R,F)}$
defined in eq.~\eqref{genprodtan}, except that it is only defined on $\Sigma$. It is thus a localized version suitable
for describing just one generalized submanifold instead of a whole foliation.

Instead of a pure spinor that is defined on the whole of $M$, associated to it is a
{\em generalized current}, defined as a linear map on the space of differentiable polyforms on $M$.
\begin{defi}
A generalized {\em real} current $j$ can be formally seen as a polyform
(which we indicate with the same symbol $j$)  such that for any smooth polyform $\phi$ we have
\eq{
j(\phi)=\int_M\langle\phi,j\rangle \ .
}
\end{defi}
Just as for polyforms we will only consider currents of definite parity.
We can associate a current $j_{(\Sigma,\calf)}$ to a generalized submanifold $(\Sigma,\calf)$ by requiring that it acts as follows on a general polyform $\phi$
\eq{
\label{sourcedef}
j_{(\Sigma,\calf)}(\phi) = \int_M\langle \phi,j_{(\Sigma,\calf)}\rangle = \int_\Sigma \phi|_\Sigma \wedge e^\calf\ .
}
This definition is inspired by the form of the Chern-Simons-like part of the D-brane action in \eqref{dbraneaction}.

\begin{figure}[t]
\centering
\definecolor{DarkGreen}{rgb}{0,0.5,0}
\psfrag{S1}[l]{\textcolor{DarkGreen}{$(\Sigma,\calf)$}}
\psfrag{S2}[l]{\color{red}$(\Sigma',\calf')$}
\psfrag{Spol}[l]{\color{blue}$(\tilde\Sigma,\tilde{\calf})$}
\psfrag{eq1}[l]{$\textcolor{DarkGreen}{\calf}=\textcolor{blue}{\tilde\calf}|_{\textcolor{DarkGreen}{\Sigma}}$}
\psfrag{eq2}[l]{\hspace{-0.5em}$\textcolor{blue}{\tilde\calf}|_{\textcolor{red}{\Sigma'}}=\textcolor{red}{\calf'}$}
\includegraphics[width=7cm]{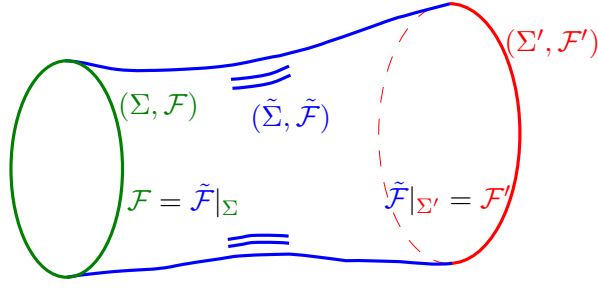}
\caption{$(\Sigma,\calf)$ and $(\Sigma',\calf')$ are in the same generalized homology class iff there exists
an interpolating D-brane $(\tilde{\Sigma},\tilde{\calf})$.}
\label{genhofig}
\end{figure}

We can define $\d_H j_{(\Sigma,\calf)}$ as follows
\eq{
(-1)^d \int_M\langle \phi,\d_H j_{(\Sigma,\calf)}\rangle= \int_M\langle \d_H\phi,j_{(\Sigma,\calf)}\rangle=\int_\Sigma \d_H\phi|_\Sigma \wedge e^\calf=\int_{\partial\Sigma} \phi|_{\partial\Sigma}\wedge e^{\calf|_{\partial\Sigma}}\ ,
}
such that it is consistent with the property \eqref{mukaiprop2} on forms. Note that we used Stokes' theorem and eq.~\eqref{fdh}.
We see that $\d_H j_{(\Sigma,\calf)}=j_{(\partial\Sigma,\calf|_{\partial\Sigma})}$ and, in particular, if $\Sigma$ is a cycle ($\partial \Sigma=0$) then
\eq{
\d_H j_{(\Sigma,\calf)}=0\ ,
\label{currentclosed}
}
and we call $(\Sigma,\calf)$ a {\em generalized cycle}.
Furthermore, we say that two generalized cycles $(\Sigma,\calf)$ and $(\Sigma^\prime,\calf^\prime)$
are in the same generalized homology class if there exists a generalized submanifold $(\tilde\Sigma,\tilde\calf)$ such that $\partial{\tilde\Sigma}=\Sigma^\prime-\Sigma$ with $\tilde\calf|_{\Sigma}=\calf$ and $\tilde\calf|_{\Sigma^\prime}=\calf^\prime $, see figure \ref{genhofig}. It is easy to see that in this case
\eq{
j_{(\Sigma^\prime,\calf^\prime)}-j_{(\Sigma,\calf)}=\d_H j_{({\tilde\Sigma},\tilde\calf)}\ .
}
Thus if $(\Sigma,\calf)\sim(\Sigma^\prime,\calf^\prime)$ in generalized homology,
then $j_{(\Sigma^\prime,\calf^\prime)}\sim j_{(\Sigma,\calf)}$ in $\d_H$-cohomology. It follows that
a generalized homology class $[(\Sigma,\calf)]$ determines a $\d_H$-cohomology class $[j_{(\Sigma,\calf)}]\in H^\bullet_{H}(M)$.

Let us now show that the generalized current is in fact a pure spinor associated to $T_{(\Sigma,\calf)}$. So
we must show that its null space is $T_{(\Sigma,\calf)}$.
Based on \eqref{mukaiprop1} we define
\eq{
(-1)^d \int_M\langle \phi,\mathbb{X}\cdot j_{(\Sigma,\calf)}\rangle = -\int_M\langle \mathbb{X}\cdot\phi,j_{(\Sigma,\calf)}\rangle\ ,
}
and find for a section $\mathbb{X}=X+\xi$ of $T_{(\Sigma,\calf)}$ and for any polyform $\phi$
\eq{
\int_M\langle \mathbb{X}\cdot\phi,j_{(\Sigma,\calf)}\rangle\ = \int_\Sigma \iota_X \left(\phi|_\Sigma \wedge e^\calf\right) = 0 \, .
}
There cannot be any other $\mathbb X$ annihilating $j_{(\Sigma,\calf)}$ because $T_{(\Sigma,\calf)}$ already has the maximal dimension
for a space of annihilators of a spinor. So it follows that the null space of the generalized current is indeed $T_{(\Sigma,\calf)}$.

Roughly for a $k$-cycle the generalized current $j_{(\Sigma,\calf)}$ looks like $e^{-\calf} \wedge \delta^{(d-k)}(\Sigma)$, where
$\delta^{(d-k)}(\Sigma)$ is the ordinary Poincar\'e current dual to the cycle $\Sigma$.

Summarizing, $j_{(\Sigma,\calf)}$ can be thought of as a sort of localized pure spinor associated to $T_{(\Sigma,\calf)}$,
seen as a maximally isotropic subbundle of $T M\oplus T^* M|_\Sigma$. Its definition as the dual current also fixes the overall
factor, and it satisfies a generalized Calabi-Yau condition \`a la Hitchin, eq.~\eqref{currentclosed}. It describes a {\em localized} D-brane.
In contrast, the  pure spinor eq.~\eqref{pureprod} can be though of as describing a foliation of generalized submanifolds, or alternatively, a {\em smeared} D-brane.

\subsection{Generalized calibrations}
\label{sec:gencal}

So after we introduced the technology for describing generalized submanifolds, let us generalize the concept of calibrations \cite{gencal,lucacal}.
Note that the terminology ``generalized calibrations'' was introduced before in \cite{papagutcal1,papagutcal2} to describe calibrations
that took into account the bulk RR-fluxes, but not yet the NSNS three-form and the world-volume gauge field.

\subsubsection*{Definition}

In order to describe D-branes in terms of calibrations we assume they are static and do not carry electric field
strengths. Suppose that apart from being extended in time, they wrap $q$ external space-like directions
in a warped compactification, given by the ansatz~\eqref{metricansatz}-\eqref{formansatz}.
Looking at \eqref{dbraneaction}, we define the energy density
\eq{
\label{braneen}
\mathcal{E}(T,\mathcal{F}) = e^{(q+1)A-\Phi} \sqrt{\det(g|_T + \calf)} - \delta_{q,3} e^{4A} \tilde{C}(T) \, ,
}
where $T$ is an oriented subspace of the tangent space and $\tilde{C}$ the external part of the RR-potentials,
defined by $\d_H (e^{4A} \tilde{C})=e^{4A} \tilde{F}$. We have then for the energy (per unit external volume)
of a static D-brane $(\Sigma,\calf)$
\eq{
E(\Sigma,\calf) = - S_{\text{D}p}/(T_p \text{Vol}_{q+1}) =  \int_{\Sigma} \mathcal{E}(\Sigma,\calf) \, ,
}
where $\mathcal{E}(\Sigma,\calf)$ is given by eq.~\eqref{braneen} with $T=T\Sigma$.
Note that if the D-brane does not completely fill all the external space directions,
it breaks the 4D Poincar\'e symmetry and cannot couple to the RR-fields, which take the
form eq.~\eqref{formansatz}. The last term in \eqref{braneen} then vanishes.

\begin{defi}
\label{gencaldef}
A {\em generalized calibration form} $\tilde{\omega}$ is a polyform on M (with definite parity)
that satisfies
\begin{enumerate}
\item an algebraic condition:
in every point $p\in M$ and for every oriented subspace $T$ of $T_p M$
and two-form $\calf$ on the subspace we have the bound
\eq{
\label{gencalbound}
\mathcal{E}(T,\mathcal{F}) \ge \left(\tilde{\omega} \wedge e^{\mathcal{F}}\right)(T) \, .
}
Furthermore, in every point there must exist a $(T,\calf)$ such
that the above bound is saturated.
\item a differential condition:
\eq{
\label{gencaldif}
\d_H \tilde{\omega}=0 \, .
}
\end{enumerate}
\end{defi}

\begin{defi}
A generalized submanifold $(\Sigma,\calf)$ is calibrated if in every point $p \in \Sigma$ the above bound
is saturated:
\eq{
\label{calcondgen}
\d^l \sigma \, \mathcal{E}(\Sigma,\mathcal{F}) = \tilde{\omega}|_{\Sigma} \wedge e^{\mathcal{F}}\big|_{l} \, .
}
\end{defi}

A generalized calibrated submanifold is now the D-brane with the lowest energy within its
generalized homology class. Indeed, suppose there is another D-brane $(\Sigma',\calf')$ within
the same generalized homology class: this means there is a $(\tilde{\Sigma},\tilde{\calf})$ such
that $\partial \tilde{\Sigma}=\Sigma' - \Sigma$ and $\tilde{\calf}|_\Sigma=\calf$, $\tilde{\calf}|_{\Sigma'}=\calf'$. We find
\eq{
\label{calboundshow}
E(\Sigma',\calf') \ge \int_{\Sigma'} \tilde{\omega}|_{\Sigma'} \wedge e^{\calf'} = \int_{\Sigma} \tilde{\omega}|_{\Sigma} \wedge e^\calf + \int_{\tilde{\Sigma}} \d_H \tilde{\omega}|_{\tilde{\Sigma}} \wedge e^{\tilde{\calf}}= \int_{\Sigma} \tilde{\omega}|_{\Sigma} \wedge e^\calf =E(\Sigma,\calf) \, ,
}
where we used the bound \eqref{gencalbound}, Stokes' theorem, the differential condition~\eqref{gencaldif} and the calibration condition~\eqref{calcondgen}.

For space-filling D-branes, the energy density $\mathcal{E}(T,\mathcal{F})$ and thus also $\tilde{\omega}$ depends on the gauge choice for the RR-potential $C$.
Therefore, in that case one often uses an alternative
definition for the calibration form $\omega$ where $\mathcal{E}(T,\mathcal{F})$ in \eqref{gencalbound} is replaced by
its Dirac-Born-Infeld part only:
\eq{
\label{gencalboundDBI}
\mathcal{E}(T,\mathcal{F})_{\text{DBI}} = e^{4A-\Phi} \sqrt{\det(g|_T + \calf)} \, ,
}
and at the same time the differential condition \eqref{gencaldif} by
\eq{
\label{gencaldifalt}
\d_H \omega= e^{4A} \tilde{F} \, .
}
The condition for a calibrated submanifold \eqref{calcondgen} is then modified to
\eq{
\label{calcondgenalt}
\d^l \sigma \, \mathcal{E}(\Sigma,\mathcal{F})_{\text{DBI}} = \omega|_{\Sigma} \wedge e^{\calf}\big|_{l} \, .
}
We find that if $\tilde{\omega}$ satisfied the original definition, $\omega$ given by
\eq{
\omega = \tilde{\omega} + e^{4A} \tilde{C} \, ,
}
satisfies the new definition.

\subsubsection*{Relation with supersymmetry of D-branes}

We will now describe the relation with the supersymmetry conditions for a D-brane embedding.
This analysis will provide us with a set of natural generalized calibration forms, which exist on
every supersymmetric background. We will be a bit sketchy and refer to \cite{gencal,lucacal} for a detailed
discussion for Minkowski compactifications, or to \cite{DbraneAdS,martucciDW} for more general backgrounds.

The supersymmetry condition for a D-brane was studied in \cite{susyDbranecederwall1,susyDbraneaganagic1,susyDbranecederwall2,susyDbranebergshoeff,susyDbraneaganagic2}.
It was found there that a D-brane embedding does not break the background supersymmetry generated by
\eq{
\epsilon = \left( \begin{array}{c} \epsilon^1\\ \epsilon^2\end{array}\right) \, ,
}
if and only if
\eq{
\label{branekappa}
\epsilon = \Gamma \epsilon \, ,
}
where $\Gamma$ is the $\kappa$-symmetry matrix of the Green-Schwarz description of branes. It satisfies
\eq{
\label{gammaprop}
\Gamma^2 = \bbone \, , \qquad \Gamma^\dagger=\Gamma \, .
}
Note that this makes $\frac{1}{2}(\bbone - \Gamma)$ into a projection operator.
For a D$p$-brane $(\Sigma,\mathcal{F})$ it is given by\footnote{We use the conventions of \cite{lucakappa}.}
\eq{
\Gamma = \left( \begin{array}{cc} \bf{0} & \hat{\Gamma} \\ \hat{\Gamma}^{-1} & \bf{0} \end{array} \right) \, ,
}
with
\eq{
\label{gammaforDbrane}
\hat{\Gamma} =
\frac{1}{\sqrt{|\det(g|_\Sigma + \mathcal{F})|}} \sum_{2n+l=p+1} \frac{1}{n! l! 2^n} \,
\epsilon^{a_1 \ldots a_{2n} b_1 \ldots b_l} \mathcal{F}_{a_1 a_2}
\ldots \mathcal{F}_{a_{2n -1} a_{2n}}  \Gamma_{b_1 \ldots b_l} \, ,
}
where $\Gamma_a = \partial_a y^M \Gamma_M$ are the 10D gamma-matrices pulled back to the D-brane.
Note that the supersymmetry condition can only be satisfied if the norms $||\epsilon^1||^2$
and $||\epsilon^2||^2$ are equal. With the compactification ansatz \eqref{adskilling} this implies $||\eta^{(1)}_+||^2=||\eta^{(2)}_+||^2$.

Using the properties \eqref{gammaprop} of the $\Gamma$-matrix we can immediately derive the following bound\footnote{Note that we use spinor bilinears
that do not transform properly under the full Lorentz group SO(9,1). As we mentioned before our description is only
valid for static D-branes in a static background, where have singled out a time direction. Therefore we only allow for space-like rotations
belonging to SO(9).}
\eq{
\epsilon^\dagger \frac{1}{2} (\bbone-\Gamma) \epsilon = \epsilon^\dagger \left(\frac{1}{2} (\bbone-\Gamma)\right)^2 \epsilon = ||\frac{1}{2} (\bbone -\Gamma) \epsilon ||^2 \ge 0 \, ,
}
or equivalently
\eq{
\epsilon^\dagger \epsilon \ge \epsilon^\dagger \Gamma \epsilon \, .
}
This bound is obviously saturated iff the supersymmetry condition \eqref{branekappa} is satisfied.
Upon expanding $\Gamma$ given by eq.~\eqref{gammaforDbrane} the bound becomes
\eq{
\sqrt{|\det(g|_\Sigma + \mathcal{F})|} \ge \frac{1}{||\epsilon^1||^2} \sum_{2n+l=p+1} \frac{1}{n! l! 2^n} \,
\epsilon^{a_1 \ldots a_{2n} b_1 \ldots b_l} \mathcal{F}_{a_1 a_2} \ldots \mathcal{F}_{a_{2n -1} a_{2n}}  \epsilon^{2\dagger} \gamma_{b_1 \ldots b_l} \epsilon^1 \, ,
}
which immediately leads to the construction of a set of calibration forms as spinor bilinears.

Indeed, plugging in the compactifications ansatz \eqref{adskilling} and using eqs.~\eqref{nps}
we find that on $\mathcal{N}=1$ supersymmetric backgrounds the following polyforms obey
the (alternative) definition of a generalized calibration form, i.e.\
definition \ref{gencaldef} with the replacements \eqref{gencalboundDBI} and for space-filling D-branes \eqref{gencaldifalt}:
\eq{\spl{
& \omega^{\text{sf}} = e^{4A-\Phi} \Re \Psi_1 \qquad \text{space-filling} \, , \\
& \omega^{\text{DW}} = e^{3A-\Phi} \Psi_2 \qquad \text{domain wall} \, , \\
& \omega^{\text{string}} = e^{2A-\Phi} \Im \Psi_1 \qquad \text{string-like} \, .
}}
As indicated, they are suitable for calibrating three types of supersymmetric D-branes, namely
the ones that fill respectively four, three or two of the external dimensions, including time.
In particular, they pick up one factor of $e^A$ for each external direction. It can be shown that in
this setup there are no supersymmetric D-branes that are point-like in the external dimensions \cite{lucacal}.
For a discussion of the calibration form for instantonic D-branes see \cite[Appendix D]{effective}.

The differential conditions \eqref{gencaldifalt} are exactly provided for by the
supersymmetry conditions \eqref{susyconds} of the background, giving a nice interpretation of these
conditions in terms of generalized calibrations.

The reader might wonder whether there is also such a correspondence for the supersymmetry conditions \eqref{susycondsAdS}
for AdS compactifications. In fact, in \cite{DbraneAdS} it was shown that this is indeed the case
although the whole story is a bit more subtle and related to the fact that AdS$_4$-space has a boundary.

Finally we note that this discussion can be extended to general static backgrounds \cite{DbraneAdS} (see
also \cite{lucajarah}).

\subsubsection*{Dissecting the calibration condition}

Let us discuss the calibration condition \eqref{calcondgenalt}
in a bit more detail. For definiteness we focus on space-filling D-branes, which are calibrated by
\eq{
\omega^{\text{sf}}=e^{4A-\Phi} \Re \Psi_1 \, ,
}
i.e. by the real part of the {\em first} pure spinor. It turns out to be natural to split the condition into:
\eq{\spl{
& \d^l \sigma \, \mathcal{E}(\Sigma,\mathcal{F})_{\text{DBI}} = e^{i\alpha} e^{4A-\Phi} \Psi_1|_{\Sigma} \wedge e^{\mathcal{F}}\big|_{l} \qquad \text{F-flatness} \, , \\
\label{Dterm} & \Im \left(\Psi_1|_{\Sigma} \wedge e^{\mathcal{F}}\big|_{l}\right)=0 \qquad \text{D-flatness} \, ,
}}
where $e^{i\alpha}$ is an arbitrary phase that can vary over $M$. It might help to think of this as the same split as in
example \ref{slagex}, namely the split into Lagrangian and ``special''. In \cite{lucasuppot}
it was shown that the first line corresponds to the F-flatness conditions in the
4D effective theory on the D-brane, while the second line is the D-flatness condition.

The F-flatness conditions correspond to the statement that the D-brane $(\Sigma,\calf)$
is a generalized complex submanifold with respect to the generalized complex structure $\mathcal{J}_2$
associated to $\Psi_2$, i.e.\ the {\em other} pure spinor \cite{gencal}. In our $\mathcal{N}=1$ background
this pure spinor is integrable (because of eq.~\eqref{susycond1}), which will make it possible to study the deformation theory (see
section \ref{sec:branedeform}).

\begin{defi}
\label{gencomplsubdef}
$(\Sigma,\calf)$ is a generalized complex submanifold with respect to $\mathcal{J}$ if
the generalized tangent space $T_{(\Sigma,\calf)}$ is stable under $\mathcal{J}$ (i.e.\ the action of $\mathcal{J}$
brings $T_{(\Sigma,\calf)}$ to itself).
\end{defi}

This is very analogous to a U($d/2$)$\times$U($d/2$)-structure, where $[\mathcal{J}_1,\mathcal{J}_2]=0$
implies that $L_1$ is stable under $\mathcal{J}_2$. Remember, that in terms of the pure spinors
we found the equivalent characterization eq.~\eqref{compcondalt} (see also exercise \ref{stabex}). In this case we find
that $T_{(\Sigma,\calf)}$ is stable under $\mathcal{J}_2$ iff
\eq{
\label{gencomplsub}
j_{(\Sigma,\calf)} \in \Gamma(V_0) \, ,
}
where $V_i$ is the filtration \eqref{decomp} associated to $\mathcal{J}_2$.
Alternatively, the condition \eqref{gencomplsub} can be rewritten as
\eq{
\label{Fflatalt}
\langle \mathbb{X} \cdot \Psi_2 , j_{(\Sigma,\calf)} \rangle = 0 \, ,
}
for all vector fields $\mathbb{X}$. Indeed, using eq.~\eqref{mukcomp} this forces $j_{(\Sigma,\calf)}|_{V_{\pm 2}}=0$, which considering
the fixed parity of $j_{(\Sigma,\calf)}$ means it must belong to $\Gamma(V_0)$.

The F-flatness conditions can be shown to follow from varying a superpotential \cite{lucasuppot}:
\eq{
\label{Dbranesuppot}
\mathcal{W}(\Sigma,\calf)-\mathcal{W}(\Sigma_0,\calf_0) = -\frac{1}{2} \int_M \langle e^{3A-\Phi} \Psi_2 , j_{(\tilde{\Sigma},\tilde{\calf})} \rangle \, ,
}
where $(\tilde{\Sigma},\tilde{\calf})$ is any D-brane such that  $\d_H j_{(\tilde{\Sigma},\tilde{\calf})} = j_{(\Sigma,\calf)}-j_{(\Sigma_0,\calf_0)}$
and $(\Sigma_0,\calf_0)$ is a fixed reference D-brane. Very roughly
\eq{
j_{(\tilde{\Sigma},\tilde{\calf})}=(\d_H)^{-1} j_{(\Sigma,\calf)} = -\frac{1}{2\kappa_{10}^2} \hat{F}_0 \, ,
}
where $\hat{F}_0$ is the part of $\hat{F}$ source by $j_{(\Sigma,\calf)}$. In this way the D-brane superpotential \eqref{Dbranesuppot} is part of the bulk
superpotential \eqref{suppot}. Using eq.~\eqref{currentdeform} one indeed obtains the F-flatness conditions \eqref{Fflatalt}
from the variation of the D-brane superpotential under $\mathbb{X}$.

\begin{ex}
Consider the warped Calabi-Yau geometry with fluxes of example \ref{CYex}
and a space-filling D-brane $(\Sigma,\calf)$ calibrated by $e^{4A-\Phi} \Re \Psi_1= e^{4A-\Phi} \Re e^{i\omega}$ (putting $e^{i\vartheta}=1$).
Compare this with example \ref{compex}. The F-flatness condition is that $(\Sigma,\calf)$ be a generalized complex submanifold
with respect to $\mathcal{J}_2$, which in this case is an ordinary complex structure (see example \ref{complex}).
One easily checks that this implies that $\Sigma$ is a complex submanifold and $\calf$ is of type $(1,1)$.
The D-flatness condition,
\eq{
\Im e^{i\omega|_\Sigma + \calf}|_{l} = 0 \, ,
}
is always satisfied for D3-branes ($l=0$) while for D7-branes ($l=4$) it
reduces to
\eq{
\omega|_\Sigma \wedge \calf = 0 \, .
}
For D5-branes it turns out to be impossible to satisfy the D-flatness conditions. Indeed, such a warped Calabi-Yau background does not
allow for supersymmetric D5-branes.
\end{ex}

For D-branes on AdS$_4$ compactifications we find that the F-flatness conditions imply the D-flatness condition (which as we mentioned
at the end of section \ref{sec:effective} is to be expected on general grounds in $\mathcal{N}=1$ supergravity theories
with non-zero vacuum expectation value for the superpotential). This can be shown as follows: from the F-flatness condition \eqref{Fflatalt}
we find:
\eq{
\int_M \langle \mathbb{X} \cdot e^{3A-\Phi}\Psi_2 , j_{(\Sigma,\calf)} \rangle = 0 \, .
}
Taking $\mathbb{X} = \d \lambda$ this becomes
\begin{multline}
0 = \int_M \langle \d \lambda \wedge e^{3A-\Phi}\Psi_2 , j_{(\Sigma,\calf)} \rangle
= - \int_M \lambda \langle  \d_H\left( e^{3A-\Phi}\Psi_2\right) , j_{(\Sigma,\calf)} \rangle \\
= - 2i W_0 \int_M \lambda \langle  e^{2A-\Phi} \Im \Psi_1 , j_{(\Sigma,\calf)} \rangle
= - 2i W_0 \int_\Sigma \lambda \, e^{2A-\Phi} \Im \Psi_1|_\Sigma \wedge e^\calf \, ,
\end{multline}
where we used eqs.~\eqref{currentclosed}, \eqref{susycond1AdS} and \eqref{sourcedef}. Since this applies for an arbitrary function
$\lambda$, the D-term condition \eqref{Dterm} follows.

\begin{ex}[Special Lagrangian D6-brane on a L\"ust-Tsimpis background]
Consider a L\"ust-Tsimpis background satisfying eqs.~(\ref{geomAdS4})-(\ref{fluxAdS4}) and eq.~\eqref{SU3IIAbound}.
For a space-filling D6-brane wrapping an internal three-cycle $(\Sigma,\calf)$ the generalized tangent bundle $T_{(\Sigma,\calf)}$ is
stable under the action of $\mathcal{J}_2$ (taking the form $\mathcal{J}_{\omega}$ of example \ref{symplex}) iff
\eq{
\omega|_\Sigma = 0 \, , \qquad \mathcal{F}=0 \, ,
}
so that the cycle is Lagrangian. In this case, the condition eq.~\eqref{geomAdS4a} implies
that
\eq{
\Re \Omega|_\Sigma = 0 \, ,
}
so that the cycle is also automatically special Lagrangian and calibrates $\Re \Psi_1=- \Im \Omega$.
That a Lagrangian cycle is automatically special in such a geometry was first noticed in \cite{ivanpapa}.
For an example of such D6-branes playing a role in the AdS$_4$/CFT$_3$ duality see \cite{slagCP3,slagCP32,slagCP33}.
\end{ex}

\begin{ex}[Coisotropic D8-brane]
\label{ex:coisotropic}
Let us now consider a supersymmetric space-filling D8-brane wrapping an internal five-cycle in a L\"ust-Tsimpis background.
From the condition that the generalized tangent bundle $T_{(\Sigma,\calf)}$
be stable under the action of $\mathcal{J}_2 = \mathcal{J}_{\omega}$ we find this time that the D-brane
must be of so-called {\em coisotropic} type \cite{coisotropic}. Coisotropic means that\footnote{We note that $T\Sigma$ is isotropic
for $\omega$, i.e.\ $\omega(X,Y)=0$ for all vectors $X,Y \in T \Sigma$, iff $\omega: T \Sigma \rightarrow \text{Ann} \, T \Sigma$. If $\Sigma$ is both isotropic and coisotropic then it is Lagrangian.}
\eq{
\omega^{-1}: \text{Ann} \, T \Sigma \rightarrow T\Sigma \, ,
}
which is indeed one of the conditions that follow from the stability of $T_{(\Sigma,\calf)}$. The stability of $T_{(\Sigma,\calf)}$
under $\mathcal{J}_2$ can however more concisely be expressed using the alternative characterization \eqref{Fflatalt}, from
which follows immediately \cite{coisotropicmarchesano}
\eq{
\left( i \omega|_{\Sigma} + \mathcal{F} \right)^2 = 0 \, .
}
Because of \eqref{geomAdS4a} and \eqref{fluxHAdS4} it implies again automatically the D-flatness condition
\eq{
\Re \Omega|_\Sigma \wedge \mathcal{F} = 0 \, .
}
When constructing solutions to these conditions, one must also keep in mind that we must satisfy eq.~\eqref{fdh}.
For examples of such calibrated coisotropic D8-branes on the family of $\mathcal{N}=1$ supersymmetric backgrounds on $\mathbb{CP}^3$
see \cite{coisotropicCP3}.
\end{ex}

\subsection{The backreaction of generalized calibrated sources}

As one can read off from \eqref{dbraneaction} the localized action for a D-brane consists of
two parts: the Dirac-Born-Infeld term and the Chern-Simons term. The latter part leads to a contribution
to the Bianchi identities as in eq.~\eqref{bianchicomp} and the former part to a contribution to
the Einstein equation and the dilaton equation of motion. Something similar happens for O-planes, for which the source terms
can be found from \eqref{oplaneaction}. Now, we have claimed in theorem \ref{theor:integr} that the
supersymmetry conditions together with the Bianchi identities imply the Einstein and dilaton equations of motion.
The question is whether this also holds in the presence of backreacting sources. The source terms
would then have to appear in a precisely related way in the Bianchi identities on the one hand and in the Einstein and dilaton
equation of motion on the other hand. It was shown in \cite{integr} that this is exactly what happens if the sources
are generalized calibrated and thus supersymmetric.

As an example of the effect of the backreaction, let us present an alternative proof of the no-go theorem \ref{nogomalnun},
which only works for the specialized case of supersymmetric
backgrounds \cite{granascan}. Suppose the sources are given by $j_{\text{tot}}=\sum_{\text{sources}} T_p j_{(\Sigma,\calf)}$ where all $T_p > 0$, which
means all the sources are D-branes. It follows that
\begin{multline}
\label{nogo}
0 \le \int_M \langle e^{4A-\Phi} \Re \Psi_1, \sum_{\text{sources}} T_p \, j_{(\Sigma,\calf)} \rangle
= -\frac{1}{2 \kappa_{10}^2} \int_M \langle e^{4A-\Phi} \Re \Psi_1, d_H \hat{F} \rangle \\
= -\frac{1}{2 \kappa_{10}^2} \int_M \langle d_H\left(e^{4A-\Phi} \Re \Psi_1 \right), \hat{F} \rangle
= -\frac{e^{4A}}{2 \kappa_{10}^2} \int_M \langle \tilde{F}, \hat{F} \rangle
= -\frac{e^{4A}}{2 \kappa_{10}^2} \int_M \hat{F} \cdot \hat{F} < 0 \, ,
\end{multline}
In the above we used the calibration condition \eqref{calcondgenalt} for each separate source
\eq{
\int_M \langle e^{4A-\Phi} \Re \Psi_1,  j_{(\Sigma,\calf)} \rangle = \int_\Sigma e^{4A-\Phi} \Re \Psi_1 \wedge e^{\calf} =\int_\Sigma \d^l \sigma \, \mathcal{E}(\Sigma,\mathcal{F})_{\text{DBI}} > 0 \, .
}
Furthermore we used the Bianchi identity \eqref{bianchicomp} in the first line. Next, we used \eqref{mukaiprop2}, \eqref{susycond0}
and \eqref{dualin6} in the second line, and found a contradiction. It follows that at least one $T_p <0$, so that we must have at least one orientifold.

\subsection{Deformation theory}
\label{sec:branedeform}

The deformation theory is similar to the deformation theory of ordinary calibrations \cite{mclean}
and was developed in \cite{deforms}. We will be very brief here and refer to the latter paper for more details.

Let us consider infinitesimal deformations of a generalized cycle $(\Sigma,\calf)$.
It turns out that they are described by the sections of the {\em generalized normal bundle}
defined as $\caln_{(\Sigma,\calf)}\equiv (T M\oplus T^* M)|_\Sigma/T_{(\Sigma,\calf)}$.
We indicate the sections of $\caln_{(\Sigma,\calf)}$ with $[\mathbb{X}]$ or equivalently,
in expressions invariant under shifts by elements of $T_{(\Sigma,\calf)}$, with representative sections $\mathbb{X}$ of $T M\oplus T^* M|_\Sigma$.
In fact, a section $[\mathbb{X}]=[N+\xi]$ of the generalized normal bundle defines
a deformation of the cycle $(\Sigma,\calf)$ which consists of two parts.
First of all it shifts $\Sigma$ defined by $y(\sigma)$ to $\Sigma'$
defined by $y(\sigma)+N(y(\sigma))$. This also acts
on $\calf$ as $\delta_N\calf=\iota_N H|_\Sigma$, see \cite{lucasuppot}.
Secondly it deforms the field-strength $\calf$ by $\delta_\xi\calf=d\xi|_\Sigma$.
In this way one can easily see that the elements of $T_{(\Sigma,\calf)}$
correspond to world-volume diffeomorphisms and must indeed be quotiented out.

In \cite{deforms} it was shown that the action of such a deformation
on the current is given by
\eq{
\label{currentdeform}
\delta_{[\mathbb{X}]}j_{(\Sigma,\calf)} = - \mathcal{L}_{\mathbb{X}} j_{(\Sigma,\calf)} = -d_H(\mathbb{X}\cdot j_{(\Sigma,\calf)}) \, ,
}
where we used eq.~\eqref{currentclosed}.

Now, from eq.~\eqref{gencomplsub} and eq.~\eqref{currentdeform} follows that this
deformation transforms a generalized complex submanifold into a generalized
complex submanifold iff
\eq{
d_H(\mathbb{X}\cdot j_{(\Sigma,\calf)}) \in \Gamma(V_0)  \, .
}
This can be rewritten as the following condition
\eq{
\label{deformcomplsub}
\partial_H(\mathbb{X}^{0,1} \cdot j_{(\Sigma,\calf)}) = 0  \, .
}
We see that two isotropic subbundles play a role in this condition:
namely $L$ associated to $\mathcal{J}_2$ and $T_{(\Sigma,\calf)}$
associated to $j_{(\Sigma,\calf)}$. $\mathbb{X}^{0,1}$ is defined
as the projection of $\mathbb{X}$ to the intersection of both, $\mathbb{X}^{0,1} \in \Gamma(L_{(\Sigma,\calf)})$
with
\eq{
L_{(\Sigma,\calf)} = L|_\Sigma \cap (T_{(\Sigma,\calf)} \otimes \mathbb{C}) \, .
}
It is not so difficult to realize (the proof is in \cite{deforms}) that the
condition \eqref{deformcomplsub} is equivalent
to
\eq{
\d_{L_{(\Sigma,\calf)}} \mathbb{X}^{0,1} = 0 \, ,
}
where $\d_{L_{(\Sigma,\calf)}}$ is the Lie algebroid derivative (see definition \ref{liealgebroidderdef}) and we should consider $\mathbb{X}^{0,1}$ as an element of $T^*_{(\Sigma,\calf)}$
using the fact that the natural metric $\mathcal{I}$
induces an isomorphism $\caln_{(\Sigma,\calf)} \rightarrow T^*_{(\Sigma,\calf)}$.

Studying the deformations of the D-flatness condition is somewhat
more difficult since it depends also on the pure spinor $\Psi_1$, and thus
indirectly on the generalized metric. In fact, through the generalized metric
on the manifold it is possible to define a natural metric $G$ on the forms
$\Lambda^\bullet L_{(\Sigma,\calf)}$. This allows to construct a Lie algebroid
codifferential
\eq{
\d^\dagger_{L_{(\Sigma,\calf)}}: \Gamma(\Lambda^k L^*_{(\Sigma,\calf)}) \rightarrow \Gamma(\Lambda^{k-1} L^*_{(\Sigma,\calf)}) \, ,
}
in the standard way, i.e. such that
\eq{
G(\d^\dagger_{(L_{(\Sigma,\calf)})} \alpha,\bar{\beta})=G(\alpha,\overline{\d_{(L_{(\Sigma,\calf)})}\beta}) \, .
}
The deformations that preserve the D-flatness conditions turn out to satisfy:
\eq{
\label{deformDflatness}
\d^\dagger_{L_{(\Sigma,\calf)}} \mathbb{X}^{0,1} = 0 \, .
}
The condition \eqref{deformDflatness} itself can be quite complicated since it depends on the generalized
metric and thus on the non-integrable pure spinor. However, while it is therefore difficult to find the exact form
of the deformations that preserve the generalized calibration condition, it still follows that they are classified
by the cohomology
\eq{
H^1(L_{(\Sigma,\calf)}) \, ,
}
since in every such cohomology class there is exactly one such harmonic
deformation satisfying \eqref{deformDflatness}. In \cite{kapustinlibranes} it was shown that the deformations
of topological D-branes are also classified by this cohomology.

\subsection{Exercises}
\begin{exc}[intermediate]
\label{excal}
Prove the statements in examples \ref{compex} and \ref{slagex}.
\end{exc}
\begin{exc}[hard]
Prove that if a generalized submanifold is calibrated with respect to $e^{4A-\Phi} \Re \Psi_1$
the F-flatness condition implies that it is a generalized complex submanifold with respect to $\mathcal{J}_2$. Note that the
results of exercise \ref{excal} can be seen as a special case. The solution can be found in \cite{gencal}.
\end{exc}
\begin{exc}[easy]
Work out the conditions for stability of $T_{(\Sigma,\calf)}$ under $\mathcal{J}_2$ for the coisotropic
D-branes of example \ref{ex:coisotropic}. For the solution see \cite[Example 7.8]{gualtieri}.
\end{exc}

\clearpage

\thispagestyle{beginsection}
\section{Outlook}

In this review we have introduced the basic concepts of Generalized Complex Geometry and
established the relation with the supersymmetry conditions for both supergravity backgrounds
and for supersymmetric D-branes in these backgrounds. We have also touched upon the more
advanced topic of deformations in each case putting the full technology into use.
To conclude let us discuss some open problems and prospects for the future.

First, the construction of flux compactifications of supergravity is in general quite complicated.
This is to a large extent because the no-go theorem of Maldacena-N\'u\~{n}ez requires the presence of orientifold-planes, which
complicates the analysis. Moreover, when we want to interpret these as string theory compactifications we have to take
care that we work in a regime where the supergravity analysis is valid. From the point of view of Generalized Complex Geometry,
backgrounds with dynamic SU(3)$\times$SU(3) are the most interesting as they require the full power of the formalism.
Partly because of the no-go theorem, examples of such backgrounds are very difficult to construct. Another reason is that so
far mathematicians are focusing on geometries that have a cleaner mathematical definition,
like generalized K\"ahler geometry, instead of the full-blown supergravity conditions.
Moreover, there is an obvious lack of solution-generating techniques as compared to case of Calabi-Yau manifolds,
where we can use techniques of algebraic geometry. The construction of such an algebraic geometry description for Generalized Complex Geometry
is a major open problem. On a positive note, as we discussed, for AdS compactifications there are by now some
non-trivial examples of dynamic SU(3)$\times$SU(3)-structures. Interestingly, they were first found though their Chern-Simons-matter
CFT-dual.

It was found that Generalized Complex Geometry can also play a role in the study of supersymmetry breaking \cite{gennonsusy,gennonsusy2} (see also \cite{grananonsusy}).
Techniques of $G$-structures and Generalized Complex Geometry have furthermore been used in the construction of non-supersymmetric AdS vacua
(see e.g.~\cite{cassred,tomasiellomassive,lusttsimpisSU3SU31,mynonsusy}), and, more interestingly from a cosmological viewpoint, dS vacua
\cite{dShaque1,mydSitter1,flaugerdS1,dShaque2,mydSitter2,wrasedS,petrinidS}.

A second open problem is the construction of the low-energy 4D effective supergravity theory that corresponds to a certain
compactification. We briefly touched upon this subject in section \ref{sec:effective}. The difficult lies here in a suitable choice of
a set of low-energy modes that survive the reduction. On general grounds one expects the low-energy theory to be a gauged supergravity.
The challenge is then to match all the possible gaugings with the 10D geometric and form field fluxes (see e.g.~\cite{granaN2part2,trigiante,samtlebenreview}).
For a complete matching it turns out that also non-geometric backgrounds \cite{hullTfold1,hmcw,stw1,stw2,hullTfold2,trigiante2} must be taken into account.

\clearpage

\thispagestyle{beginsection}
\section*{Acknowledgements}
I would like to thank the organizers of the Modave School for the opportunity to give these lectures,
and also Jeong-Hyuck Park and the Center for Quantum Space-Time at Sogang University in Seoul, where the first
version of these lectures was presented. Further thanks goes to Davide Cassani, Claudio Caviezel,
Simon K\"ors, Hwasung Mars Lee, Luca Martucci, Michela Petrini, Dimitrios Tsimpis and Thomas Van Riet for discussions and comments.

The author is a Postdoctoral Fellow of the FWO -- Vlaanderen. His work is further supported in part by the FWO -- Vlaanderen
project G.0235.05 and in part by the Federal Office for Scientific, Technical and Cultural Affairs through the
`Interuniversity Attraction Poles Programme Belgian Science Policy'
P6/11-P.

\vspace{0.9cm}





\clearpage
\section*{Appendix}

\setcounter{section}{0}
\renewcommand{\thesection}{\Alph{section}}

\thispagestyle{beginsection}
\section{Conventions}

The components of an $l$-form $\omega$ are defined by the expansion
\eq{
\omega=\frac{1}{l!} \, \omega_{i_1\ldots i_l}\d y^{i_1}\wedge\cdots\wedge\d y^{i_l}\ .
}
Sometimes, for instance in our expressions for the supersymmetry variation \eqref{susyvar}, we will separate one index
as follows
\eq{
\omega_j= \iota_{\!\frac{\partial\,}{\partial y^j}} \omega = \frac{1}{(l-1)!} \, \omega_{ji_2\ldots i_{l}}\d y^{i_2}\wedge\cdots\wedge\d y^{i_{l}}\ .
}
The contraction of $\omega$ with gamma-matrices is indicated as follows:
\eq{
\slashchar{\omega} = \frac{1}{l!} \, \omega_{i_1 \ldots i_l} \gamma^{i_1 \ldots i_l} =
\frac{1}{l!} \, \omega_{i_1 \ldots i_l} e^{i_1}{}_{a_1} \cdots e^{i_l}{}_{a_l} \gamma^{a_1 \ldots a_l} \, .
}
We often use the operator $\sigma$, reversing the order of indices, i.e.\
\eq{
\sigma(\omega)= \frac{1}{l!} \, \omega_{i_1\ldots i_l}\d y^{i_l}\wedge \cdots\wedge \d y^{i_1}\ .
\label{reversal}
}
Given a submanifold $\Sigma$, described by a parametrization $\sigma \rightarrow y^i(\sigma)$, the pullback is given by
\eq{
\label{pullback}
\omega|_\Sigma = \frac{1}{l!} \, \omega_{i_1\ldots i_l} \, \frac{\partial y^{i_1}}{\,\partial \sigma^{a_1}} \cdots \frac{\partial y^{i_l}}{\,\partial \sigma^{a_l}} \, \d \sigma^{a_1} \wedge\cdots\wedge \d \sigma^{a_l} \, .
}
All these operations can be trivially extended to polyforms, which are sums of forms of different dimensions. For a polyform $\omega$
we will indicate the projection on the $l$-form part by
\eq{
\omega|_l \, : \quad \text{projection $l$-form part} \, ,
}
and the projection on the top-form part, i.e.\ the $d$-form part where $d$ is the dimension of $M$, by
\eq{
\omega|_{\text{top}} = \omega|_d
}
The Mukai pairing is then defined by
\eq{
\langle\omega,\chi\rangle = \omega\wedge\sigma({\chi})|_{\text{top}} \, ,
\label{mukai}
}
for any pair of polyforms $\omega$ and $\chi$.

\noindent The Hodge-star operator in $d$ dimensions $*_{d}$ is defined as
\eq{\label{hodgestar}
*_{d}\,\omega = \frac{1}{l!(d-l)!} \sqrt{|g|} \, \epsilon_{i_1\ldots i_{d}}\omega^{i_{d-l+1}\ldots i_{d}} \d y^{i_1}\wedge \cdots\wedge \d y^{i_{d-l}} \, .
}
Furthermore we define the inner product on forms as follows
\eq{
\omega\cdot\chi = \frac{1}{l!}\,\omega_{i_1\ldots i_l}\chi^{i_1\ldots i_l}
\label{cdot}\ .
}
where the indices are raised with the inverse of the metric. If $\omega$ and $\chi$ are polyforms, then
\eq{
\omega\cdot\chi\, =\sum_l\omega_l\cdot\chi_l\ .
\label{cdot2}
}

\clearpage

\thispagestyle{beginsection}
\section{Type II supergravity}
\label{sugra}

In this review we used Generalized Complex Geometry to study type II supergravity \cite{schwarzIIB,schwarzwestIIB,howewestIIB,romansIIA}.
In this appendix we will briefly describe this theory, which is also the low-energy limit of type II string theory.

Type II supergravity lives in ten dimensions
and has 32 supersymmetries generated by {\em two} 16-dimensional Majorana-Weyl spinors --- hence
the name {\em type II}. It comes in two flavours, depending on whether
these two supersymmetry generators $\epsilon^{1,2}$ have the same chirality, type IIB,
or opposite chirality, type IIA. The bosonic content common to both type IIA and type IIB consists of a metric $g$, a scalar $\Phi$ --- called the {\em dilaton} -- and a three-form $H$.
This sector is called the NSNS-sector, since these field originate
from states in the string theory that obey Neveu-Schwarz-Neveu-Schwarz (NSNS) boundary conditions.
Away from NS5-branes, the $H$-field satisfies the Bianchi identity
\eq{
\d H = 0\ .
}
This allows for the introduction of the nilpotent $H$-twisted exterior derivative acting on polyforms
\eq{
\d_H = \d +H\wedge\ \, , \quad \text{with} \quad \d_H^2 = 0 \, .
}
Furthermore, there are Ramond-Ramond(RR)-fields, which are form-fields
$F_{n}$, with $n=0,2,4$ in type IIA and $n=1,3,5$ in type IIB. Actually, for type IIA, $F_{0}$ does not have propagating degrees of
freedom and corresponds to a constant $m=F_0$, called the Romans mass \cite{romansIIA}. For type IIB, $F_{5}$ is not completely free, but must satisfy
a self-duality condition. In fact, we will use the democratic formalism of \cite{democratic}, where the number
of RR-fields is doubled, so that $n$ runs over $0,2,4,6,8,10$ in IIA and over $1,3,5,7,9$ in type IIB.
The redundancy is then compensated by introducing duality conditions for {\em all} the RR-fields.
These duality conditions read
\eq{
\label{Fduality}
F_{n} = (-1)^{\frac{(n-1)(n-2)}{2}} *_{10} F_{10-n} \, ,
}
and should be imposed by hand after the equations of motion are derived from the action. Since the duality conditions do not follow
from the action, the action is only a {\em pseudo-action}. We will often assemble the different RR-forms into one polyform $F=\sum_n F_{n}$.
The duality condition can then be concisely written as
\eq{
\label{Fduality2}
F = *_{10} \, \sigma(F) \, .
}
We have also doubled the RR-potentials, and collectively denote them by $C=\sum_n C_{n-1}$. The RR-fields
satisfy Bianchi identities that are twisted with the NSNS three-form $H$ and as such the relation to the potentials is also twisted
by $H$. In polyform notation, we have
\eq{
F=\d_H C \, ,
}
for type IIB and
\eq{
F=\d_H C + m e^{-B} \, ,
}
for type IIA. The reader can check that with these expressions
the Bianchi identities \eqref{bianchis} (away from the sources) are automatically satisfied.

The fermionic content consists
of a doublet of gravitino's $\psi_{M}$ and a doublet of dilatino's $\lambda$.
The components of the doublet are of different chirality
in type IIA and of the same chirality in type IIB.
The supersymmetry variation of the gravitino and dilatino doublet are then given by
\subeq{\label{susyvar}
\al{
\delta\psi^{1}_M & = (D_M \epsilon)^{1} \equiv \left(\nabla_M+\frac14\slashchar{H_M}\right)\epsilon^{1}+\frac{1}{16}e^\Phi
\slashchar{F}\,\Gamma_M\Gamma_{(10)}\epsilon^2\ , \\
\delta\psi^{2}_M & = (D_M \epsilon)^{2} \equiv \left(\nabla_M-\frac14\slashchar{H_M}\right)\epsilon^{2}-\frac{1}{16}e^\Phi
\sigma(\slashchar{F})\,\Gamma_M\Gamma_{(10)}\epsilon^1\ , \\
\delta\lambda^{1} & = \left(\slashchar{\partial}\Phi+\frac12\slashchar{H}\right)\epsilon^1+\frac{1}{16}e^{\Phi}\Gamma^M\slashchar{F}\,\Gamma_M\Gamma_{(10)}\epsilon^2\ , \\
\delta\lambda^{2} & = \left(\slashchar{\partial}\Phi-\frac12\slashchar{H}\right)\epsilon^2-\frac{1}{16}e^{\Phi}\Gamma^M\sigma(\slashchar{F})\,\Gamma_M\Gamma_{(10)}\epsilon^1\ ,
}}
where $\Gamma_M$ are the 10D gamma-matrices, and $\Gamma_{(10)}$ is the 10D chirality operator.
Notice also that one has the following modified dilatino equations, which do not contain the RR-fields
\subeq{\label{modified}\al{
\Gamma^M\delta\psi^{1}_M-\delta\lambda^{1} & = \left(\slashchar{\nabla}-\slashchar{\partial}\Phi+\frac14\slashchar{H}\right)\epsilon^1\ , \\
\Gamma^M\delta\psi^{2}_M-\delta\lambda^{2} & = \left(\slashchar{\nabla}-\slashchar{\partial}\Phi-\frac14\slashchar{H}\right)\epsilon^2\ .
}}

The pseudo-action of the democratic formalism is given by
\eq{\label{sfaction}
S = \frac{1}{2\kappa^2_{10}}\int\d^{10}x\sqrt{-g}\Big\{e^{-2\Phi} \big[R+4 \,\d\Phi \cdot \d\Phi -\frac12 H \cdot H] -\frac14 F \cdot F \Big\}+S_{\text{loc}} \, ,
}
where $S_{\text{loc}}$ is the action associated to localized sources, which in type II string theory might be D-branes, orientifolds, NS5-branes, fundamental
strings and KK-monopoles.
Note that, as opposed to the supergravity actions in the normal formalism, which one can find e.g.\ in \cite{polchinskiII},
this action does not contain Chern-Simons terms. Nevertheless, the equations of motion are, upon manually imposing the duality constraints, equivalent
to the equations of motion in the normal formalism. Since the scalar curvature $R$ is multiplied by $e^{-\Phi}$, this action does not have a standard
Einstein-Hilbert term. However, redefining the metric as follows
\eq{
g = e^{\Phi/2} g_{\E} \, ,
}
one obtains a standard Einstein-Hilbert term. This frame is then called the 10D {\em Einstein frame}, while the original action \eqref{sfaction} is in
the {\em string frame}. In this review, we will always use the string frame, while upon compactification to four dimensions, people
often use the {\em 4D} Einstein frame (where the effective 4D action has standard Einstein-Hilbert term).

As for the sources contributing to $S_{\text{loc}}$ we will only consider D-branes and orientifold planes.
The action for a single D-brane wrapping $\Sigma$ and supporting world-volume gauge field $\mathcal{F}$,
and a single orientifold source wrapping $\Sigma$ are respectively\footnote{The integrand of the Chern-Simons
term must satisfy: $\d (\text{integrand}) = F \wedge e^{\mathcal{F}}$. For type IIA with non-zero Romans mass, there is then an extra
contribution to the Chern-Simons term, which is more subtle \cite{hullmCS}. This does not affect the analysis in this review since the Bianchi identities
\eqref{bianchis} are correct.}
\subeq{\al{
\label{dbraneaction}
S_{\text{D}p} & = - T_p \int_\Sigma e^{-\Phi} \sqrt{g|_\Sigma+\mathcal{F}} + T_p \int_\Sigma C|_\Sigma \wedge e^\mathcal{F} \, , \\
\label{oplaneaction}
S_{\text{O}p} & = - T_{\text{O}p} \int_\Sigma e^{-\Phi} \sqrt{g|_\Sigma} + T_{\text{O}p} \int_\Sigma C|_\Sigma \, ,
}}
with $T_{\text{O}p} = -2^{p-5} T_p$.

The equations of motion and Bianchi identities for the RR-fields are then
\subeq{
\label{eomB}
\al{
\d_{-H} *_{10} F & = 2 \kappa_{10}^2 \, \sigma(j_{\text{total}}) \, , \\
\label{bianchis} \d_{H} F & = - 2 \kappa_{10}^2 \,  j_{\text{total}} \, ,
}}
and the equation of motion for $H$
\eq{
\label{eomH}
\d (e^{-2\Phi} *_{10} \! H) - \frac{1}{2} \sum_n *_{10} F_{(n)} \wedge F_{(n-2)} -  2 \kappa_{10}^2 \frac{\delta S_{\text{loc}}}{\delta B}= 0 \, ,
}
where $\frac{\delta S_{\text{loc}}}{\delta B}$ is a term depending on the D-brane sources described in \cite{integr}.
We will not need the Einstein and dilaton equations of motion in this review, but present them here for completeness:
\subeq{\al{
\text{dilaton e.o.m.} & : \qquad \nabla^2\Phi- \d\Phi \cdot \d\Phi +\frac{1}{4} R-\frac{1}{8} H\cdot H-\frac14\frac{\kappa^2_{10}e^{2\Phi}}{\sqrt{-g}}\frac{\delta S_{\rm loc}}{\delta\Phi}\, =\, 0 \, , \\
\text{modified Einstein} & : \qquad R_{MN}+2\nabla_{M}\nabla_{N}\Phi-\frac{1}{2} H_M\cdot H_N-\frac{1}{4}e^{2\Phi} F_M\cdot F_N \nonumber \\
& \qquad -\kappa^2_{10}e^{2\Phi}\Big( T_{MN \,\text{loc}}  +\frac{g_{MN}}{2\sqrt{-g}}\frac{\delta S_{\text{loc}}}{\delta \Phi}\Big)\, =\, 0 \, ,
}}
where $T_{MN \, \text{loc}}$ is the energy-momentum tensor of the sources
\eq{
T_{MN \, \text{loc}}  =  -\frac{2}{{\sqrt{-\det g}}}\frac{\delta S_{\rm loc}}{\delta g^{MN}} \, ,
}
and the modified Einstein equation is a simplifying linear combination of the Einstein equation, its trace and the dilaton equation of motion \cite{gennonsusy}.

\cleardoublepage

\bibliographystyle{JHEP}

\addcontentsline{toc}{section}{\sffamily Bibliography}

\lhead[\fancyplain{}{}]{\fancyplain{}{}}

\rhead[\fancyplain{}{}]{\fancyplain{}{}}

\chead{\fancyplain{}{\bf BIBLIOGRAPHY}}


\bibliography{gengeomreviewfinal}

\end{document}